\documentclass[prc,twocolumn,showpacs,showkeys,preprintnumbers,superscriptaddress,amsmath,amssymb,nofootinbib]{revtex4}

\usepackage[dvips,xdvi]{graphicx,psfrag}
\usepackage{dcolumn}
\usepackage{bm}
\usepackage{amsmath, amssymb}

\newcommand{\Slash}[1]{\ooalign{\hfil/\hfil\crcr$#1$}}

\def\be{\begin{equation}}
\def\ee{\end{equation}}
\def\bc{\begin{center}}
\def\ec{\end{center}}
\def\dsp{\displaystyle}
\def\trace{\text{tr}}

\def\LamFOF{\Lambda \text{(1405)}}
\def\FE{F_{\text{E}}}
\def\FM{F_{\text{M}}}
\def\FB{F_{\text{B}}}
\def\FS{F_{\text{S}}}
\def\Feff{F^{\text{eff}}}
\def\FEeff{F_{\text{E}}^{\text{eff}}}
\def\FMeff{F_{\text{M}}^{\text{eff}}}
\def\Rho{\text{P}}
\def\PE{\Rho _{\text{E}}}
\def\PM{\text{P}_{\text{M}}}
\def\PB{\text{P}_{\text{B}}}
\def\PS{\text{P}_{\text{S}}}
\def\PEeff{\text{P}_{\text{E}}^{\text{eff}}}
\def\PMeff{\text{P}_{\text{M}}^{\text{eff}}}
\def\MSR{\langle r^{2} \rangle}

\def\EMSR{\langle r^{2} \rangle _{\text{E}}}
\def\MMSR{\langle r^{2} \rangle _{\text{M}}}
\def\BMSR{\langle r^{2} \rangle _{\text{B}}}
\def\SMSR{\langle r^{2} \rangle _{\text{S}}}

\def\MSD{\langle x^{2} \rangle}
\def\FCFF{F_{\text{CFF}}}

\def\KbarN{\bar{K} N}
\def\Kmp{K^{-} p}

\def\mev{\text{ MeV}}
\def\gev{\text{ GeV}}
\def\fm{\text{ fm}}

\usepackage[normalem]{ulem}  
\usepackage[dvips]{color} 

\renewcommand\sout{\bgroup \color{red} \ULdepth=-.5ex \ULset}

\begin{document}

\preprint{KUNS-2317, YITP-10-99}

\title{Internal structure of resonant $\bm{\LamFOF}$ state in chiral dynamics}

\author{Takayasu Sekihara}
\affiliation{Department of Physics, Kyoto University, Kyoto 606-8502, Japan}
\affiliation{Yukawa Institute for Theoretical Physics, 
  Kyoto University, Kyoto 606-8502, Japan}

\author{Tetsuo Hyodo}
\affiliation{Department of Physics, Tokyo Institute of Technology, 
  Tokyo 152-8551, Japan}

\author{Daisuke Jido}
\affiliation{Yukawa Institute for Theoretical Physics, 
  Kyoto University, Kyoto 606-8502, Japan}

\date{\today}

\begin{abstract}
  The internal structure of the resonant $\LamFOF$ state is
  investigated based on meson-baryon coupled-channels chiral dynamics,
  by evaluating density distributions obtained from the form factors of
  the $\LamFOF$ state.  The form factors 
  are defined as an extension of the ordinary stable particles and 
  are directly evaluated from the 
  current-coupled 
  meson-baryon scattering amplitude, 
  paying attention to the charge conservation of the probe 
  interactions.
  For the resonant $\LamFOF$ state we calculate the density
  distributions in two ways. One is on the pole position of the
  $\LamFOF$ in the complex energy plane, which evaluates the resonant
  $\LamFOF$ structure without contamination from nonresonant
  backgrounds, and another on the real energy axis around the $\LamFOF$
  resonance energy, which may be achieved in experiments.
  Using 
  several
  probe interactions and channel decomposition, 
  we separate the various 
  contributions to the internal structure of the $\LamFOF$. 
  As a result, we find that the resonant $\LamFOF$ state is composed
  of widely spread $\bar{K}$ around $N$, which gives dominant
  component inside the $\LamFOF$, with escaping $\pi \Sigma$
  component.
  Furthermore, we consider $\KbarN$ bound state without decay
  channels, with which we can observe the internal structure of the
  bound state within real numbers.  We also study the dependence of
  the form factors on the binding energy and meson mass.  This
  verifies that the form factor defined through the 
  current-coupled 
  scattering
  amplitude serves as a natural generalization of the form factor
  for the resonance state.  The relation between the interaction strength
  and the meson mass shows that the physical kaon mass appears to be
  within the suitable range to form a molecular bound state with the
  nucleon through the chiral SU(3) interaction.
\end{abstract}


\maketitle

\section{Introduction}
\label{sec:Intro}


\hyphenation{anti-kaon}

The $\Lambda(1405)$ is a baryonic resonance state with spin-parity
$J^{P}=1/2^{-}$, isospin $I=0$, and strangeness $S=-1$, and is located
just below the threshold of antikaon ($\bar{K}$) and nucleon ($N$).
This resonance has been considered as a quasi-bound molecule state of
the $\bar KN$ system~\cite{Dalitz:1960du,Dalitz:1967fp,Wyld:1967zz}
before the establishment of QCD.  In simple constituent quark models,
the $\Lambda(1405)$ is classified into the 70 dimensional
representation of the spin-flavor SU(6) with excitation of one of the
quarks to the $p$-state~\cite{Isgur:1978xj}, but it was hard to
explain the lighter mass of the $\Lambda(1405)$ than the nucleon
resonance $N(1535)$ in the same representation and the larger
spin-orbit splitting between $\Lambda(1405)$ and $\Lambda(1520)$ than
the nucleon partners.  Recently the $\LamFOF$ was investigated with
the unitarized coupled-channels method based on chiral
dynamics~\cite{Kaiser:1995eg,Oset:1997it,Oller:2000fj,
  Lutz:2001yb,Hyodo:2008xr,Jido:2003cb}, in which the $\Lambda(1405)$ is
dynamically generated in meson-baryon scattering without introducing
explicit pole terms.  In this method, one can successfully reproduce
the cross sections of $K^{-}p$ to various channels together with the
mass spectrum of the resonant $\LamFOF$ state below the $\KbarN$
threshold~\cite{Kaiser:1995eg,Oset:1997it,Oller:2000fj,Lutz:2001yb,
  Oset:2001cn,Jido:2002zk,Hyodo:2002pk,Hyodo:2003qa,GarciaRecio:2002td,
  Borasoy:2004kk,Borasoy:2005ie}.

One of the important consequences on the structure in the
coupled-channels approach is that the $\Lambda(1405)$ is composed of
two resonance
states~\cite{Oller:2000fj,Jido:2002yz,Jido:2003cb}.\footnote{The
  two-state nature of $\Lambda(1405)$ was pointed out first in
  Ref.~\cite{Fink:1989uk} in a different model.}  These two states
have different coupling nature to the meson-baryon
states~\cite{Jido:2002yz,Jido:2003cb}, and the $\LamFOF$ state which
dominantly couples to the $\bar KN$ is located at $1420 \mev$ instead of
the nominal 1405 MeV. These difference may be important for the $\bar
KN$ system, as the $\LamFOF$ resonance position is measured from the
$\bar KN$ threshold.  The double-pole structure also suggests that the
$\Lambda(1405)$ resonance position in the $\pi\Sigma$ invariant-mass
spectra depends on the production mechanism of the $\Lambda(1405)$.
Some experimental indications of this structure are found in
Refs.~\cite{Hyodo:2003jw,Prakhov:2004an,Magas:2005vu,Jido:2009jf,Jido:2010rx}.
Physical origin of the two poles is attributed to the attractive
forces both in the $\bar{K}N$ and the $\pi\Sigma$
channels~\cite{Hyodo:2007jq}.

Recently, the importance of the meson-baryon dynamical component in
the structure of the $\Lambda(1405)$ has been discussed in the chiral
unitary framework.  It was revealed in an analysis of the
phenomenologically obtained scattering amplitude that the $\LamFOF$ is
described predominantly by the meson-baryon
dynamics~\cite{Hyodo:2008xr}.  The study of the $\LamFOF$ based on the
$N_{c}$ scaling in chiral unitary framework 
suggested the dominance of the non-$qqq$
component in the $\LamFOF$~\cite{Hyodo:2007np,Roca:2008kr}.  The
electromagnetic mean squared radii were shown to be much larger than
that of ground state baryons~\cite{Sekihara:2008qk}.\footnote{The
  electromagnetic mean squared radii of the $\LamFOF$ was evaluated
  also in~\cite{Schat:1994gm} within the framework of the bound-state
  soliton model. }  Relationship between the couplings to the 
channels and the wave function of the $\LamFOF$ was clarified
in Ref.~\cite{YamagataSekihara:2010pj}.  The spatial structure of the
$\Lambda(1405)$ was also discussed based on the $\bar KN$ molecular
picture with 
the chiral $\KbarN$ 
potential~\cite{Dote:2008in,Dote:2008hw}.  
The molecular-like
structure of the $\LamFOF$ suggests further few-body nuclear systems
with kaon~\cite{Akaishi:2002bg}, such as $\bar
KNN$~\cite{Akaishi:2002bg,Yamazaki:2007cs,Shevchenko:2006xy,
Shevchenko:2007zz,Ikeda:2007nz,Ikeda:2008ub,Ikeda:2010tk,
Dote:2008in,Dote:2008hw}, 
$\bar KKN$~\cite{Jido:2008kp,MartinezTorres:2008kh,MartinezTorres:2010zv}, 
and $\bar{K}\bar{K}N$~\cite{KanadaEn'yo:2008wm}. 
Although the dominant component of the $\LamFOF$ was found to be the
meson-baryon molecule structure in these studies, the internal
structure of the $\LamFOF$ resonance in terms of the hadronic
constituents was not directly extracted from the meson-baryon
scattering amplitude.

If the $\LamFOF$ is dominated by the meson-baryon quasi-bound
molecule, it may have a spatially larger size, which will lead to the
softer form factor than that of typical baryons dominated by the
genuine quark component.  The soft form factor could result in some
experimental consequences, for instance in the nontrivial energy
dependence of the $\LamFOF$ photoproduction cross section recently
reported in Ref.~\cite{Niiyama:2008rt}.  The information of the size
of hadrons is also important for the production yield in the heavy ion
collisions estimated by the coalescence model~\cite{Cho:2010db}.  One
of the standard methods to study the structure of a particle is to use
the vector currents, such as the photon, as probes.  The properties of
the $\LamFOF$ in relation to electromagnetic dynamics have been
investigated in the chiral unitary approach in
Refs.~\cite{Kaiser:1996js,Nacher:1998mi,Nacher:1999ni,Jido:2002yz,Hyodo:2004vt,Geng:2007hz,Doring:2010rd}.
In the molecular picture of the $\LamFOF$, the photon coupling to the
$\LamFOF$ is introduced through the photon couplings to its
constituent meson and baryon.  Among others, the form factor of the
$\LamFOF$ will tell us about the intuitive ``size'' of the particle.
Therefore, for both theoretical and experimental understandings of the
$\LamFOF$, it is interesting to investigate the internal structure of
the $\LamFOF$ through the evaluation of its form factor.

In this paper we study the form factor of the $\LamFOF$ in chiral
unitary approach, in which the $\LamFOF$ is described as the
dynamically generated resonance state.  We emphasize that this is the
first study to extract the form factor of the $\LamFOF$ directly from
the scattering amplitude involving the resonance state as a response
to the external probe current 
in a microscopic way.  We have already reported the result of
the electromagnetic mean squared radii of the $\LamFOF$ in
Ref.~\cite{Sekihara:2008qk}. Here we extend this scheme to evaluate
the momentum dependence of the form factors, 
using the chiral effective theory of the octet mesons and baryons 
with the external currents for the elementary interaction between 
the constituent hadrons and the currents.  
From the form factors the density
distributions can be obtained through the Fourier transformation. This
allows us to visualize the spatial distribution of the $\Lambda(1405)$
in coordinate space.  We also include the finite size effect of the
constituent hadrons.

There is one subtlety in the formulation because the $\LamFOF$ is a
hadron resonance and decays \textit{via} strong interaction. In this
case, the form factors (or coupling constants in general) are obtained
as complex numbers, whose interpretation is not as straightforward as
in the case of the stable particle.  Such a difficulty about the
complex form factors generally shows up for the unstable states, as
seen in the $\Delta (1232)$ electromagnetic form factors studied in
Refs.~\cite{Heller:1986gn,Kumano:1988wu} as complex numbers. Here we
carefully define the form factor of the resonance particle as an
extension of the ordinary stable particles, and explain the methods to
extract the form factors out of the 
photon-coupled meson-baryon scattering amplitude in a
gauge invariant way.  In order to achieve the intuitive understanding
of the size of the resonance, we evaluate the form factor with several
different model set-ups and draw a conclusion through the comparison
of the results.

This paper is organized as follows. In Sec.~\ref{sec:formulation} we
explain our scheme to calculate the form factors and the density
distributions of the $\LamFOF$ based on the chiral dynamics, taking
account of how to extract the information of the ``size'' of the
resonance state.  The coupling of the $\LamFOF$ to the external current 
is obtained by the couplings of its constituent meson and baryon. 
In Sec.~\ref{sec:Result} we show our numerical results
of the calculations for the resonant $\LamFOF$ state, on the resonance
pole position in the complex energy plane and on the real axis.
To obtain the intuitive understanding of the $\LamFOF$ as a
quasi-bound state, we consider the $\KbarN$ bound state without decay
width and study its structure by varying the parameters such as the
binding energy and the 
meson 
mass in Sec.~\ref{sec:Discussion}.
Section~\ref{sec:conclusion} is devoted to the conclusion of this
study.

\section{Formulation}
\label{sec:formulation}

\subsection{How to probe structure of resonance state}
\label{subsec:How_to}

First of all, we discuss the method to probe the properties of a
short-living resonance state. Let us consider an observable represented by
a Hermitian operator $\mathcal{O}$. One of the simplest and
model-independent ways to define resonance properties is given by 
the evaluation of the matrix element of the
operator $\mathcal{O}$ for the resonance state vectors
\be
\langle Z_{\text{R}} (\bm{P}^{\prime}) ^{(-)} | 
\mathcal{O} | Z_{\text{R}} (\bm{P})^{(+)} \rangle .
\ee
This is an extension of the matrix element of the stable particle such as 
nucleon, 
$\langle N^{(-)} | \mathcal{O} | N^{(+)} \rangle$.
The labels ``$+$'' and ``$-$'' represent ``in'' and ``out'' states, 
respectively. Here the 
ket 
$| Z_{\text{R}} (\bm{P})^{(+)} \rangle$ and 
bra 
$\langle Z_{\text{R}} (\bm{P})^{(-)}|$ vectors represent 
the resonance state in the generalized Hilbert space, whose eigenvalues of the energy-momentum operator $\hat{P}^{\mu}$ are complex
as\footnote{For the thorough description of the eigenvector for the resonance 
state, see, \textit{e.g.}, Ref.~\cite{Bohm:1981pv}.  See also 
Ref.~\cite{Bohm:2001qm}. }, 
\begin{align}
& \hat{P}^{\mu} | Z_{\text{R}} (\bm{P})^{(+)} \rangle 
= \left ( E_{\text{R}} , \, \bm{P} \right ) | Z_{\text{R}} (\bm{P})^{(+)} \rangle , 
\\
& \langle Z_{\text{R}} (\bm{P})^{(-)} | \hat{P}^{\mu} 
= \left ( E_{\text{R}} , \, \bm{P} \right ) \langle Z_{\text{R}} (\bm{P})^{(-)} | , 
\end{align}
\be
E_{\text{R}} = 
\sqrt{ \left ( M_{\text{R}} - i \frac{\Gamma _{\text{R}}}{2} \right )^{2} 
+ \bm{P}^{2}} . 
\ee
Here $M_{\text{R}}$ and $\Gamma _{\text{R}}$ are the mass and decay
width of the resonance state, respectively. It is noted that this
definition leads to complex values of the matrix elements such as the
radius and charge distribution for the resonance state, because the
Hermitian operators are allowed to have complex eigenvalues for the
resonance state in the generalized Hilbert space. 
When the decay width of the resonance state is sufficiently 
small, the imaginary part of the eigenvalues of the Hermitian 
operators is also small and the matrix elements can be interpreted 
as physical osbervables in the ordinary sence. 
In this paper, we mainly discuss
the real part of the eigenvalues of the Hermitian operators as 
``observables'', assuming that the significance of the imaginary part
is sufficiently small.

In order to scan the internal structure of the resonance with
different probes, we evaluate the matrix elements of 
various conserved vector currents, 
the electromagnetic transition current
$J_{\text{EM}}^{\mu}$, baryonic current $J_{\text{B}}^{\mu}$ and
strangeness current $J_{\text{S}}^{\mu}$, stemming from the U(1)
symmetry of the underlying theory.  Electromagnetic current is
obtained from the gauged U(1) symmetry with appropriate charges of
quarks, baryonic current from the simultaneous U(1) phase
transformation for all the quarks, and the strangeness current from
the phase rotation for strange quark only. The other currents may also
be used as the probes of the structure, as long as the currents are
well defined.

The structure of the resonance state is reflected in the form factors,
as in the same way with stable particles such as the nucleon. For
resonance states with baryon number $1$ and spin $1/2$, we define the
form factors $F^{\mu} (Q^{2})$ as the matrix elements of the probe
current $J^{\mu} (x)$,
\be
\langle Z_{\text{R}} (\bm{P}^{\prime} )^{(-)} | 
J^{\mu} (x) | Z_{\text{R}} (\bm{P} )^{(+)} \rangle 
\equiv e^{i (P^{\prime} - P)^{\nu} x_{\nu} } F^{\mu} (Q^{2}) , 
\label{eq:def1}
\ee
where $Q^{2}=-(P^{\prime} - P)^{\mu} (P^{\prime} - P)_{\mu}$. Especially 
in the Breit frame the matrix 
element can be written as~\cite{Jido:2002yz}: 
\begin{align}
& \langle Z_{\text{R}} (\bm{q}/2 )^{(-)} | 
J^{\mu} (x) | Z_{\text{R}} (-\bm{q}/{2} )^{(+)} \rangle 
\nonumber \\
& = 
e^{- i \bm{q} \cdot \bm{x}} 
\chi _{Z_{\text{R}}}^{\prime \dagger} \left ( F_{\text{time}} (Q^{2}), \, 
F_{\text{space}} (Q^{2}) 
\frac{i \bm{\sigma} \times \bm{q}}{2 M_{\text{p}}} 
\right ) \chi _{Z_{\text{R}}} ,
\label{eq:definition}
\end{align}
with the form factors of the time component $F_{\text{time}} (Q^{2})$ and the 
space one $F_{\text{space}} (Q^{2})$, the momentum transfer
$q^{\mu}=(0,\bm{q})$, $Q^{2}=\bm{q}^{2}$, the spinor for the resonance state 
$\chi_{Z_{\text{R}}}^{(\prime)}$, and the Pauli matrices 
$\sigma ^{a}\, (a=1, \, 2, \, 3) $ for the spin space. 
For the electromagnetic current, 
the time component form factor is identified as the electric form factor 
$\FE (Q^{2})$, whereas the space component form factor as the magnetic form 
factor 
$\FM (Q^{2})$. The magnetic form factor $\FM (Q^{2})$ is normalized as the 
nuclear magneton $\mu _{\text{N}}=e/(2 M_{\text{p}})$ with the proton charge 
$e$ and mass $M_{\text{p}}$. We define 
the baryonic form factor $\FB (Q^{2})$ and the strangeness form factor 
$\FS (Q^{2})$ as the time component of the matrix elements of the baryonic and strangeness currents, respectively.

The form factors $F(Q^{2})$ contain the information of the structure of 
the system in response to the probe interaction, being 
related to the ``classical density'' of the system (such as charge density) 
\textit{via} the Fourier transformation as, 
\be
\rho (r) 
= \int \frac{d^{3} Q}{(2 \pi)^{3}} e^{- i \bm{q} \cdot \bm{r}} 
\frac{M}{E(Q^{2})} F(Q^{2}) , 
\ee
where $r$ is the radial coordinate from the center of mass of the system, and 
$M$ and $E(Q^{2})=\sqrt{M^{2}+Q^{2}}$ are the mass and the energy of the state, 
respectively. 
Taking the nonrelativistic reduction ($M/E \to 1$), we have, 
\be
\rho (r) 
= \int \frac{d^{3} Q}{(2 \pi)^{3}} e^{- i \bm{q} \cdot \bm{r}} 
F(Q^{2}) . 
\label{eq:Density}
\ee 
We can also obtain the mean squared radius: 
\be
\MSR 
= \int d^{3} r \, r^{2} \rho (r)
= - 6 \left . 
\frac{d F}{d Q^{2}} \right | _{Q^{2}=0} . 
\label{eq:DefMSR}
\ee
For the magnetic mean squared radius, we take conventional normalization by 
dividing the right-hand side of Eq.~\eqref{eq:DefMSR} by the factor $\FM (0)$. 
The mean squared radius is well defined when the density $\rho
  (r)$ sufficiently falls off at large $r$ and the integration of
  $r^{2} \rho (r)$ converges, which is the case on the poles for the
  bound states as well as the resonance states.

\subsection{Matrix elements of resonance state in scattering amplitudes}
\label{subsec:matrix_elements}
Here we explain the method to extract the matrix elements of the resonance 
state from the scattering amplitudes of dynamical approaches~\cite{Jido:2002yz,
Sekihara:2008qk}. Once the matrix elements are extracted, 
following the prescription given in the preceding subsection, we can 
evaluate the form factors [from Eq.~\eqref{eq:definition}] and density 
distributions [from Eq.~\eqref{eq:Density}].

\begin{figure}[t]
\centering
 \begin{tabular*}{8.6cm}{@{\extracolsep{\fill}}cc}
    \includegraphics[scale=0.18]{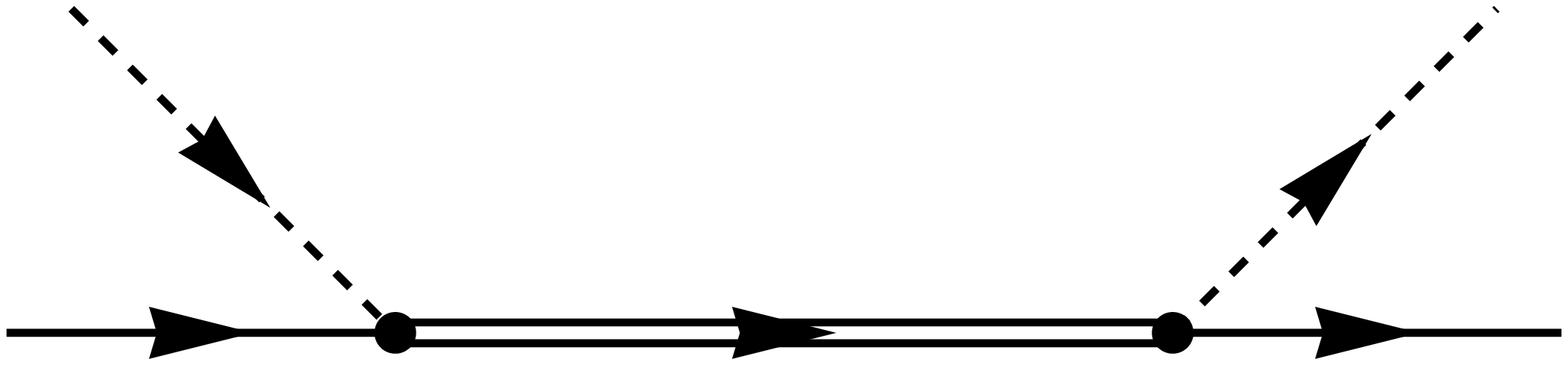} &
    \includegraphics[scale=0.18]{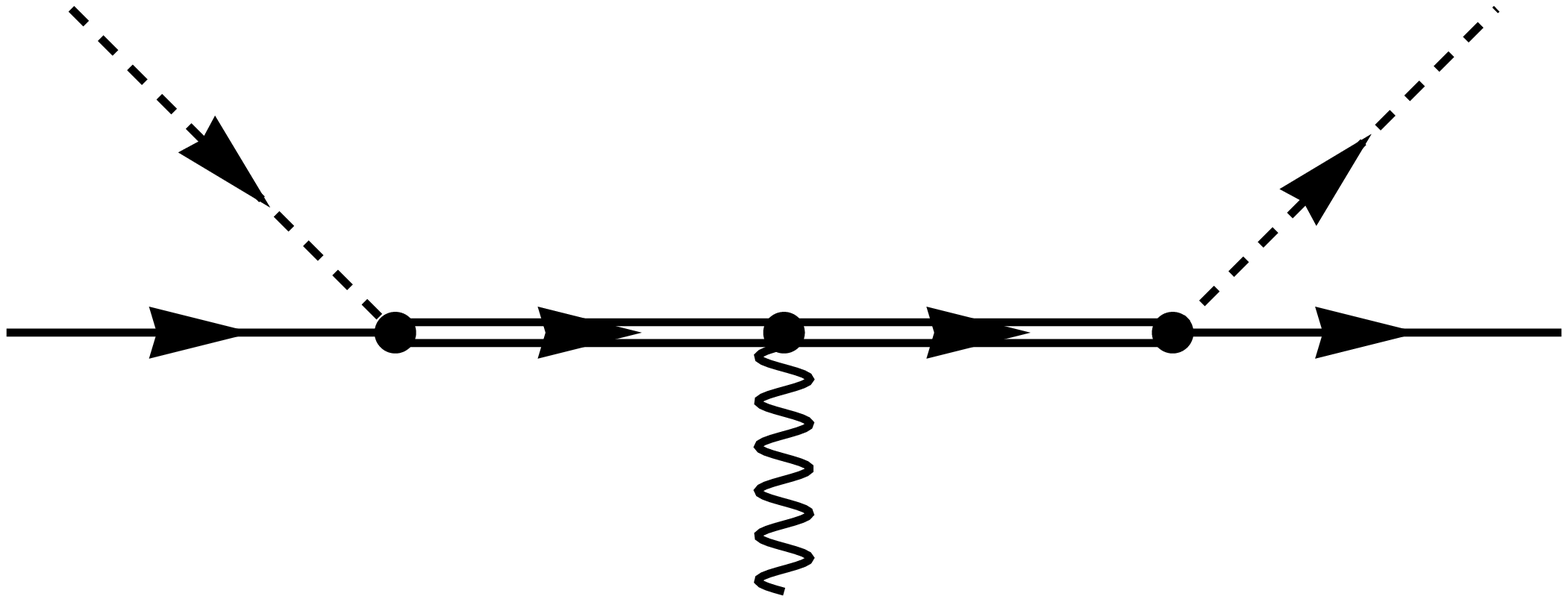}\\
    (a) $T_{ij}$ & (b) $T_{\gamma ij}^{\mu}$ 
 \end{tabular*}
 \caption{(a) Feynman diagram for meson-baryon scattering amplitudes 
   close to the pole of the excited baryon, 
   (b) Feynamn diagram for meson-baryon scattering amplitudes with the
   probe current attached to the excited baryon. The double, dashed,
   solid and wiggly lines correspond to the excited baryon, the ground
   state meson, the ground state baryon and the probe current,
   respectively.}
\label{fig:resonance}
\end{figure}

\begin{figure}[t]
  \centering
  \begin{tabular*}{8.6cm}{@{\extracolsep{\fill}}ccc}
    \includegraphics[scale=0.18]{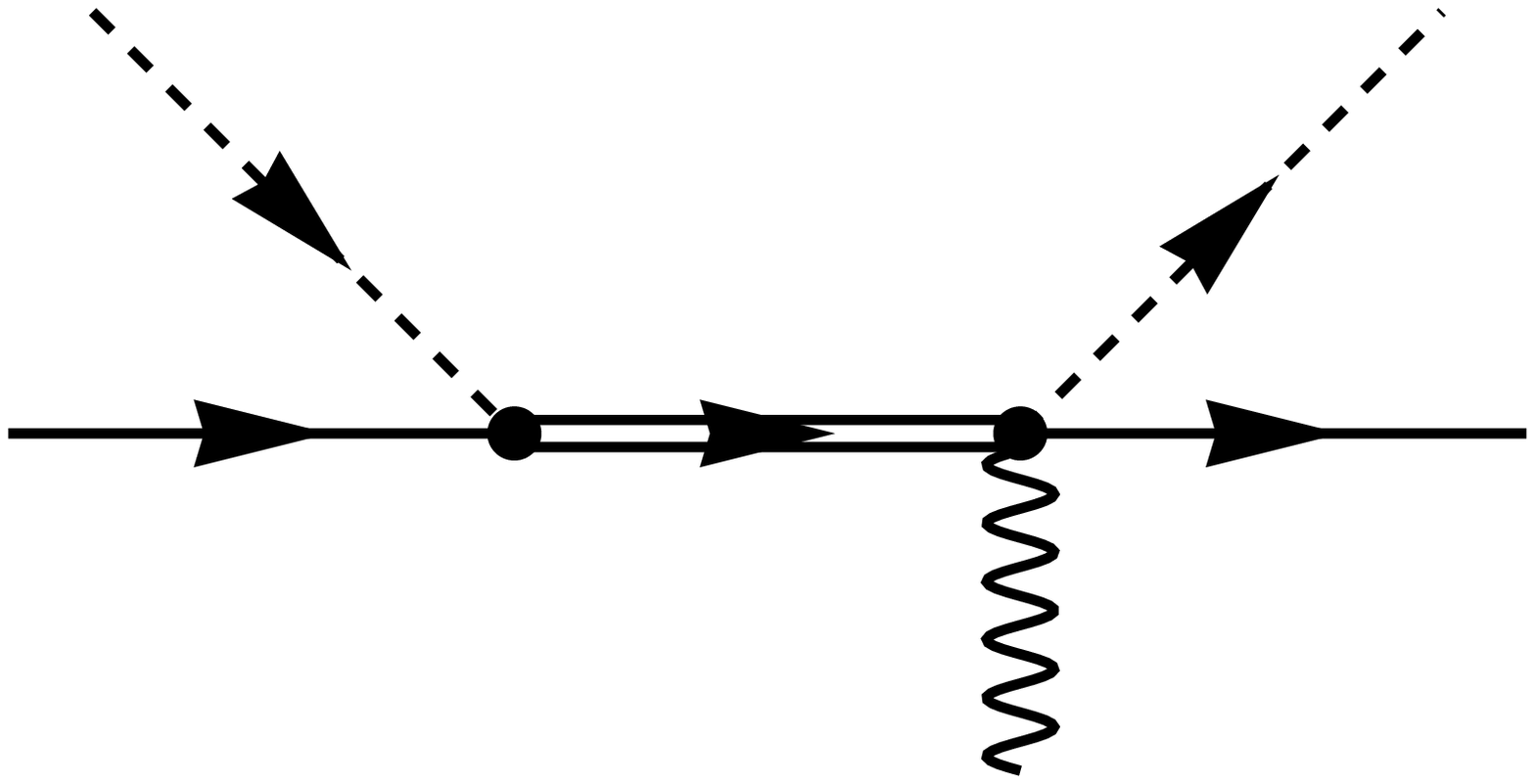} &
    \includegraphics[scale=0.18]{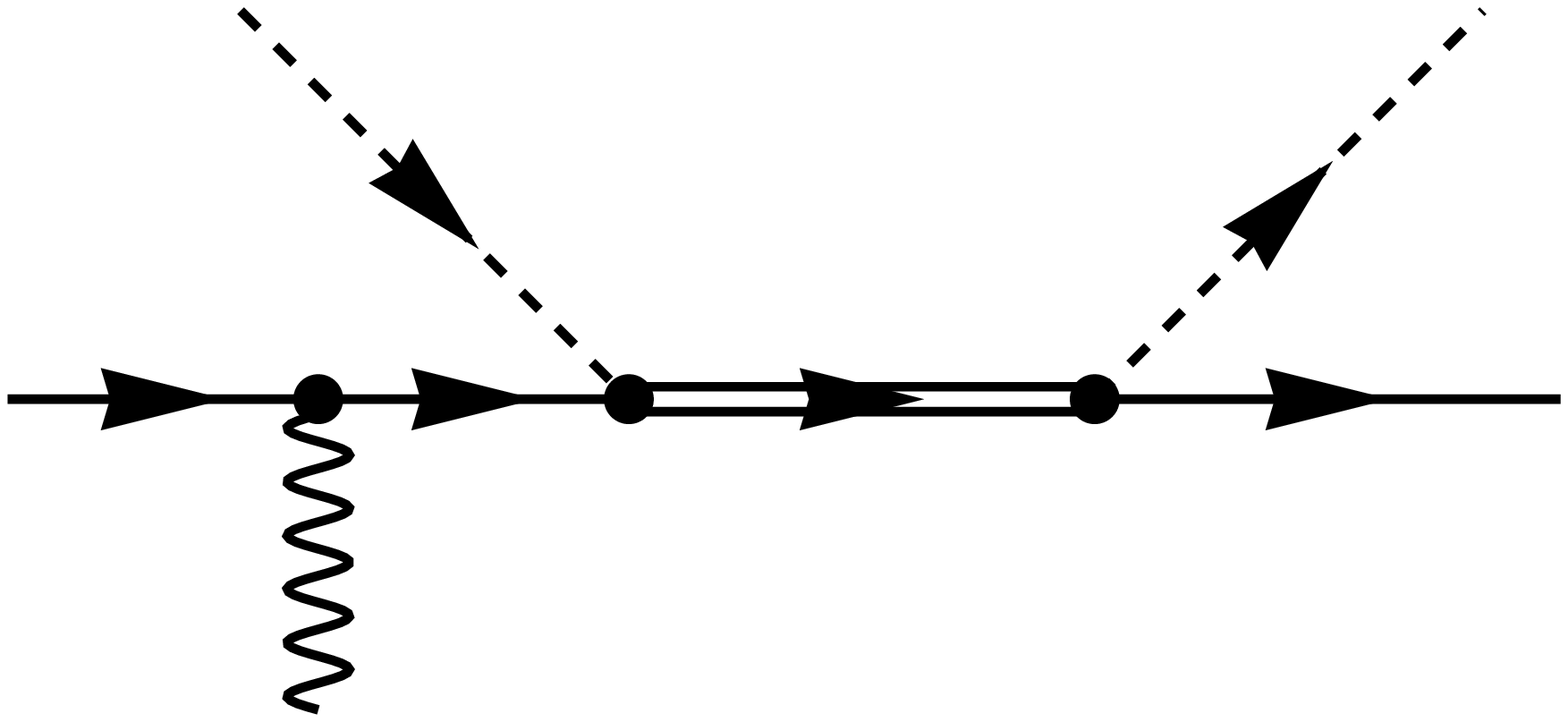} &
    \includegraphics[scale=0.18]{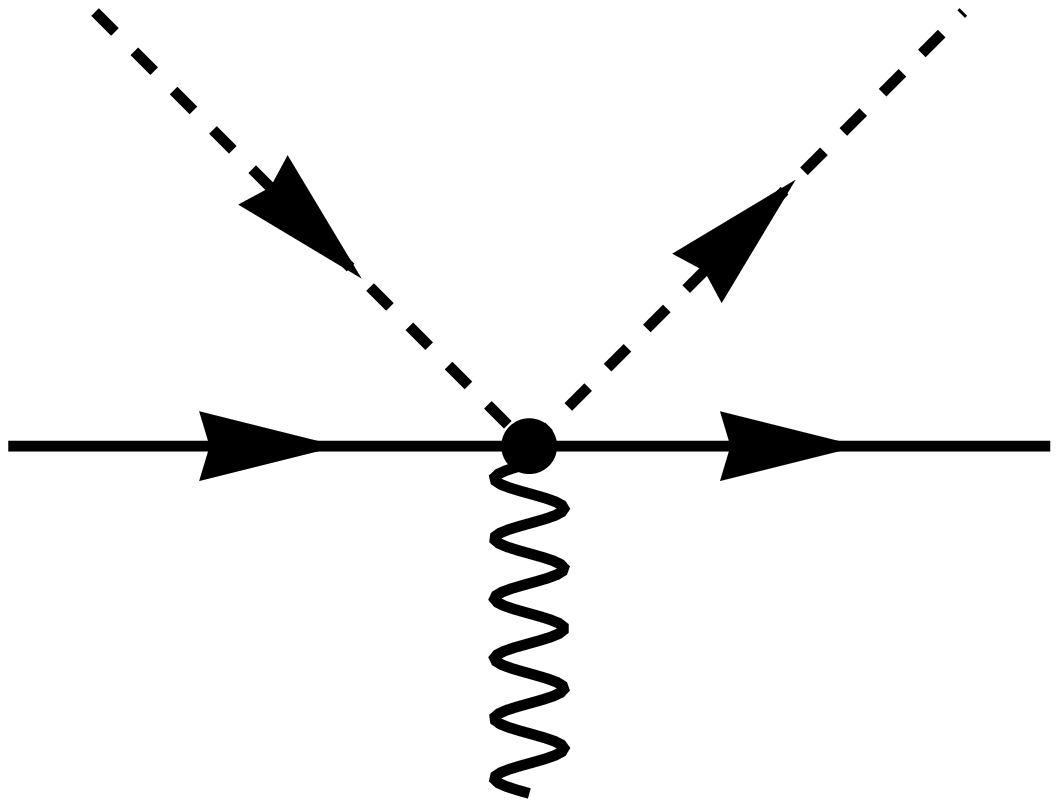} 
  \end{tabular*}
  \caption{Examples of the Feynman diagrams for the scattering
    amplitudes $T_{\gamma}^{\mu}$ in which the probe current does 
    not couple to the excited baryon.  The double,
    dashed, solid and wiggly lines correspond to the excited baryon,
    the ground state meson, the ground state baryon and the probe
    current, respectively.}
  \label{fig:nonresonance}
\end{figure}

Let us here consider $s$-wave meson-baryon scattering ($MB \to
M^{\prime} B^{\prime}$) amplitude $T(\sqrt{s})$ with the total energy
$\sqrt{s}$, in which the excited baryon appears as a resonance
state.
Performing the analytic continuation of the energy variable to the
complex plane $\sqrt{s}\to z$, we obtain the scattering amplitude in
the complex energy plane.  The resonance state is expressed as a pole
of the scattering amplitude in the second Riemann sheet. 
We define the ``pole contribution'' of the amplitude as
\be
\left. - i T_{ij} (z) \right|_{\rm pole} = 
i g_{i} \frac{i}{z -  Z_{\text{R}}} i g_{j} , 
\label{eq:T_mat}
\ee
where the pole position is expressed as the resonance mass and width as
$Z_{\text{R}}=M_{\text{R}}-i\Gamma _{\text{R}}/2$
and the indices $i$ and $j$ represent the final and initial meson-baryon 
channels, respectively.
The residue $g_{i}g_{j}$ at the pole position 
can be interpreted as 
the product of the coupling strengths to the
meson-baryon channels $i$ and $j$. 
On the real energy axis, the full scattering amplitude can be written as
\be
- i T_{ij} (\sqrt{s}) = 
i g_{i} \frac{i}{\sqrt{s} -  Z_{\text{R}}} i g_{j} 
- i T_{ij}^{\text{BG}}(\sqrt{s}) . 
\label{eq:T_matreal}
\ee
The contributions which are not represented by the pole term is
summarized as a nonresonant background term $T_{ij}^{\text{BG}}$ 
(see also~\cite{Doring:2009bi}).

Next we relate the meson-baryon scattering amplitude with the matrix
element of the resonance state vectors introduced in the preceding
subsection. The pole contribution in the scattering matrix is written
as
\begin{align}
&
 S_{ij} (P^{\prime},P) |_{\text{pole}} 
\nonumber \\ 
=& 
\int \frac{d^{4}P^{\prime\prime}}{(2\pi)^{4}} 
\langle \Phi_{i}(P^{\prime})^{(-)} | Z_{\text{R}}(P^{\prime\prime})^{(+)} \rangle 
\frac{i}{z^{\prime\prime} - Z_{\text{R}}} \nonumber \\
&\times \langle Z_{\text{R}}(P^{\prime\prime})^{(-)} 
| \Phi_{j}(P)^{(+)} \rangle 
\label{eq:resonant-prop}
\\
=& -i (2\pi)^{4} \delta^{4}(P^{\mu}-P^{\prime \mu}) \frac{g_{i}g_{j}}{z- Z_{\text{R}}}
\end{align}
with
\begin{align}
\langle \Phi_{i}(P^{\prime})^{(-)} | Z_{\text{R}}(P)^{(+)} \rangle 
=&
\langle Z_{\text{R}}(P^{\prime})^{(-)} | \Phi_{i}(P)^{(+)} \rangle 
\nonumber \\
=&
i g_{i} (2\pi)^{4} \delta^{4}(P^{\mu}-P^{\prime \mu}),
\label{eq:coupling}
\end{align}
where $| \Phi_{i}(P^{\prime})^{(-)} \rangle$ ($| \Phi_{j}(P)^{(+)}
\rangle $) is the out (in) $s$-wave meson-baryon scattering state with
channel $i$ ($j$) and momentum $P^{\prime \mu}$ ($P^{\mu}$).  The
total energies are given by $z^{\prime\prime}=\sqrt{P^{\prime\prime
    \mu}P_{\mu}^{\prime\prime}}$ and $z=\sqrt{P^{\mu}P_{\mu}}$. 
In this way, the residue of the pole in the
scattering amplitude is related to the 
inner product of the resonance state
with the scattering state, representing the coupling strength. It is
important that
\begin{equation}
 P_{Z_{\text R}}  \equiv \int \frac{d^{4}P}{(2\pi)^{4}} 
 |Z_{\text{R}}(P)^{(+)} \rangle \frac{i}{z - Z_{\text{R}}}  
 \langle Z_{\text{R}}(P)^{(-)} | 
 \label{eq:projection}
\end{equation}
in Eq.~\eqref{eq:resonant-prop} can be understood as 
the ``projection operator'' to the resonance state of the mass $Z_{\text{R}}$, 
which is a generalization of that of the usual stable particle. 
This resonance component of the amplitude, $T_{ij}(z)|_{\text{pole}}$, is 
represented diagrammatically in Fig.~\ref{fig:resonance}(a). 

In the same way, we also consider the scattering amplitude 
$T_{\gamma ij}^{\mu}$
for the $MB \gamma ^{\ast} \to M^{\prime} B^{\prime}$ process,
where $\gamma ^{\ast}$ stands for a probe such as photon for the 
electromagnetic current. The Fourier component of the matrix element of 
the probe current in the $s$-wave meson-baryon state, 
$S_{\gamma ij}^{\mu}$, is given as: 
\begin{align}
& S_{\gamma ij}^{\mu}(P^{\prime},\, P;\, Q^{2}) 
\nonumber \\
& = \int d^{4} x \, e^{- i q^{\nu} x_{\nu}} 
\langle \Phi _{i}(P^{\prime}) ^{(-)} | i J^{\mu} (x) | \Phi _{j}(P) ^{(+)} \rangle , 
\end{align}
with incoming and outgoing momenta of the meson-baryon system
$P^{\mu}$ and $P^{\prime \mu}$, respectively, and $Q^{2}=-q^{\mu}q_{\mu}$. 
The resonance contribution to the matrix element is obtained by 
inserting the ``projection operator''~\eqref{eq:projection} for the 
resonance state on both sides of the current operator: 
\begin{align}
& S_{\gamma ij}^{\mu} (P^{\prime} , \, P ; \, Q^{2}) |_{\text{pole}}
\nonumber \\ 
=& \int d^{4} x \, e^{- i q^{\nu} x_{\nu}} 
\int \frac{d^{4}P^{\prime\prime}}{(2\pi)^{4}} 
\int \frac{d^{4}P^{\prime\prime\prime}}{(2\pi)^{4}} \nonumber \\
&\times \langle \Phi_{i}(P^{\prime})^{(-)} | Z_{\text{R}}(P^{\prime\prime\prime})^{(+)} \rangle 
\frac{i}{z^{\prime\prime\prime} - Z_{\text{R}}} 
\nonumber \\
& \times \langle Z_{\text{R}}(P^{\prime\prime\prime})^{(-)} | i J^{\mu} | Z_{\text{R}}(P^{\prime\prime})^{(+)}
 \rangle \frac{i}{z^{\prime\prime} - Z_{\text{R}}} \nonumber \\
&\times\langle Z_{\text{R}}(P^{\prime\prime})^{(-)} | \Phi_{j}(P)^{(+)} \rangle 
\nonumber \\ 
=& i (2\pi)^{4} \delta^{4}(P^{\prime \nu}-P^{\nu}-q^{\nu}) 
g_{i} \frac{1}{z^{\prime} - Z_{\text{R}}} F^{\mu} (Q^{2}) 
\frac{1}{z - Z_{\text{R}}} 
g_{j} ,
\label{eq:Sgamma-pole}
\end{align}
where we have used Eqs.~\eqref{eq:def1} and \eqref{eq:coupling}.
Hence, the pole contribution to the scattering amplitude of $MB \gamma ^{\ast} \to M^{\prime} B^{\prime}$ process is 
\be
T_{\gamma ij} ( z^{\prime}, \, z; \, Q^{2} ) |_{\text{pole}} 
= - g_{i} \frac{1}{z^{\prime} - Z_{\text{R}}} F (Q^{2}) 
\frac{1}{z - Z_{\text{R}}} g_{j} , 
\label{eq:T_gamma_mat}
\ee
which is represented diagrammatically in Fig.~\ref{fig:resonance}(b) 
(here and following 
in this subsection 
we omit the superscript $\mu$ in $T_{\gamma}$ and $F$). 
It is important that this term contains the matrix element
of the probe current with the resonance state as 
the residue of double pole at $z = z^{\prime} = Z_{\text{R}} $. 

The rest contributions of the amplitude $T_{\gamma ij}$
has less singularity at the resonance position $Z_{\text{R}}$.
We show several examples of these contributions diagrammatically 
in Fig.~\ref{fig:nonresonance}, in which 
the current does not couple to the intermediate 
resonance propagator. 

Combining Eqs.~\eqref{eq:T_mat} and \eqref{eq:T_gamma_mat}, 
the matrix elements in the Breit frame can be evaluated as the residue 
of the 
single
pole at $z=z^{\prime}=Z_{\text{R}}$ of 
$- T_{\gamma ij} / T_{ij}$ 
in the complex energy plane: 
\begin{align}
F (Q^{2}) |_{\text{Breit}} 
& =  
- \frac{(z^{\prime}-Z_R)T_{\gamma ij}
(z^{\prime}, \, z; \, Q^{2} )}{T_{ij} (z)} 
\Bigg |_{z \to Z_{\text{R}}}
\Bigg |_{z^{\prime} \to Z_{\text{R}}} .
\label{eq:Res_scheme}
\end{align}
As one can see from the above derivation, this 
scheme indicates that only the amplitude 
$T_{ij}$ ($T_{\gamma ij}$) 
with the 
single- (double-) pole
terms 
contributes to the 
right-hand side, whereas both nonresonant 
background term $T_{ij}^{\text{BG}}$ 
in Eq.~\eqref{eq:T_matreal} for $T_{ij}(z)$ and the less singular terms for 
$T_{\gamma ij} (z^{\prime},\, z; \, Q^{2})$ automatically 
drop and have no effect to the form factors for the resonance state. 
As confirmed in 
Ref.~\cite{Sekihara:2008qk}, the form factor obtained 
in Eq.~\eqref{eq:Res_scheme} is a gauge invariant quantity. 

Equation~\eqref{eq:Res_scheme} should be evaluated in the complex 
energy plane for the resonance state. It is an interesting issue how
the matrix elements of the currents for the resonance can be obtained 
on the real axis, in which amplitudes are in principle observed experimentally.
In order to keep closer connection to experimental measurements,
we 
discuss 
a method to evaluate the form factor of the 
resonance state on the real axis, as 
developed 
in Ref.~\cite{Jido:2002yz}.

As we have seen above, the matrix element of the current for 
the resonance state is expressed as the residue of the amplitude 
$T_{\gamma}(z^{\prime},\, z; \, Q^{2})$ at the resonance pole position
($z^{\prime}=z=Z_{\text{R}}$). 
If we take the Breit frame of the resonance ($z^{\prime}=z$), 
the form factor $F(Q^{2})$ of the resonance can be
written on the real axis as, 
\be
T_{\gamma ij} 
= - \frac{g_{i}}{\sqrt{s} - Z_{\text{R}}}\Big[ F (Q^{2}) \Big]
\frac{g_{j}}{\sqrt{s} - Z_{\text{R}}}
+ T^{\text{less}}_{\gamma ij}( \sqrt{s} ) 
\label{eq:gammaij}
\ee
where $T^{\text{less}}_{\gamma}$ represents less singular terms than the 
double-pole contribution given in the first term. 
It is important to note that, if the decay width of the resonance
state is small and its pole position is close to the real energy axis, 
the double-pole amplitudes will give dominant contribution to 
the $MB \gamma ^{\ast} \to M^{\prime} B^{\prime}$ process than the other 
contributions in the resonance energy region.

Motivated by Eq.~\eqref{eq:gammaij}, we define an effective form factor on the 
real axis as
\begin{align}
 \Feff (Q^{2} ; \, \sqrt{s}) 
& \equiv 
\frac{T_{\gamma ij} 
( \sqrt{s} , \, \sqrt{s} ; \, Q^{2} ) 
}{d T_{ij} / d \sqrt{s}} .
\label{eq:Feffective}
\end{align}
As discussed in Appendix~\ref{sec:normalization}, the effective form 
factor~\eqref{eq:Feffective} is correctly 
normalized so as to give the corresponding charge of the system at 
$Q^{2}=0$ independently 
of $\sqrt s$, if we consider all of the appropriate diagrams 
giving both the double-pole and less singular terms. 
 
Let us study the relation between the effective form
factor~\eqref{eq:Feffective} on the real axis and the form factor at
the pole position~\eqref{eq:Res_scheme}.  Around the resonance energy
region, where the pole contribution dominates the amplitude, the
derivative of the $MB$ scattering amplitude~\eqref{eq:T_matreal} with
respect to $\sqrt s$ is written as
\be
\frac{d}{d \sqrt{s}} T_{ij} = 
- \frac{g_{i} g_{j}}{(\sqrt{s} - Z_{\text{R}})^{2}} 
+ \frac{d}{d \sqrt{s}} T_{ij}^{\text{BG}} .  
\label{eq:delT}
\ee
Assuming that the background contribution is smoothly changing with respect 
to the energy $\sqrt s$ around the resonance energy and, thus, neglecting 
the second term of Eq.~\eqref{eq:delT}, we can write the effective form factor 
approximately as
\begin{align}
 \Feff (Q^{2} ; \, \sqrt{s}) 
& \approx  F (Q^{2})
- T^{\text{less}}_{\gamma ij}( \sqrt{s}) 
\frac{( \sqrt{s} - Z_{\text{R}} )^{2}}{g_{i} g_{j}} .  
\label{eq:effFFapprox}
\end{align}
In this way, the effective form factor is related to the resonance form
factor $F(Q^2)$. The deviation mainly comes from
the less singular terms in
$T_{\gamma ij}$ which contribute to the effective form factor $\Feff$.
Nevertheless, as seen in Eq.~\eqref{eq:effFFapprox}, if we take the
energy $\sqrt s$ close to the resonance mass and choose the channel
which strongly couples to the resonance, we can reduce the
contamination from the less singular terms. In addition $\Feff
(Q^{2})$ will coincide with $F(Q^{2})$ evaluated by
Eq.~\eqref{eq:Res_scheme} when we take $\sqrt{s}\to Z_{\text{R}}$ 
by analytic continuation, since in this case both 
the less singular contribution $T_{\gamma}^{\text{less}}$ in 
Eq.~\eqref{eq:effFFapprox} and the nonresonant background contribution 
in Eq.~\eqref{eq:delT} 
automatically drop at $\sqrt{s}\to Z_{\text{R}}$.

\subsection{$\bm{\LamFOF}$ in chiral dynamics}
\label{subsec:dynamics}
In this subsection, we briefly review our formulation of the $\LamFOF$
resonance generated dynamically in the $MB \to M^{\prime} B^{\prime}$ 
process using chiral unitary model (ChUM) developed in 
Refs.~\cite{Kaiser:1995eg,Oset:1997it,Oller:2000fj,Lutz:2001yb}. 
It 
turns
out that explicit pole terms are not necessary 
in the elementary interaction 
for the description of the $\LamFOF$ in ChUM~\cite{Hyodo:2008xr}.

The starting point of our formulation is the important fact that
chiral symmetry and its spontaneous breakdown in QCD constrain the
form of low-energy interactions including Nambu-Goldstone bosons. This
is systematically expressed in chiral perturbation
theory~\cite{Weinberg:1978kz,Gasser:1984gg,Pich:1995bw}, in which the
effective Lagrangian is sorted out according to chiral expansion. From
the lowest order meson-baryon chiral Lagrangian, the tree-level
$s$-wave meson-baryon interaction known as the Weinberg-Tomozawa term
can be obtained as,
\begin{align}
V_{ij} 
&= - \frac{C_{ij}}{4 f^{2}} (\Slash{k} + 
\Slash{k}^{\prime}) 
\label{eq:WToriginal} \\
&\simeq - \frac{C_{ij}}{4 f^{2}}(2 \sqrt{s} - M_{i} - M_{j}), 
\label{eq:Vij}
\end{align}
with the incoming and outgoing meson momenta, $k^{\mu}$ and $k^{\prime \mu}$, the
meson decay constant $f$, the Clebsch-Gordan coefficient $C_{ij}$ fixed by the 
$\text{SU}(3)$ 
group structure of the interaction, and masses of outgoing and incoming baryon, 
$M_{i}$ and $M_{j}$, respectively. 
The last form in Eq.~\eqref{eq:Vij} is obtained
by applying the nonrelativistic reduction for the baryons. The explicit value of the
coefficient $C_{ij}$ for the $\KbarN$ scattering is given in 
Ref.~\cite{Oset:1997it}. 

Only this lowest-order Weinberg-Tomozawa interaction, however, is not
sufficient for the description of the scattering amplitude, especially
for three flavors. The main reason is the existence of the baryonic
resonance state $\LamFOF$ in $I=0$ and $S=-1$ channel, just below the
$\KbarN$ threshold, which spoils any perturbative expansion around
the threshold. Therefore, in order to reproduce the $\KbarN$
scattering amplitude, some nonperturbative and coupled-channels
treatment for the $\KbarN$ scattering is needed.

One valuable approach to take into account the nonperturbative effect is to 
formulate a scattering amplitude fulfilling exact unitarity with the N/D method~\cite{Chew:1960iv}. 
Following Ref.~\cite{Oller:2000fj}, with assumption that the intermediate states 
of the scattering are composed of only one octet meson and one octet baryon and 
with neglect of the left-hand cuts, one can write the inverse of the general 
form of the scattering amplitudes fulfilling unitarity as, 
\be
T_{ij}^{-1} (\sqrt{s}) 
= - \delta _{ij} G_{i} (\sqrt{s}) + \mathcal{T}_{ij}^{-1} (\sqrt{s}) ,
\label{eq:NoverD}
\ee
where, 
\begin{align}
& G_{i} (\sqrt{s}) = 
- \tilde{a}_{i}(s_{0}) - \frac{s - s_{0}}{2 \pi}
\int _{s_{i}^{+}}^{\infty} d s^{\prime} 
\frac{\rho _{i} (s^{\prime})}{(s^{\prime} - s) (s^{\prime} - s_{0})} , \\
& \rho _{i} (s) = \frac{2 M_{i} \tilde{q}_i}{4 \pi \sqrt{s}} . 
\end{align}
Here $s_{0}$ denotes the subtraction point, $s_{i}^{+}$ is the threshold value 
of $s$ of the channel $i$ and $\tilde{q}_{i}$ center-of-mass momentum in the 
channel $i$
\be
\tilde{q}_{i} \equiv 
\sqrt{\frac{(s-M_{i}^{2}+m_{i}^{2})^{2} - 4 s m_{i}^{2}}{4 s}} ,
\label{eq:q_k}
\ee
where $m_{i}$ represents the meson mass in the channel $i$. 

The function $G_{i}$ takes the same form as, except for an infinite constant, 
the ordinary meson-baryon loop integral, 
\begin{align}
& G_{i} (\sqrt{s}) 
= i \int \frac{d ^{4} q_{1}}{(2 \pi)^{4}} 
\frac{1}{q_{1}^{2} - m_{i}^{2} + i \epsilon} 
\frac{2 M_{i}}{(P - q_{1})^{2} - M_{i}^{2} + i \epsilon} , 
\end{align}
where 
$\epsilon$ is an infinitesimal real constant to specify the boundary
condition. With dimensional regularization scheme, which keeps
analytic properties of the loop function, 
the finite part of 
the loop integral $G_{i}$ can be written as,
\begin{widetext}
\begin{align}
G_{i} (\sqrt{s}) 
&= \frac{2 M_{i}}{16 \pi ^{2}} \Bigg[ a_{i} (\mu _{\text{reg}})
 + \ln \frac{M_{i}^{2}}{\mu _{\text{reg}}^{2}} 
 + \frac{m_{i}^{2} - M_{i}^{2} + s}{2 s} 
\ln \frac{m_{i}^{2}}{M_{i}^{2}}
 \nonumber \\
& \phantom{= \frac{2 M_{i}}{16 \pi ^{2}}  a} 
+ \frac{\tilde{q}_{i}}{\sqrt{s}} 
\Bigl( \ln (s - M_{i}^{2} + m_{i}^{2} + 2 \tilde{q}_{i} \sqrt{s}) 
+ \ln (s + M_{i}^{2} - m_{i}^{2} + 2 \tilde{q}_{i} \sqrt{s}) 
\nonumber \\ & 
\phantom{=\frac{2 M_{i}}{16 \pi ^{2}} a + (\frac{q_{i}}{\sqrt{s}})} 
- \ln (- s + M_{i}^{2} - m_{i}^{2} + 2 \tilde{q}_{i} \sqrt{s}) 
- \ln (- s - M_{i}^{2} + m_{i}^{2} + 2 \tilde{q}_{i} \sqrt{s})\Bigr)
\Bigg] , 
\label{eq:loop}
\end{align}
\end{widetext}
with the regularization scale $\mu _{\text{reg}}$ and the subtraction constant 
$a_{i}=-(4\pi )^{2} \tilde{a}_{i} / (2 M_{i})$.

For the $\mathcal{T}_{ij}$ in Eq.~\eqref{eq:NoverD}, we adopt the
matching scheme developed in Ref.~\cite{Oller:2000fj}, which tells
that the identification of $\mathcal{T}_{ij}$ with the tree-level
amplitudes by chiral perturbation theory is valid up to
$O(p^{2})$. Hence in our approach we identify the Weinberg-Tomozawa
interaction $V_{ij}$ in Eq.~\eqref{eq:Vij} as $\mathcal{T}_{ij}$. In
this way, the $T$-matrix is written in matrix form as,
\be
T (\sqrt{s}) 
= [V^{-1}
- G]^{-1} .
\label{eq:amplitude}
\ee
This amplitude satisfy the following equation
\be
T_{ij} (\sqrt{s}) = V_{ij} 
+ \sum _{k} V_{ik} G_{k} T_{kj} 
= V_{ij} + \sum _{k} T_{ik} G_{k} V_{kj} ,
\label{eq:BSEq}
\ee
which corresponds to the Bethe-Salpeter equation in algebraic form. 
We hence refer to this $T_{ij}$ as the BS amplitude.  
In the present approach we have 10 meson-baryon channels, 
$i,j=K^{-} p$, $\bar{K}^{0} n$, 
$\pi ^{0} \Lambda$, $\pi ^{0} \Sigma ^{0}$, 
$\eta \Lambda$ , $\eta \Sigma ^{0}$, $\pi ^{+} \Sigma ^{-}$, 
$\pi ^{-} \Sigma ^{+}$, $K^{+} \Xi ^{-}$, and $K^{0} \Xi ^{0}$. 

Now we fix the parameters in our approach; we have the masses 
of the ground state mesons and baryons, the meson decay 
constant $f$, and the subtraction constant $a_{i} (\mu _{\text{reg}})$ with the 
regularization scale $\mu _{\text{reg}}$. We use the isospin-averaged masses for 
the mesons and baryons: 
\begin{align}
  & m_{\pi} = m_{\pi ^{0}} = m_{\pi ^{+}} = m_{\pi ^{-}} 
  & = & \phantom{0} 138.04 
  \mev , \nonumber \\ 
  & m_{K} = m_{K^{-}} = m_{\bar{K}^{0}} = m_{K^{+}} = m_{K^{0}} 
  \! \! \! \! \! & = & \phantom{0} 495.67 
  \mev , \nonumber \\ 
  & m_{\eta} & = & \phantom{0} 547.45 
  \mev , \nonumber \\ 
  & M_{N} = M_{\text{p}} = M_{\text{n}} & = & \phantom{0} 938.92 
  \mev , \nonumber \\ 
  & M_{\Lambda} & = & 1115.68 
  \mev , \nonumber \\ 
  & M_{\Sigma} = M_{\Sigma ^{0}} = M_{\Sigma ^{+}} = M_{\Sigma ^{-}} 
  & = & 1193.12 
  \mev , \nonumber \\ 
  & M_{\Xi} = M_{\Xi ^{-}} = M_{\Xi ^{0}} & = & 1318.11 
  \mev . \nonumber 
\end{align}
For the meson decay constant $f$, we choose an averaged value 
$f = 1.123 f_{\pi}$ with 
$f_{\pi}=93.0 \mev$, which is one of the typical values used 
in ChUM. 
Finally we choose the subtraction constant $a_{i}$ so as to reproduce the 
threshold properties of $\Kmp$ observed in stopped $K^{-}$ capture on 
hydrogen~\cite{Tovee:1971ga,Nowak:1978au}, 
as done in Ref.~\cite{Kaiser:1995eg},
\be
\begin{split}
& a_{\bar{K} N} = -1.84, \phantom{{}_{\eta \Lambda}} \! \! 
a_{\pi \Sigma} = -2.00, \phantom{{}_{\eta \Sigma}} \! \! 
a_{\pi \Lambda} = -1.83, \\ &
a_{\eta \Lambda} = -2.25, \phantom{{}_{\bar{K} N}} \! \! 
a_{\eta \Sigma} = -2.38, \phantom{{}_{\pi \Lambda}} \! \! 
a_{K \Xi} = -2.67, 
\end{split}
\ee
with the regularization scale $\mu _{\text{reg}} = 630 \mev$~\cite{Oset:2001cn}. 

In the present model, the $\LamFOF$ is dynamically generated in the
obtained BS scattering amplitude. With above parameters, the model
well reproduces the $\LamFOF$ mass spectrum below the $\KbarN$
threshold~\cite{Jido:2009jf}.  Furthermore, in the present model the
$\LamFOF$ is expressed by two poles of the scattering amplitude in
complex energy plane, as ($Z_{1} = 1391-66i \mev$) and
($Z_{2} = 1426-17i \mev$)~\cite{Jido:2003cb}. The study of
the renormalization condition reveals that the $\LamFOF$ resonance in
this approach is dominated by the meson-baryon molecule state and the
effect from the possible seed of the resonance in the subtraction
constant is found to be small~\cite{Hyodo:2008xr}. It has been
reported in Refs.~\cite{Borasoy:2005ie,Hyodo:2007jq} that the position
of the lower pole $Z_{1}$ is largely dependent on details of the model
parameters, whereas that of the higher pole $Z_{2}$ shows little
dependence.

The residues of the meson-baryon scattering amplitude $T_{ij}(\sqrt s)$
at the pole position express coupling strengths of the resonance to
the meson-baryon channels as seen in Eq.~\eqref{eq:T_mat}. The
coupling strengths of the two poles $Z_{1}$ and $Z_{2}$ obtained in the
present formulation are listed in
Table~\ref{tab:residua}.\footnote{The coupling strengths are shown in the
  particle basis. To compare with the coupling strengths from the amplitude in
  the isospin basis as in Ref.~\cite{Jido:2003cb}, a factor
  $\sqrt{N_I}$ should be multiplied with the isospin multiplicity
  $N_I=2$ for $\bar{K}N$ and $K\Xi$, and $N_I=3$ for $\pi\Sigma$.} From
this Table, one can see that the lower pole $Z_{1}$ strongly couples
to the $\pi \Sigma$ state, whereas the higher pole $Z_{2}$ dominantly
couples to the $\bar KN$ state.  Among the poles, the higher one is
considered to be originated from the $\KbarN$ bound
state~\cite{Hyodo:2007jq}.  Since we are interested in the structure
of the $\LamFOF$ in the $\KbarN$ bound state picture, we mainly
consider the contribution from the higher pole $Z_{2}$ and regard the
lower pole $Z_{1}$ as the background.

\begin{table}
  \caption{Coupling strengths of the $\LamFOF$ resonances,
    $Z_{1} = 1391-66i \mev$ (upper) and 
    $Z_{2} = 1426-17i \mev$ (lower), to the meson-baryon state 
    obtained as the residues of the meson-baryon scattering 
    in the particle basis.}
  \begin{ruledtabular}
    \begin{tabular}{lc}
      Channel & coupling ($Z_{1}$)  \\
      \hline
      \rule[0pt]{0pt}{10pt}$\bar{K}N$ ($K^{-} p$, $\bar{K}^{0}n$) 
      & $-0.86 + 1.26 i$  
      \\
      $\pi \Sigma$ ($\pi ^{0} \Sigma^{0}$, $\pi ^{+} \Sigma ^{-}$,
      $\pi ^{-} \Sigma ^{+}$) 
      & $-1.42 + 0.88 i$  
      \\
      $\eta \Lambda$ 
      & $-0.01 + 0.79 i$  
      \\
      $K \Xi$ ($K ^{0} \Xi ^{0}$, $K^{+} \Xi ^{-}$) 
      & $-0.33 + 0.30 i$  
      \\
      \hline
      \hline
      \\
      \hline
      \hline      
      Channel & coupling ($Z_{2}$) \\
      \hline
      \rule[0pt]{0pt}{10pt}$\bar{K}N$ ($K^{-} p$, $\bar{K}^{0}n$) 
      & $\phantom{-} 1.84 + 0.67 i$ 
      \\
      $\pi \Sigma$ ($\pi ^{0} \Sigma^{0}$, $\pi ^{+} \Sigma ^{-}$,
      $\pi ^{-} \Sigma ^{+}$) 
      & $\phantom{-} 0.26 + 0.85 i$ 
      \\
      $\eta \Lambda$ 
      & $\phantom{-} 1.44 + 0.21 i$ 
      \\
      $K \Xi$ ($K ^{0} \Xi ^{0}$, $K^{+} \Xi ^{-}$) 
      & $\phantom{-} 0.09 + 0.24 i$ 
      \\
    \end{tabular}
  \end{ruledtabular}
  \label{tab:residua}
\end{table}

\subsection{Electromagnetic interactions in chiral dynamics}
\label{subsec:EMint}

As discussed in Sec.~\ref{subsec:matrix_elements}, photon-coupling to
the resonance state is obtained as the residue of the second-rank pole
in the $MB \gamma ^{\ast} \to M^{\prime} B^{\prime}$ amplitude.  This
process can be calculated by making a photon couple to the scattering
process $MB \to M^{\prime} B^{\prime}$ which has been obtained in the
previous subsection to describe the $\LamFOF$. For the calculation of
the $MB \gamma ^{\ast} \to M^{\prime} B^{\prime}$ process, we take a
picture that the photon couples to the resonance state only through
its constituent mesons and baryons~\cite{Jido:2002yz,Sekihara:2008qk}, 
which may be valid in view of the dominance of the meson-baryon
component in the $\LamFOF$~\cite{Hyodo:2008xr}.  Thus, we need only
electromagnetic interactions of the constituent mesons and baryons in our
approach.

Due to the requirement of the gauge invariance, the elementary
couplings of the photon to the mesons and baryons should be given by
gauging the chiral effective Lagrangian in a consistent way with the
description of the $\LamFOF$. The photon couplings to the meson and
baryon appearing in the BS amplitude are derived in the minimal
coupling scheme. In addition, the anomalous magnetic couplings in the
$B B^{\prime} \gamma$ and $MB M^{\prime} B^{\prime} \gamma$ vertices
are given by the chiral perturbation theory as done in
Ref.~\cite{Jido:2002yz}. The sum of the above two contributions
determines the elementary electromagnetic couplings
$V_{\text{M}_{i}}^{\mu}$ for $M M^{\prime} \gamma$, $
V_{\text{B}_{i}}^{\mu}$ for $B B^{\prime} \gamma$, and $\Gamma
_{ij}^{\mu}$ for $M B M^{\prime} B^{\prime} \gamma$, respectively.

Now let us consider the minimal coupling scheme. The photon coupling to the 
meson, $V_{\text{M}_{i}}^{\mu}$, is given by 
\be
- i V_{\text{M}_{i}}^{\mu}(k, \, k^{\prime})
= i Q_{\text{M}_{i}} (k + k^{\prime})^{\mu} , 
\label{eq:Fermi-meson-photon} 
\ee
with the meson charge $Q_{\text{M}_{i}}$, the incoming and outgoing meson momenta 
$k^{\mu}$ and $k^{\prime \mu}$. The minimal coupling of the photon to the baryon 
is given by 
\begin{align}
& - i V_{\text{B}_{i}}^{\text{(N)}, \, \mu} (p, \, p^{\prime}) 
\nonumber \\
& = \left( i Q_{\text{B}_{i}} \frac{(p + p^{\prime})^{0}}{2 M_{i}} , \, 
i Q_{\text{B}_{i}} \frac{\bm{p} + \bm{p}^{\prime}}{2 M_{i}} 
+ i Q_{\text{B}_{i}} 
\frac{i \bm{\sigma} \times \bm{q}}{2 M_{\text{p}}} 
\right) ,
\label{eq:Fermi-baryon-photon}
\end{align}
with the baryon charge $Q_{\text{B}_{i}}$ and the incoming and
outgoing baryon momenta $p^{\mu}$ and $p^{\prime \mu}$. These two
couplings~\eqref{eq:Fermi-meson-photon} and
\eqref{eq:Fermi-baryon-photon} are appropriate with the propagators in
the loop function \eqref{eq:loop}. For the $M B M^{\prime} B^{\prime}
\gamma$ coupling, we use the following vertex, 
which is required by the Ward-Takahashi identity at tree-level 
with the Weinberg-Tomozawa interaction~\eqref{eq:Vij}:
\begin{align}
& - i \Gamma _{ij}^{\text{(N)}, \, \mu}  (P, P^{\prime})
= i \frac{C_{ij}}{4 f^{2}} 
\frac{P^{\mu} + P^{\prime \mu}}{\sqrt{\mathstrut s} + \sqrt{\mathstrut s^{\prime}}}
(Q_{\text{T}_{i}} + Q_{\text{T}_{j}}) , 
\label{eq:FermiWTphoton}
\end{align}
with the incoming and outgoing meson-baryon total momenta $P^{\mu}$
and $P^{\prime \mu}$, respectively, 
and $Q_{\text{T}_{i}}\equiv Q_{\text{M}_{i}} + Q_{\text{B}_{i}}$.  
We note that this $M B M^{\prime} B^{\prime}\gamma$ coupling does 
not contain the magnetic part.  
Actually for the electromagnetic
properties of the neutral excited baryon this term does not contribute
due to $Q_{\text{EM}}=Q_{\text{M}}+Q_{\text{B}}=0$. 
This is a different point compared with Ref.~\cite{Jido:2002yz}, 
where the normal magnetic part coming from the original form of the 
Weinberg-Tomozawa term~\eqref{eq:WToriginal}, proportional to 
$Q_{\text{M}_{i}}+Q_{\text{M}_{j}}$, was introduced. 

For the anomalous $BB^{\prime}\gamma$ and $M B M^{\prime} B^{\prime} \gamma$ couplings which are 
gauge invariant by themselves, we use the interaction Lagrangian appearing in 
the chiral perturbation theory~\cite{Meissner:1997hn}:
\begin{align}
\mathcal{L}_{\text{int}} = & - \frac{i}{4 M_{p}} b_{6}^{\text{F}} 
\text{Tr} \left( \overline{B} [ S^{\mu}, S^{\nu}][F_{\mu \nu}^{+}, B]
   \right) \nonumber \\ &
   -\frac{i}{4 M_{p}} b_{6}^{\text{D}} \text{Tr}
   \left( \overline{B} [ S^{\mu}, S^{\nu}] \{ F_{\mu \nu}^{+}, B \}
   \right) , 
\label{eq:Pauliint}
\end{align}
with
\be
   F^{+}_{\mu \nu} = - e \left( u^{\dagger} \hat{Q} F_{\mu \nu} u 
   + u \hat{Q} F_{\mu \nu} u^{\dagger} \right), 
\ee
the electromagnetic field tensor $F_{\mu \nu}$, the charge matrix $\hat{Q}$, 
the spin 
matrix $S_{\mu}$, the SU(3) matrix of the baryon octet field $B$, and the 
chiral field $u^{2}= U = \exp(i \sqrt{2} \Phi /f)$ where
$\Phi$ is the SU(3) matrix of the Nambu-Goldstone boson field. This interaction 
Lagrangian gives us spatial components of both the $BB^{\prime}\gamma$ and the 
$M B M^{\prime} B^{\prime} \gamma$ vertices ($a=1, \, 2, \, 3$):
\begin{align}
& - i V_{\text{B}_{i}}^{\text{(A)}, \, a} = 
i K_{i} \left( 
\frac{i \bm{\sigma} \times \bm{q}}{2 M_{\text{p}}} 
\right) ^{a} , \\
& - i \Gamma _{ij}^{\text{(A)}, \, a} = i A_{ij} \left(
\frac{i \bm{\sigma} \times \bm{q}}{2 M_{\text{p}}} 
\right) ^{a} , 
\label{eq:GBa} 
\end{align}
where we have made nonrelativistic reduction. Here 
the anomalous magnetic moment for the baryon $K_{i}$ and the matrix 
$A_{ij}$ are given 
by, 
\be
K_{i} = b_{6}^{\text{D}} d_{i} + b_{6}^{\text{F}} f_{i} , 
\label{eq:anomalous}
\ee
\be
A_{ij} =  \frac{b_{6}^{\text{D}} X_{ij} + 
 b_{6}^{\text{F}} Y_{ij}}{2 f^{2}} , 
\label{eq:Amat}
\ee
where the coefficients $d_{i}$, $f_{i}$, $X_{ij}$, and $Y_{ij}$ 
are fixed by the flavor SU(3) symmetry and their explicit values 
are found in Ref.~\cite{Jido:2002yz}. 
The values of the coefficients $b_{6}^{\text{D}}$ and
$b_{6}^{\text{F}}$ are 
determined to be 
$b_{6}^{\text{D}}=2.40$ and
$b_{6}^{\text{F}}=0.82$ so as to reproduce the observed anomalous
magnetic moments of the baryons\footnote{
  Here we note that in Ref.~\cite{Jido:2002yz} the additional 
    contribution ($\Delta b_{6}^{\text{F}}=1$) was introduced in order to 
    take into account the normal magnetic part of the gauged Weinberg-Tomozawa 
    interaction.  This is unnecessary in our calculation, where 
    $b_{6}^{\text{F}}=0.82$, 
    because Eq.~\eqref{eq:FermiWTphoton} does not produce 
    the magnetic part. }.  In the
calculation, these values are used for the $M B M^{\prime} B^{\prime}
\gamma$ vertices \eqref{eq:GBa}, while for the baryon anomalous
magnetic moments we use the experimental values instead of
$K_{i}$~\eqref{eq:anomalous}.  For the unobserved $\Sigma ^{0}$
magnetic moment, we use the SU(3) flavor relation $\mu_{\Sigma ^{0}} =
(\mu_{\Sigma ^{+}} + \mu_{\Sigma ^{-}})/2$, which is consistent with
quark models.  The transition magnetic coupling $\Sigma^{0} \to 
\Lambda \gamma$ does not contribute in the isospin symmetric limit,
because this interaction changes the isospin of the excited baryon $0$
to $1$.  Now the total $BB^{\prime}\gamma$ coupling
$V_{\text{B}_{i}}^{\mu}$ and the $M B M^{\prime} B^{\prime} \gamma$
vertex $\Gamma _{ij}^{\mu}$ are given by the sum of the normal and
anomalous contributions:
\begin{align}
  & V_{\text{B}_{i}}^{\mu} (p, \, p^{\prime})
\nonumber \\
  & = \left( - Q_{\text{B}_{i}} \frac{(p + p^{\prime})^{0}}{2 M_{i}} , \, 
    - Q_{\text{B}_{i}} \frac{\bm{p} + \bm{p}^{\prime}}{2 M_{i}} 
    - \mu_{\text{B}_{i}} 
\frac{i \bm{\sigma} \times \bm{q}}{2 M_{\text{p}}} 
\right) ,
  \label{eq:baryon-photon} \\
&  \Gamma _{ij}^{\mu}  (P, P^{\prime})
\nonumber \\ & 
= \left ( - \frac{C_{ij}}{4 f^{2}} 
\frac{P^{0} + P^{\prime 0}}{\sqrt{\mathstrut s} + \sqrt{\mathstrut s^{\prime}}}
(Q_{\text{T}_{i}} + Q_{\text{T}_{j}}) , \right. 
\nonumber \\ & 
\left. - \frac{C_{ij}}{4 f^{2}} 
\frac{\bm{P} + \bm{P}^{\prime}}
{\sqrt{\mathstrut s} + \sqrt{\mathstrut s^{\prime}}}
(Q_{\text{T}_{i}} + Q_{\text{T}_{j}}) 
- A_{ij} \frac{i \bm{\sigma} \times \bm{q}}{2 M_{\text{p}}} \right ) ,
  \label{eq:Gamma}
\end{align}
where $\mu _{\text{B}_{i}}$ is the observed magnetic moment of the baryon.

Here we note that the magnetic interactions of the excited baryons are
obtained only from the baryonic dynamics, since in our approach the
spinless meson is bound by the baryon in $s$-wave channel and the
spatial component of the $M M^{\prime} \gamma$ couplings do not
contribute to the magnetic interactions in the Breit frame of the
excited baryons.

In order to study the internal structure of the $\LamFOF$ theoretically,
we also consider the form factors probed with the baryonic and strangeness 
currents. 
For the baryonic and strangeness current interactions, we replace the meson and 
baryon electric charges with the corresponding quantum numbers, namely,
the baryon number and strangeness of the mesons and baryons:
\be
Q_{\text{M}} =0 , \quad Q_{\text{B}}=1, 
\label{eq:BCh}
\ee
for the baryonic current and
\begin{align}
 & Q_{\pi}=Q_{\eta}=Q_{N}=0 , \nonumber \\
 & Q_{\bar K}=Q_{\Lambda}=Q_{\Sigma} = -1 , \label{eq:SCh} \\
 & Q_{K}=1 , \quad Q_{\Xi} = -2 , \nonumber
\end{align}
for the strangeness current. 
Then we consider the time components as the form factors for 
the baryonic and strangeness current 
interactions, as described below Eq.~\eqref{eq:definition}. 
Since the baryonic and strangeness form factors are calculated by
the time component of the current, we do not need the counterparts of 
the matrices $X_{ij}$ and $Y_{ij}$ given in Eq.~\eqref{eq:Amat},
which contribute to the spatial component.

\subsection{Calculation of photon-coupled meson-baryon amplitudes}
\label{subsec:photon-coupled}

\begin{figure*}[!Ht]
 \centering
 \begin{tabular*}{\textwidth}{@{\extracolsep{\fill}}ccc}
    \includegraphics[scale=0.17]{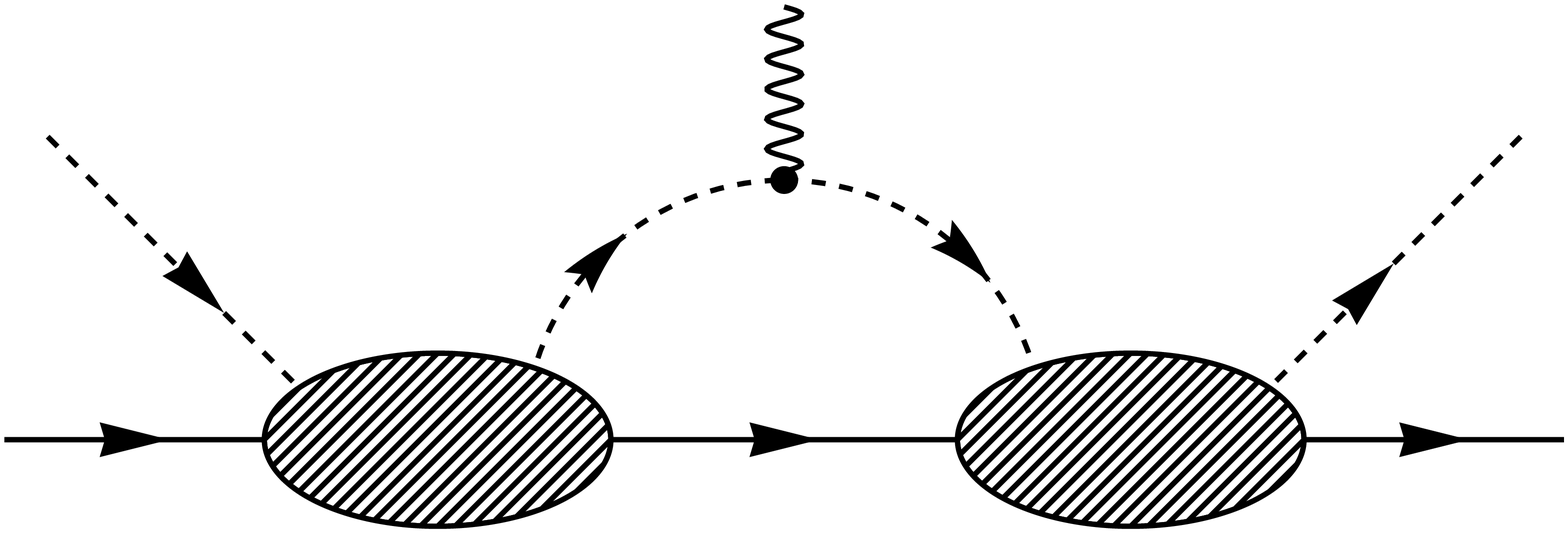} &
    \includegraphics[scale=0.17]{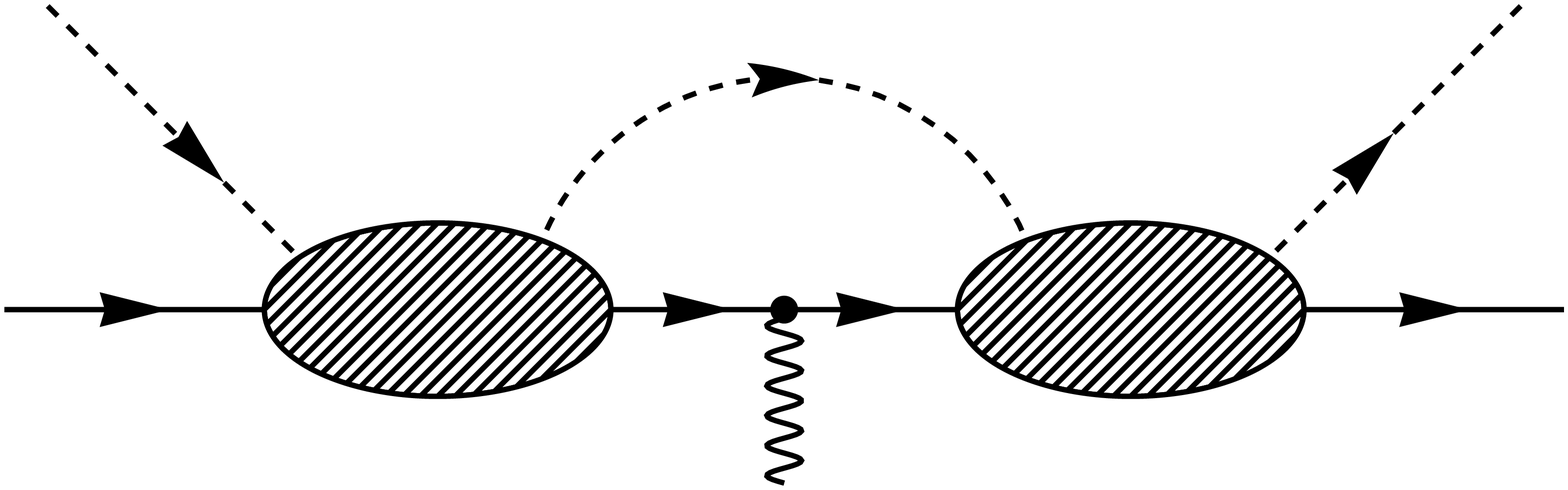} &
    \includegraphics[scale=0.17]{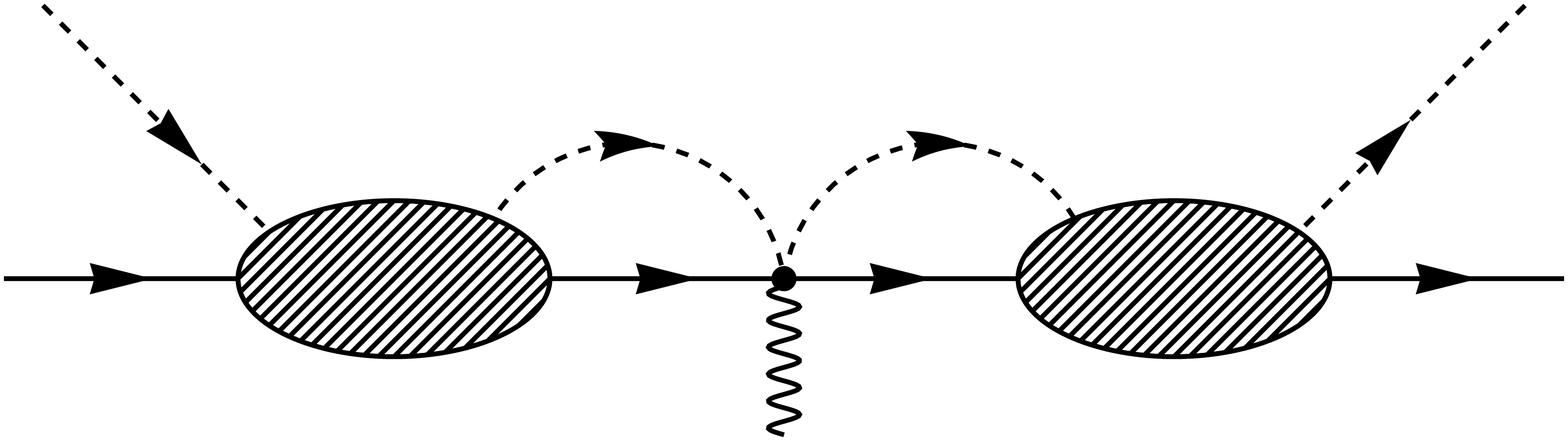} \\
    $T_{\gamma (1)}^{\mu}$ &
    $T_{\gamma (2)}^{\mu}$ &
    $T_{\gamma (3)}^{\mu}$
 \end{tabular*}
 \caption{Diagrams for the $T_{\gamma}^{\mu}$ 
   which contain the 
   double-pole terms of the excited
   baryon~\cite{Sekihara:2008qk,Nacher:1999ni,Borasoy:2005zg}.  The
   shaded ellipses represent the BS amplitude.  The dashed, solid and
   wiggly lines correspond to the ground state meson, the ground state
   baryon and the probe current, respectively.}
 \label{fig:Tgamma}
\end{figure*}%

\begin{figure*}[!Ht]
  \centering
  \begin{tabular*}{12.0cm}
    {@{\extracolsep{\fill}}ccc}
    \includegraphics[scale=0.17]{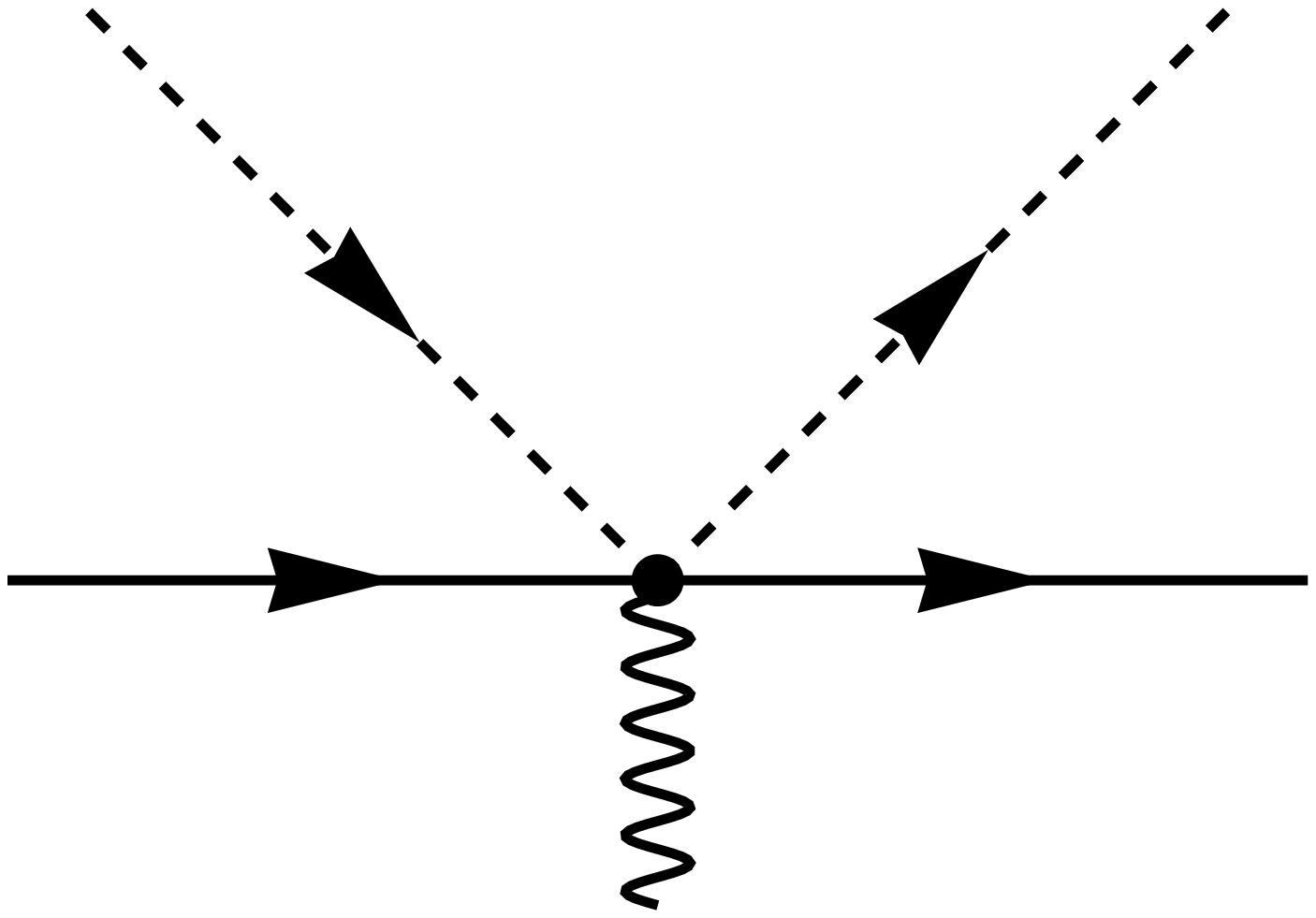} &
    \includegraphics[scale=0.17]{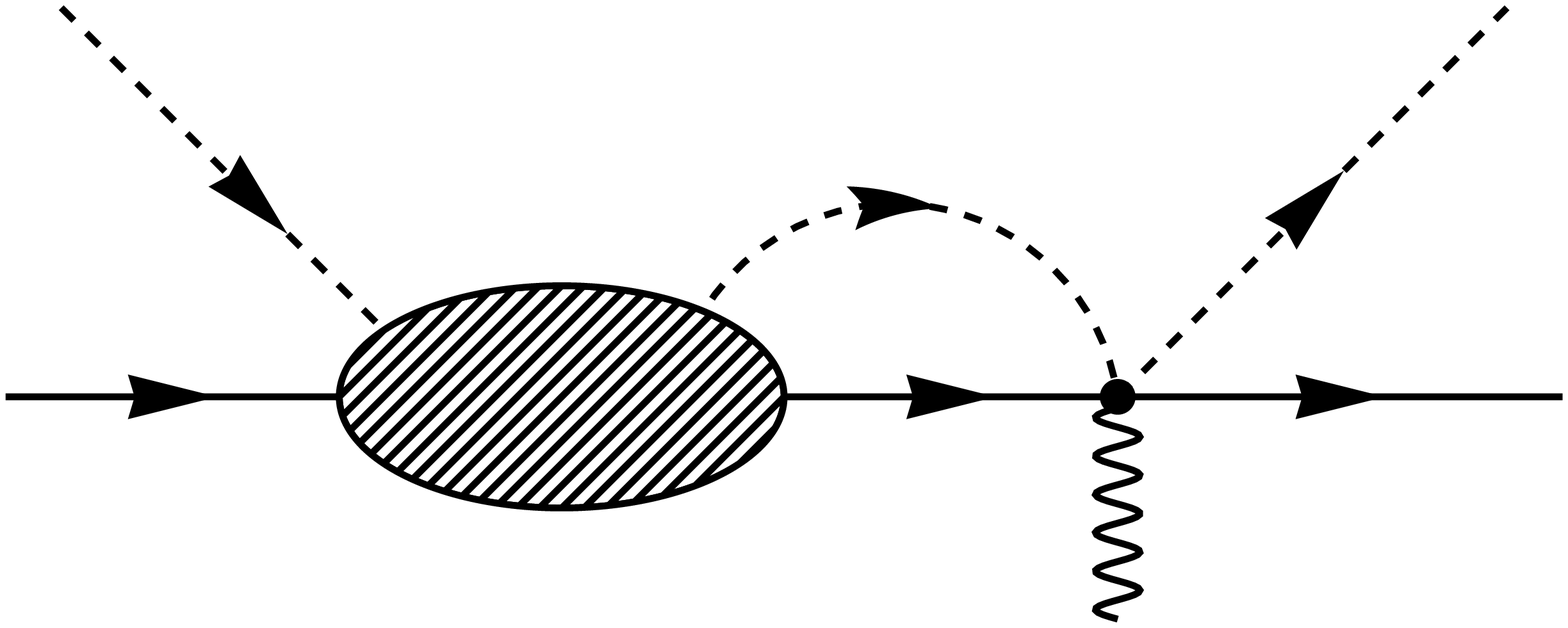} &
    \includegraphics[scale=0.17]{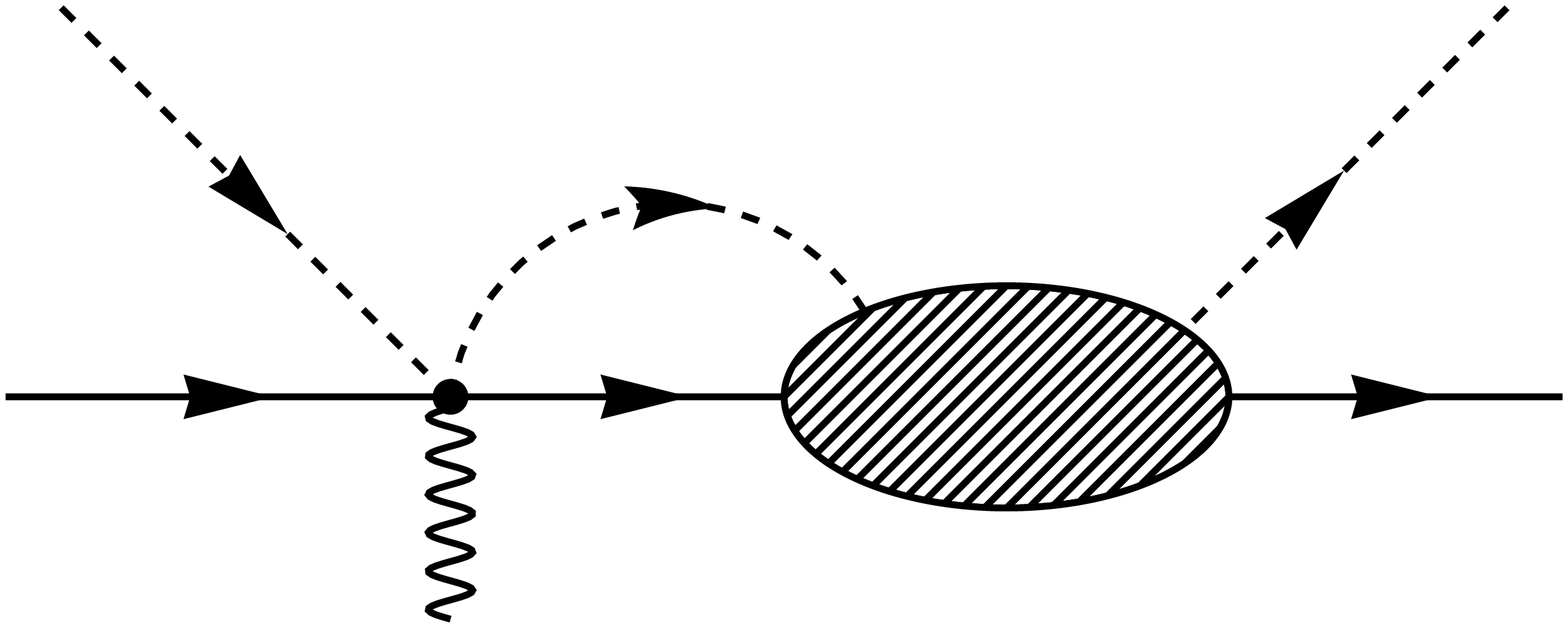} \\
    $T_{\gamma (4)}^{\mu}$ & 
    $T_{\gamma (5)}^{\mu}$  \vspace{10pt} &
    $T_{\gamma (6)}^{\mu}$ 
  \end{tabular*}
  \begin{tabular*}{14.0cm}
    {@{\extracolsep{\fill}}cccc}
    \includegraphics[scale=0.17]{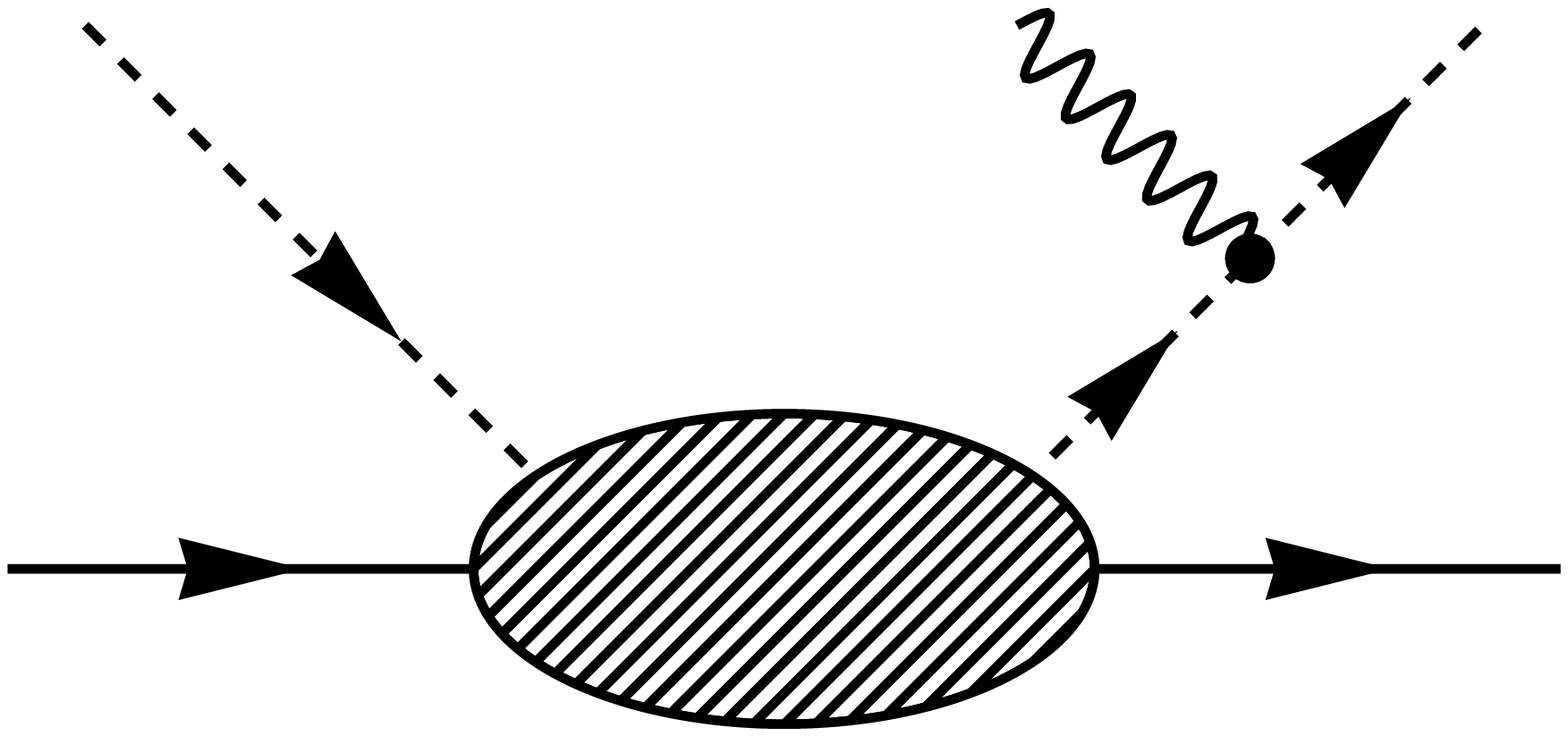} &
    \includegraphics[scale=0.17]{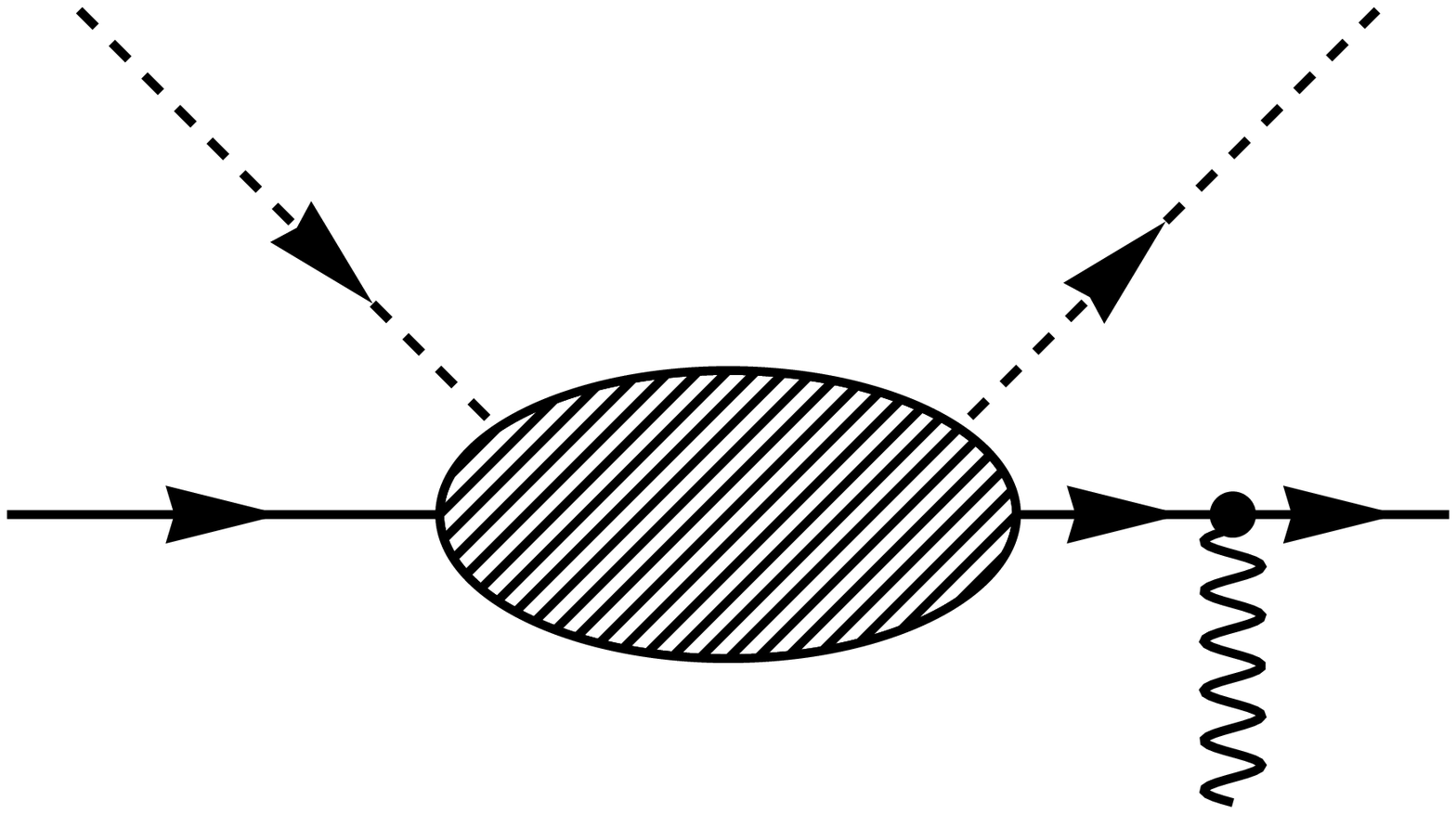} &
    \includegraphics[scale=0.17]{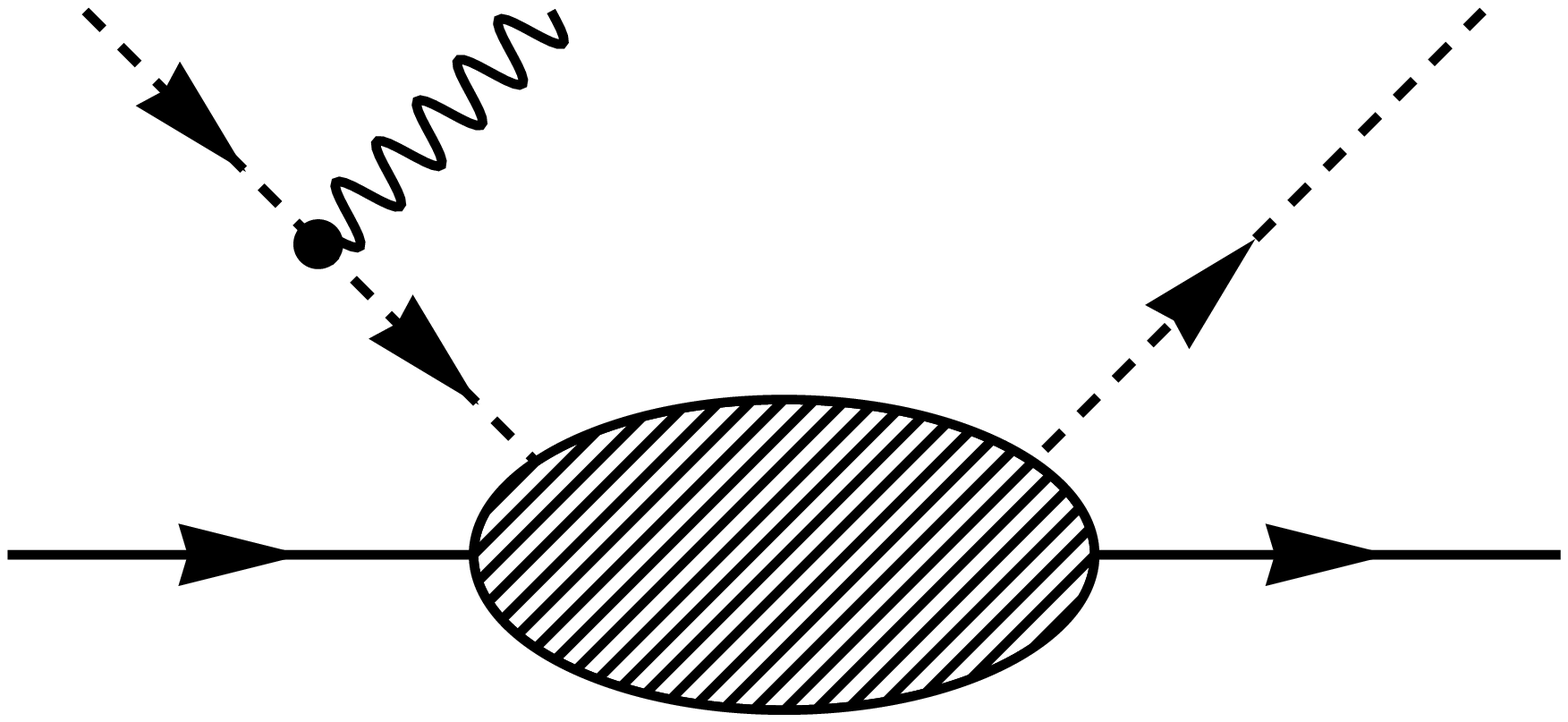} &
    \includegraphics[scale=0.17]{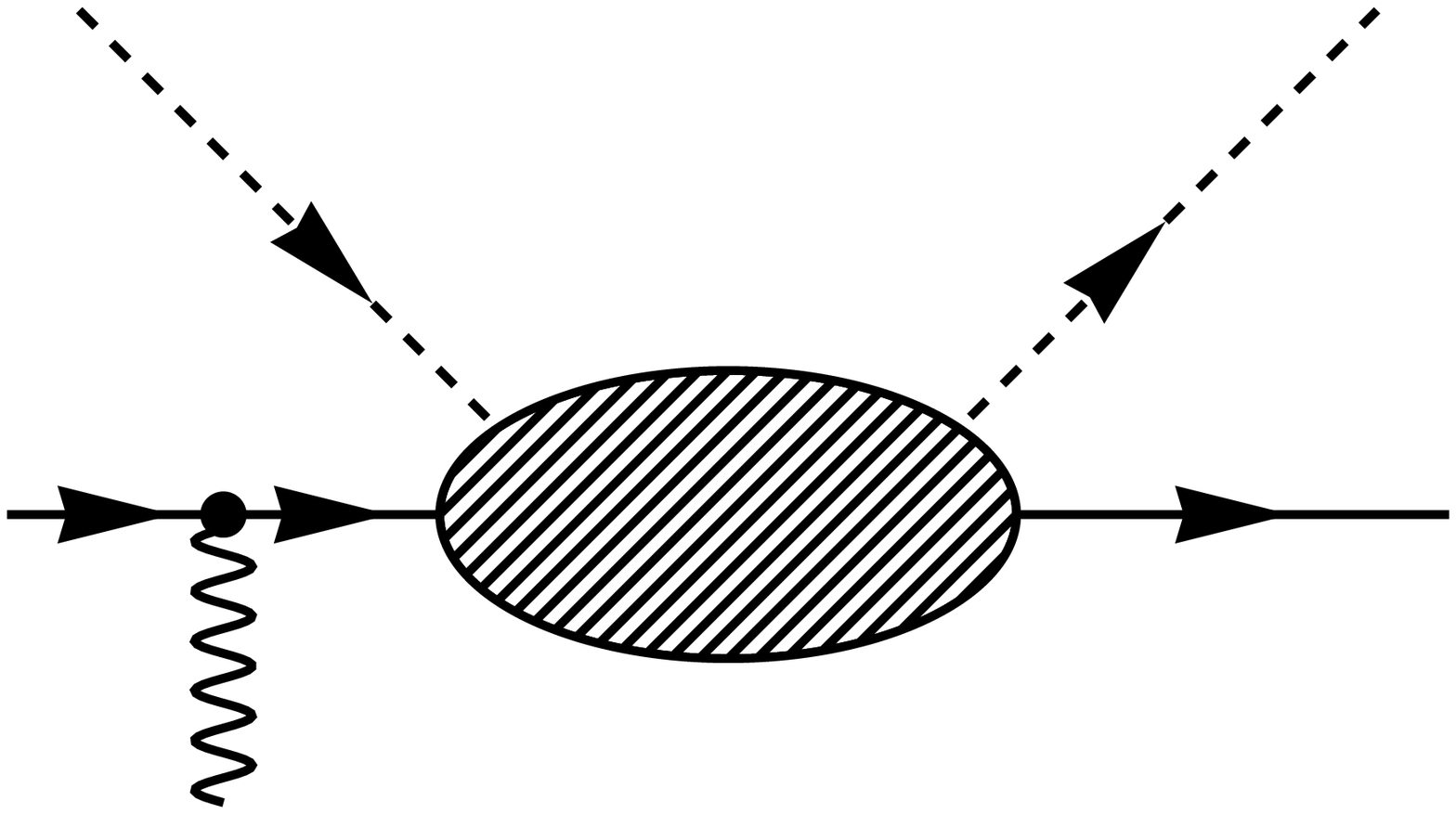} \\
    $T_{\gamma (7)}^{\mu}$ &
    $T_{\gamma (8)}^{\mu}$ &
    $T_{\gamma (9)}^{\mu}$ &
    $T_{\gamma (10)}^{\mu}$ 
  \end{tabular*}
  \caption{Supplemental diagrams for the charge conservation 
    of the $T_{\gamma}^{\mu}$ 
    in addition to the diagrams in Fig.~\ref{fig:Tgamma}~\cite{Borasoy:2005zg}. 
    The shaded ellipses represent the BS amplitude.
    The dashed, solid and wiggly lines correspond to the ground state meson, the ground state baryon and the probe current, respectively.
     }
  \label{fig:Tgamma-other}
\end{figure*}

Now we discuss the details of the calculation of the scattering
amplitude of the $M B \gamma^{\ast} \to M^{\prime} B^{\prime}$ process
in the chiral unitary approach, in which the amplitude for the $MB
\rightarrow M^{\prime}B^{\prime}$ is given by multiple scattering of
the meson and baryon.

One of the most important issues we take account of is the charge
conservation in the calculation of the scattering amplitude for the $M
B \gamma^{\ast} \to M^{\prime} B^{\prime}$ process. This ensures the
correct normalization of the form factor of the excited baryon, $\FE
(Q^{2}=0)=Q_{\text{EM}}$, $\FB (Q^{2}=0)=B=1$, and $\FS
(Q^{2}=0)=S$. Following the method proposed in
Refs.~\cite{Nacher:1999ni,Borasoy:2005zg,Sekihara:2008qk}, 
to calculate the form
factors we take three relevant diagrams shown in
Fig.~\ref{fig:Tgamma}, which contain the double-pole 
terms for the resonance
states.  Although charge conservation requires seven other diagrams as
shown in Fig.~\ref{fig:Tgamma-other} for the general amplitude
$T^{\mu}_{\gamma}$~\cite{Borasoy:2005zg}, these diagrams cannot
contribute to the matrix elements at the resonance pole calculated by
Eq.~\eqref{eq:Res_scheme}, since these terms have at most a single
pole~\cite{Sekihara:2008qk}. This means that, on the resonance pole,
the charge conservation is maintained by only three diagrams shown in
Fig.~\ref{fig:Tgamma}. Summing up the diagrams in
Fig.~\ref{fig:Tgamma}, we obtain the relevant amplitude for the
evaluation of the form factors:
\be
T_{\gamma ij}^{\mu} 
\equiv T_{\gamma (1) ij}^{\mu} + T_{\gamma (2) ij}^{\mu} + T_{\gamma (3) ij}^{\mu} .
\label{eq:Tgamma}
\ee
These contributions can be expressed by the combination of the BS amplitude and 
the elementary couplings discussed before, according to the Feynman diagrams 
given in Fig.~\ref{fig:Tgamma}. In the Breit frame, in which the momenta of 
the photon and the $\LamFOF$ before photon coupling 
are expressed as $q^{\mu} = (0, \, \bm{q})$ and 
$P^{\mu} = \left( \sqrt{s + {\bm{q}^{2}}/{4}}, \, -{\bm{q}}/{2} \right)$, 
respectively, their explicit forms are written as:
\begin{align}
& T_{\gamma (1) ij}^{\mu} = \sum _{k} 
T_{ik} (\sqrt{s}) D_{\text{M}_{k}}^{\mu} ( \sqrt{s};\, Q^2 ) T_{kj} (\sqrt{s}) , 
\label{eq:Tgamma1} \\
& T_{\gamma (2) ij}^{\mu} = \sum _{k} 
T_{ik} (\sqrt{s}) D_{\text{B}_{k}}^{\mu} ( \sqrt{s};\, Q^2 ) T_{kj} (\sqrt{s}) , 
\label{eq:Tgamma2} \\
& T_{\gamma (3) ij}^{\mu} = \sum _{k,l}
T_{ik} (\sqrt{s}) G_{k} (\sqrt{s}) 
\Gamma _{kl}^{\mu} ( \sqrt{s};\, Q^2 ) G_{l} (\sqrt{s}) T_{lj} (\sqrt{s}) ,
\label{eq:Tgamma3} 
\end{align}
where the vertex $\Gamma _{kl}^{\mu}$ is given in Eq.~\eqref{eq:Gamma}
and the loop integrals with the photon couplings to the meson and
baryon are given by, 
\begin{widetext}
\begin{align}
& D_{\text{M}_{k}}^{\mu} ( \sqrt{s};\, Q^2 ) \equiv i
\int \frac{d^4 q_{1}}{(2 \pi )^4} 
\frac{2 M_{k}}{(P - q_{1})^2 - M_{k}^2 + i \epsilon }
\frac{1}{(q_{1} + q)^{2} - m_{k}^{2} + i \epsilon } 
\big[ V_{\text{M}_{k}}^{\mu} 
(q_{1}, \, q_{1} + q) \big]
\frac{1}{q_{1}^2 - m_{k}^2 + i \epsilon } , 
\label{eq:DloopM} 
\\
& D_{\text{B}_{k}}^{\mu} ( \sqrt{s}; \, Q^2 ) \equiv i 
\int \frac{d^4 q_{1}}{(2 \pi )^4} 
\frac{1}{q_{1}^2 - m_{k}^2 + i \epsilon } 
\frac{2 M_{k}} {(P + q - q_{1})^2 - M_{k}^2 + i \epsilon } 
\big[ V_{\text{B}_{k}}^{\mu} 
(P - q_{1}, \, P - q_{1} + q) \big]
\frac{2 M_{k}}{(P - q_{1})^2 - M_{k}^2 + i \epsilon } . 
\label{eq:DloopB} 
\end{align}
\end{widetext}
In the Breit frame $s = P^{\mu} P_{\mu} = (P+q)^{\mu} (P+q)_{\mu}$ and 
$2P^{\mu} q_{\mu} = \bm{q}^{2}= Q^{2} \geq 0$, and $D_{\text{M}}$, 
$D_{\text{B}}$, and $\Gamma$ are functions of $\sqrt{s}$ and $Q^{2}$. 
The function $D_{\text{M}}$ ($D_{\text{B}}$) at $Q^{2}=0$ is related to
the loop integral $G$ given in \eqref{eq:loop} as, 
\be
D_{\text{M}_{k}}^{0} ( \sqrt{s}; \, Q^{2}=0 ) 
= Q_{\text{M}_{k}} \frac{d G_{k}}{d \sqrt{s}} , 
\label{eq:DMGM}
\ee
\be
D_{\text{B}_{k}}^{0} ( \sqrt{s}; \, Q^{2}=0 ) 
= Q_{\text{B}_{k}} \frac{d G_{k}}{d \sqrt{s}} , 
\label{eq:DBGB}
\ee which can be easily proved by calculating the derivative of the
loop integral $G$ and using the vertices~\eqref{eq:Fermi-meson-photon}
and \eqref{eq:baryon-photon}. Other analytic properties of these loop
integrals, $D_{\text{M}}$ and $D_{\text{B}}$, are discussed in
Appendix~\ref{sec:Loop}.  \footnote{If one adopts the cut-off
    procedure for the meson-baryon loop integral $G$, one must
    calculate the photon-coupled meson-baryon loop integrals
    $D_{\text{M}}$ and $D_{\text{B}}$ in a consistent way with the
    same cut-off.  Form factors of $\LamFOF$ in such a cut-off
    procedure were evaluated in Ref.~\cite{YamagataSekihara:2010pj}.
    The two approaches, the dimensional regularization in our
    procudure and the cut-off procedure, would not make much
    differences in form factors except for the high momentum region
    compared to the cut-off scale.}

Now we introduce the finite size effects of the constituent hadrons
in our scheme.  The ground state mesons and baryons have
spatial structures, while we have implicitly assumed that they are
pointlike particles in the effective Lagrangian approach. Therefore we 
should include the finite size effects of the constituent
hadrons in a gauge invariant way. Here we simply multiply a common
form factor (CFF) $\FCFF (Q^{2})$ of the constituent hadrons to each
photon vertex,
\begin{align}
& D_{\text{M}_{k}}^{\mu}(\sqrt{s}; \, Q^2) 
\to D_{\text{M}_{k}}^{\mu}(\sqrt{s}; \, Q^2) \FCFF (Q^{2}) , \\
&  D_{\text{B}_{k}}^{\mu}(\sqrt{s}; \, Q^2)
\to  D_{\text{B}_{k}}^{\mu}(\sqrt{s}; \, Q^2) \FCFF (Q^{2}) ,
 \\
& \Gamma _{kl}^{\mu} (\sqrt{s}; \, Q^2)
\to \Gamma _{kl}^{\mu} (\sqrt{s}; \, Q^2) \FCFF (Q^{2}) . 
\end{align}
Note that the $\FCFF$ only depends on $Q^2$ so it can be factorized out 
from the 
loop integrals of Eqs.~\eqref{eq:DloopM} and \eqref{eq:DloopB}. In this study 
we employ the dipole type form factor as, 
\be
\FCFF (Q^{2}) 
= \left( \frac{\Lambda ^{2}}{\Lambda ^{2} + Q^{2}} \right)^{2} . 
\label{eq:dipole}
\ee
We take $\Lambda ^{2}=0.71 \, \text{GeV}^{2}$, which reproduces nucleon 
form factors well. This CFF corresponds to the hadron density 
$\sim \exp (- \Lambda r)$ with radial coordinate $r$ of each hadron and 
mean squared radius $\MSR =12/\Lambda ^{2} \simeq 0.66 \, \text{fm}^{2}$.

Here we comment on the normalization of the effective form factor
$\Feff$ obtained on the real axis.  As discussed in
Appendix~\ref{sec:normalization}, the effective form factor is
correctly normalized so as to give the corresponding charge of the
system at $Q^{2}=0$ independently of the energy $\sqrt{s}$, if we
consider all of the diagrams shown in Figs.~\ref{fig:Tgamma} and
\ref{fig:Tgamma-other}.  Among them, the three double-pole diagrams in
Fig.~\ref{fig:Tgamma} give dominant contributions at the energies
close to the resonance pole position. 
Therefore, even if we take into account only the three diagrams 
in Eq.~\eqref{eq:Tgamma}, 
deviation from the correct normalization for
the effective form factor
is considered to be small. 
We also note that a part of the less singular terms 
in $T_{\gamma ij}^{\mu}$ (hence $\Feff$) is included in these three 
diagrams because off the pole
position the nonresonant background in $T_{ij}$ [see
Eq.~\eqref{eq:T_matreal}] generates less singular parts in
Eqs.~\eqref{eq:Tgamma1}--\eqref{eq:Tgamma3}.  Here we choose the
$\KbarN (I=0) \gamma ^{\ast} \to \KbarN (I=0)$ channel for the
evaluation of $\Feff$ for the $\LamFOF$, which reduces the contribution
from the less singular terms thanks to the large coupling strength
$g_{\KbarN}$, as demonstrated in Eq.~\eqref{eq:effFFapprox}.

At last, let us show a relation among electric ($\FE$), baryonic ($\FB$), 
and strangeness ($\FS$) form factors for the $\LamFOF$ based on isospin 
symmetry, in which we have a generalized Gell-Mann-Nishijima relation 
for the probe current, 
\be
J_{\text{E}}^{\mu} = 
J_{I_{z}}^{\mu} 
+ \frac{1}{2} [ J_{\text{B}}^{\mu} + J_{\text{S}}^{\mu} ] , 
\ee
with the current for the third component of the isospin, $J_{I_{z}}^{\mu}$. 
Taking the matrix element for the $\LamFOF$ with $I=0$, 
we obtain\footnote{At $Q^{2}=0$, we obtain the usual 
Gell-Mann-Nishijima relation, $Q_{\text{EM}}=(B+S)/2$, 
for $I_{z}=0$ hadrons}, 
\be
  \FE(Q^{2}) 
  = \frac{1}{2} [F_{\text{B}}(Q^{2}) + F_{\text{S}}(Q^{2}) ] ,
  \label{eq:genGN}
\ee
where the matrix element of $J_{I_{z}}^{\mu}$ vanishes for the $I=0$ state. 
From the relation~\eqref{eq:genGN}, we have the relation for the 
spatial densities, 
\be
  \rho _{\text{E}} (r)
  = \frac{1}{2} [\rho _{\text{B}} (r) + \rho _{\text{S}} (r) ] , 
\ee
with radial coordinate $r$ as well.

\section{Numerical Results} 
\label{sec:Result}

In this section, we discuss the internal structure of the resonant
$\LamFOF$ state.  We will show the numerical results of the $\LamFOF$
form factors measured by the electromagnetic, baryon number, and
strangeness currents in momentum space.  We will also show the spatial
density distributions in coordinate space which are obtained by
performing Fourier transformation of the form factors.

The calculation of the form factor is performed in two ways; one is on
the $\LamFOF$ pole position (Sec.~\ref{subsec:on_pole} and the other
is to evaluate the effective form factor~\eqref{eq:Feffective} on the
real energy axis around the resonance energy region $\sqrt{s}\sim 1420
\mev$ (Sec.~\ref{subsec:on_real}.  On the $\LamFOF$ pole position,
the internal structure of the resonance can be obtained in a way to
keep charge conservation without nonresonant background
contributions. The results, however, may not be directly compared with
experimental observables. The effective form factor on the real energy
axis, on the other hand, may be 
determined 
in experiments, but the
obtained form factors have both the resonant and nonresonant
contributions.

For a reference of the electric size of the typical neutral baryon, we
will compare the electric $\LamFOF$ form factor with 
an experimental fit of neutron electric form
factor~\cite{Platchkov:1989ch},
\begin{align}
& F_{\text{E}}^{\text{n}} (Q^{2}) 
= - \frac{a \mu _{\text{n}} \tau }{1 + b \tau} 
\left( \frac{\Lambda ^{2}}{\Lambda ^{2} + Q^{2}} \right)^{2} , 
\quad 
\tau = \frac{Q^{2}}{4 M_{\text{n}}^{2}} 
\label{eq:FEneutron} 
\end{align}
with $a=1.25$, $b=18.3$, $\Lambda ^{2}=0.71 \, \text{GeV}^{2}$, the
neutron mass $M_{\text{n}}$, and the neutron magnetic moment $\mu
_{\text{n}}=-1.913 ~ \mu_{\text{N}}$, where $\mu_{\text{N}}$ is the
nuclear magneton.

We also compare the results of magnetic, baryonic and strangeness form factors 
with a dipole type form factor, 
\be
F_{\text{dipole}} (Q^{2}) 
= c \times 
\left( \frac{\Lambda ^{2}}{\Lambda ^{2} + Q^{2}} \right)^{2} , 
\label{eq:Gdipolefit}
\ee 
with $\Lambda ^{2}=0.71 \, \text{GeV}^{2}$, which reproduces well 
the observed nucleon magnetic form factor~\cite{Lachniet:2008qf}.  
The overall factor $c$ will
be adjusted to the normalization of the form factor of the
$\Lambda(1405)$ in question.

\subsection{Form factors on the resonance pole}
\label{subsec:on_pole}

Here we discuss the internal structure of the $\LamFOF$ using the form
factors obtained at the pole position in the complex energy plane. We
evaluate the form factors of the higher $\LamFOF$ state, $Z_{2}$, out
of two $\LamFOF$ states, since this state gives the dominant
contribution to the spectrum and is considered to be originated from
the $\KbarN$ bound state. The form factors of the $\LamFOF$ at the
resonance position are obtained by Eq.~\eqref{eq:Res_scheme} together
with the amplitudes calculated in Eqs.~\eqref{eq:BSEq} and
\eqref{eq:Tgamma}.

\subsubsection{Electromagnetic, baryonic, and strangeness  structures}

\begin{figure*}[!Ht]
  \centering
  \begin{tabular*}{\textwidth}{@{\extracolsep{\fill}}cc}
    \includegraphics[width=8.6cm]{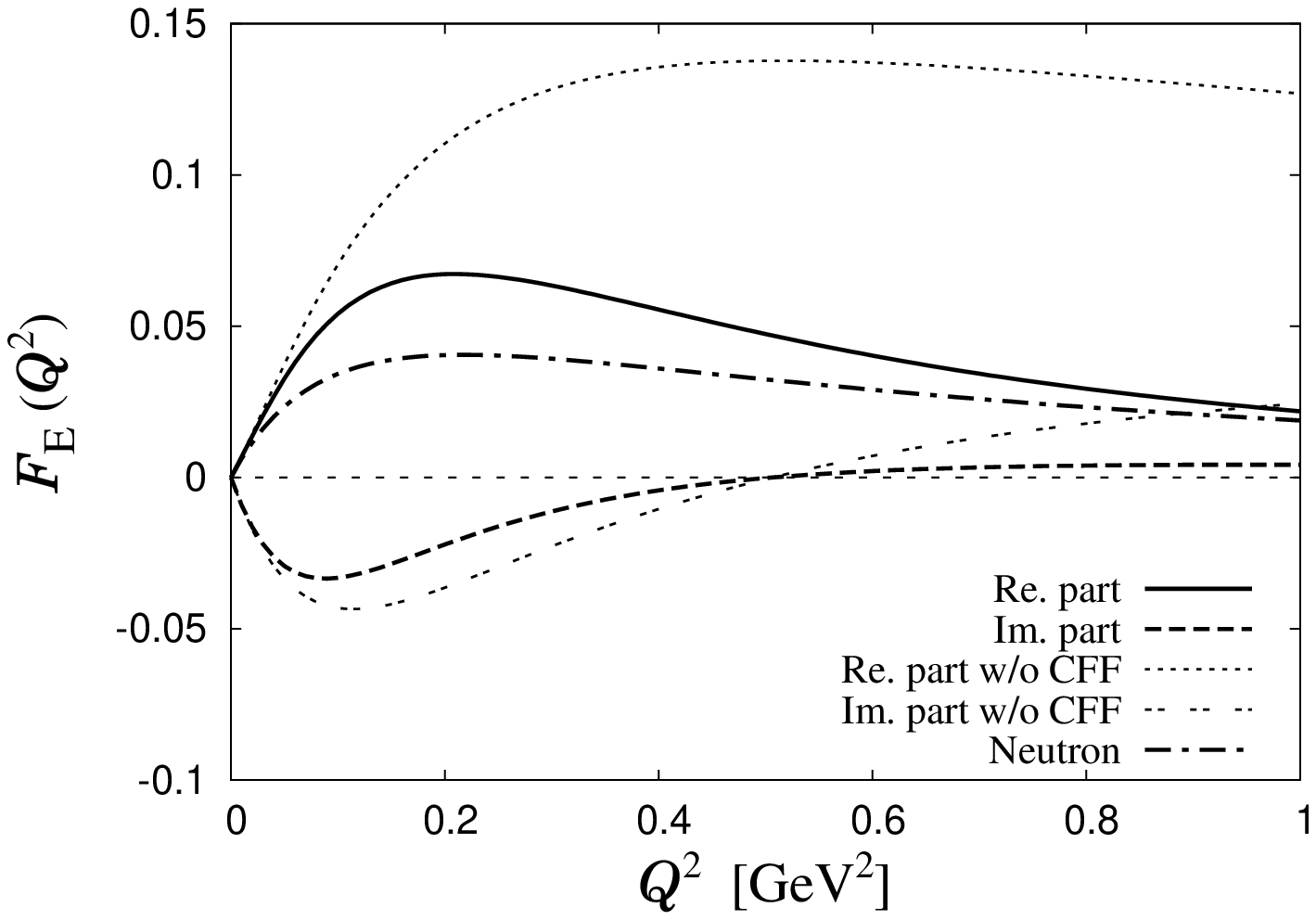} &
    \includegraphics[width=8.6cm]{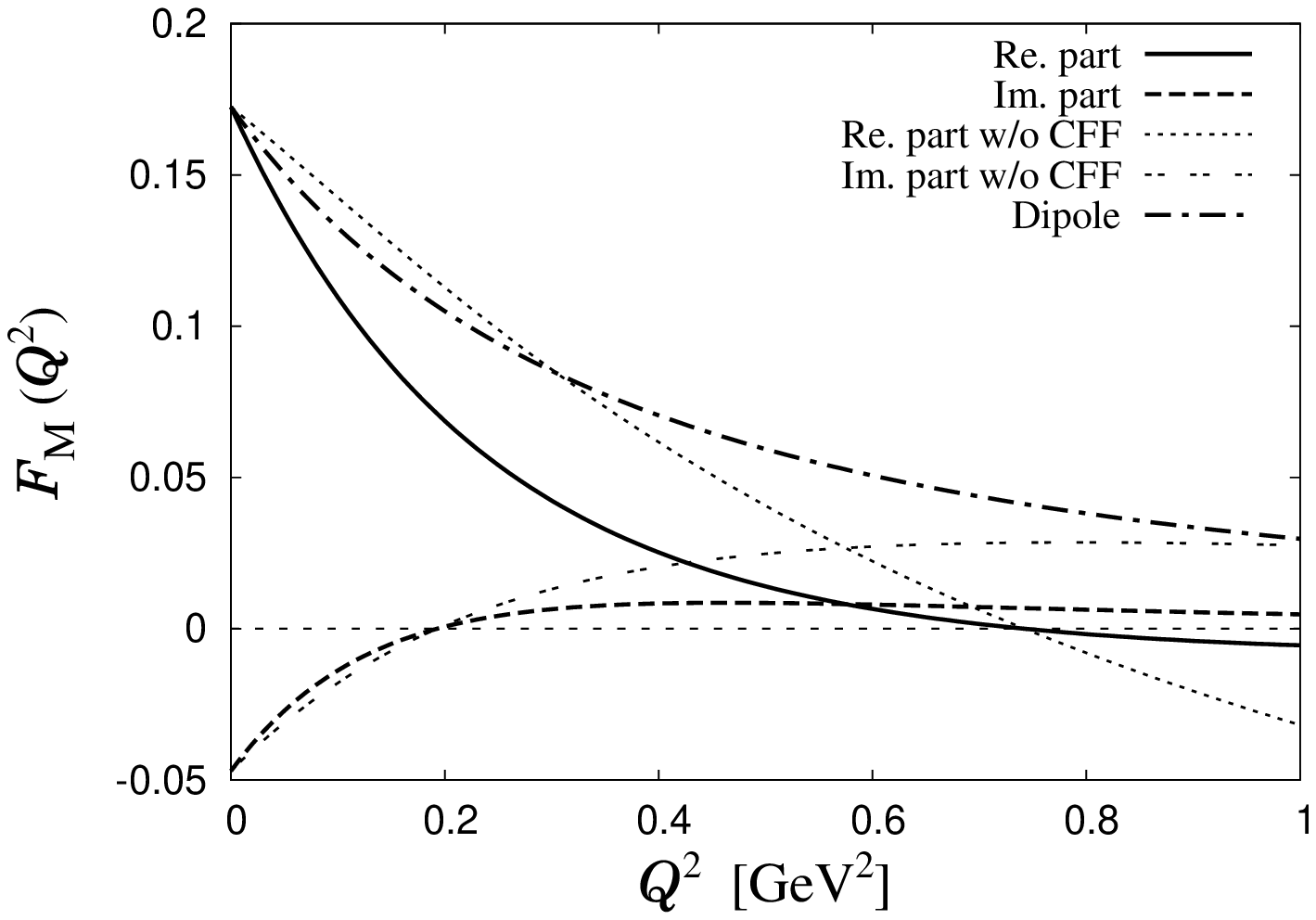} 
  \end{tabular*}
  \caption{
    Electromagnetic form factors of the 
    $\LamFOF$ state on the higher pole position $Z_{2}$, together with the 
    empirical form factors of the neutron. Left (right) panel shows the 
    electric (magnetic) form factor $\FE$ ($\FM$). The label ``w/o CFF'' 
    represents the result without inclusion of the common form factor 
    in  Eq.~\eqref{eq:dipole}. The parameter $c$ in the dipole form factor is 
    chosen to be $c=\text{Re} [\FM (Q^{2}=0)]$, the real part of the magnetic 
    moment of the $\LamFOF$.}
  \label{fig:FFEM_pole}
\end{figure*}

First of all, we show our results of the electric and magnetic form
factors in Fig.~\ref{fig:FFEM_pole} together with the empirical form
factors of the neutron given in Eqs.~\eqref{eq:FEneutron} and
\eqref{eq:Gdipolefit}.  The normalization parameter is given by the
real part of the magnetic moment of the $\LamFOF$, $c=\text{Re} [\FM
(Q^2=0)]$.  Here, in order to see the finite size effects of the
constituent hadrons, we also show results without the common
form factor (CFF) introduced in Eq.~\eqref{eq:dipole}.  The finite
size effects make the magnitude of the form factors reduced, especially
in the large $Q^{2}$ region. Hereafter, we show only the results with
CFF unless explicit mentionings. 

Now let us discuss the electromagnetic form factors of the resonant
$\LamFOF$ shown in Fig.~\ref{fig:FFEM_pole}.  
The form factors $\FE$ and $\FM$ contain the imaginary parts,
since they are evaluated on the resonance pole position in the complex
energy plane. However, we obtain the imaginary parts in smaller
magnitude than the real parts. This 
is 
the consequence of the
relatively small imaginary part of the pole position of $Z_{2}$, since
the form factors 
are 
real numbers in the limit of zero imaginary part
of the pole position.
For the charge neutral
$\LamFOF$, deviation from zero in the electric form factor indicates
that the $\LamFOF$ has a nontrivial charge
distribution as seen in the neutron form factor. Comparing the real
part of the $\LamFOF$ form factor and the empirical neutron form
factor, we find that 
the $\LamFOF$ form factor has larger magnitude than that of 
the neutron, especially at low $Q^{2}$. This indicates that the 
spatial structure of the $\LamFOF$ is larger than the neutron. 
For the magnetic
form factor of the $\LamFOF$, to which only the baryon components
contribute as seen in Sec.~\ref{subsec:EMint}, the real part of the
result shows faster decreasing than the dipole fit to the
nucleon. These results of $\FE$ and $\FM$ suggest the peculiar
electromagnetic structure of the $\LamFOF$ compared with the typical
neutral baryon such as the neutron.

\begin{figure}[!t]
  \centering
  \begin{tabular*}{8.6cm}{@{\extracolsep{\fill}}c}
    \includegraphics[width=8.6cm]{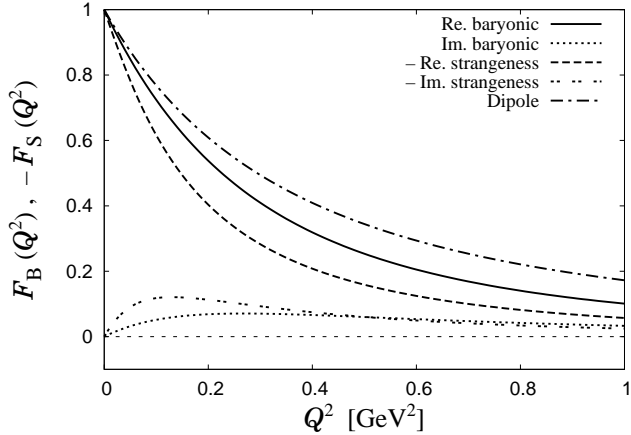} 
  \end{tabular*}
  \caption{Baryonic ($\FB$) and strangeness ($\FS$) form factors of the 
    $\LamFOF$ state on the higher pole position $Z_{2}$. The strangeness 
    form factor is presented with the opposite 
    sign for comparison. The parameter $c$ in the dipole form factor 
    is chosen to be $c=1$. 
  }
  \label{fig:FFBS_pole}
\end{figure}

Next we discuss the baryonic and strangeness form factors of the
$\LamFOF$ using the baryonic and strangeness currents as external
probes.  The calculation is done in the same way as the electric form
factor 
using the baryonic charge~\eqref{eq:BCh} and the
strangeness charge~\eqref{eq:SCh} instead of the electric charge.  The
baryonic and strangeness form factors, $\FB$ and $\FS$, are plotted in
Fig.~\ref{fig:FFBS_pole}. For comparison, the strangeness form factor
is presented with the opposite sign.  Because of the baryon number and
strangeness conservation in our formulation the form factors are
correctly normalized as $\FB (Q^{2}=0)=1$ and $\FS (Q^{2}=0)=-1$,
respectively.  From Fig.~\ref{fig:FFBS_pole} we find that the
imaginary parts of the total baryonic and strangeness form factors are
small compared with their real parts, as in the case of the
electromagnetic form factors.  We can also see that both the baryonic
and strangeness form factors give steeper derivative at $Q^{2}=0$
compared with the nucleon form factor, and that the strangeness form
factor shows faster decreasing than the baryonic one. Hence, the
baryonic and strangeness components also imply the peculiar structure
of the resonant $\LamFOF$ among the ordinary low-lying hadrons.

\begin{figure*}[t]
  \centering
  \begin{tabular*}{\textwidth}{@{\extracolsep{\fill}}cc}
    \includegraphics[width=8.6cm]{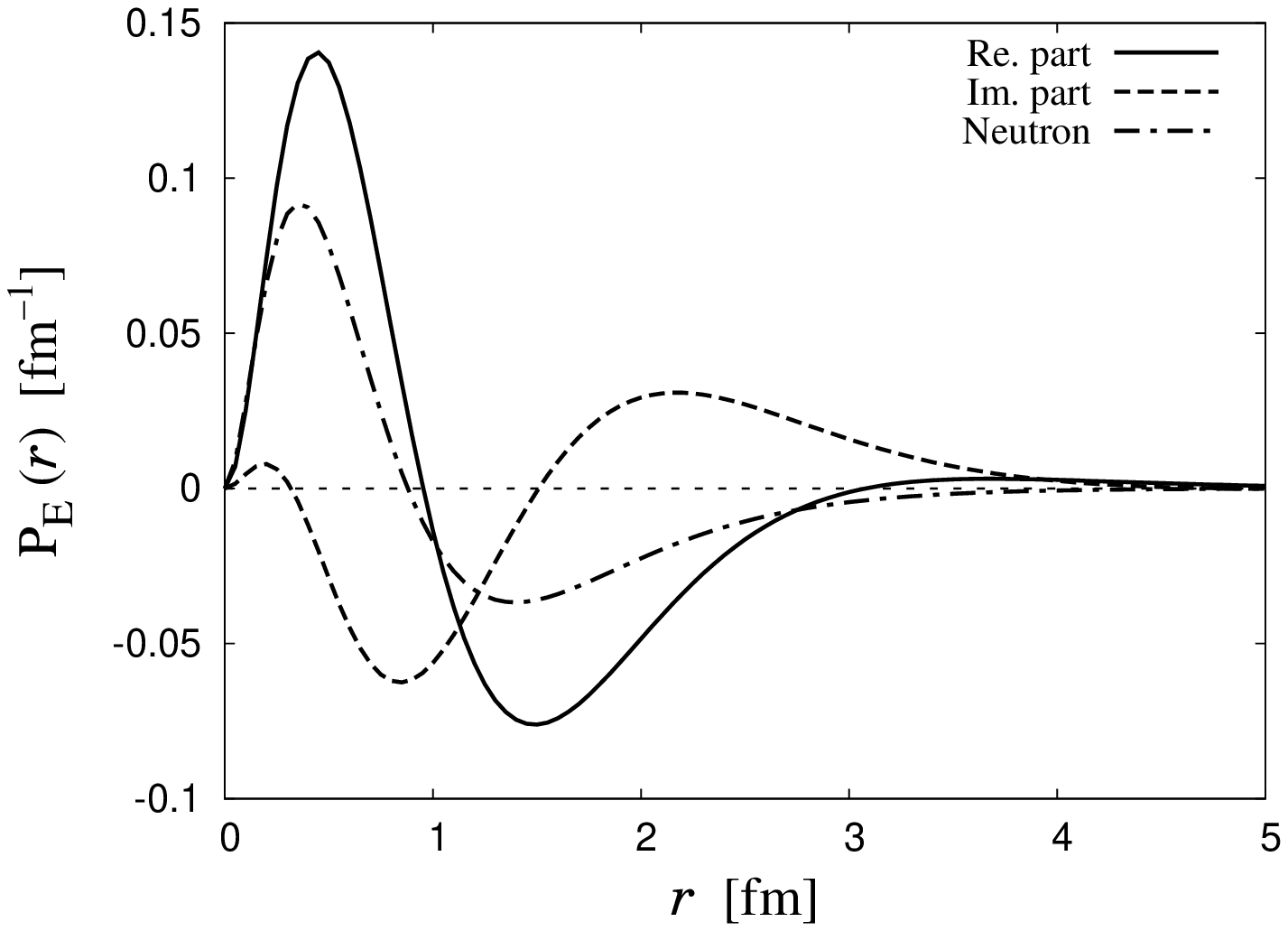} &
    \includegraphics[width=8.6cm]{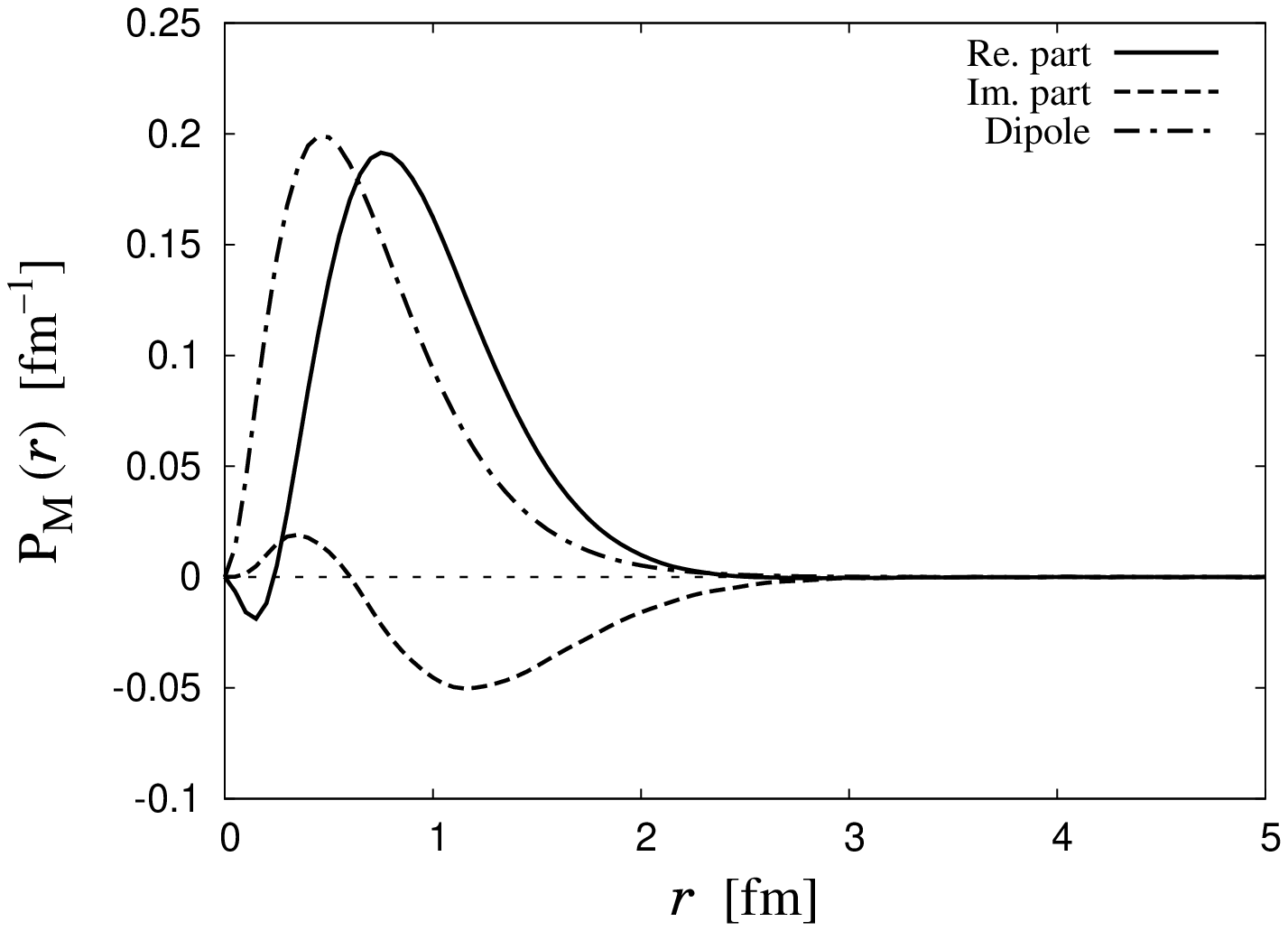} 
  \end{tabular*}
  \caption{
    Normalized distributions 
    $4\pi r^2\rho(r)$ of charge ($\PE$, left) and magnetic moment ($\PM$, 
    right) densities 
    of the $\LamFOF$ state on the higher pole position $Z_{2}$. 
    Empirical charge distribution in the neutron 
    is evaluated by Eq.~\eqref{eq:FEneutron}. 
    Line denoted as ``Dipole'' in magnetic moment density 
    is evaluated by Eq.~\eqref{eq:Gdipolefit} with 
    $c=\text{Re} [\FM (Q^{2}=0)]$. 
  }
  \label{fig:RhoEM_pole}
\end{figure*}

Now it is interesting to visualize the spatial structure of the
resonant $\LamFOF$ in coordinate space as the density distributions
and the mean squared radii obtained from the form factors using
Eqs.~\eqref{eq:Density} and \eqref{eq:DefMSR}.  We introduce a
normalized density distribution $\text{P} (r) \equiv 4 \pi r^{2} \rho
(r)$ with the density distribution $\rho (r)$.  With this definition,
integrating $\text{P}(r)$ from $0$ to $\infty$, one gets the total
charge (or magnetic moment) of the system:
\be
\int _{0}^{\infty} d r \, \text{P} (r) = Q . 
\ee
In Fig.~\ref{fig:RhoEM_pole} we plot the electromagnetic density
distributions of the $\LamFOF$. As we can see from the left panel, the
real part of the charge distribution $\PE$ has large magnitude
compared with that of the neutron. This may indicate 
smaller overlap
between positively and negatively charged components in the $\LamFOF$
than in the neutron.  It is also found that the charge distribution
has positive values in the inner part ($r\lesssim 1$ fm) and negative
values in the outer part ($r\gtrsim 1$ fm). Since the $\LamFOF$
($Z_{2}$) strongly couples to the $\KbarN$ channel, the 
charge form factor of the $\LamFOF$ is expected to be dominated by the
$K^{-}p$ component.  Hence, our result 
implies that the lighter $K^{-}$ surrounds 
the heavier $p$.  From the
magnetic distribution $\PM$ (Fig.~\ref{fig:RhoEM_pole}, right) we can
see the spatially larger structure of the $\Lambda(1405)$ compared
with the neutron.  Since the magnetic density distribution is
contributed mainly from the magnetic moment of the baryon in the
$\LamFOF$, the magnetic structure reflects the baryonic component.

\begin{figure}[!t]
  \centering
  \begin{tabular*}{8.6cm}{@{\extracolsep{\fill}}c}
    \includegraphics[width=8.6cm]{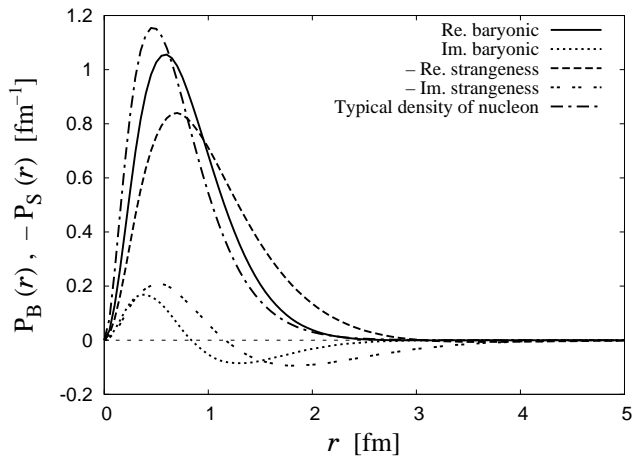} \\
  \end{tabular*}
  \caption{
    Baryonic ($\PB$) and strangeness ($\PS$) density 
    distributions of the 
    $\LamFOF$ state on the higher pole position $Z_{2}$. The strangeness 
    density distribution is presented with the 
    opposite sign for comparison. Typical density of nucleon 
    is evaluated by Eq.~\eqref{eq:Gdipolefit} with $c=1$. 
  }
  \label{fig:RhoBS_pole}
\end{figure}

In the same way, the baryonic and strangeness density
distributions are plotted in Fig.~\ref{fig:RhoBS_pole} 
with the typical density distribution of the
nucleon, which is evaluated by the Fourier transformation of the
dipole form factor in Eq.~\eqref{eq:Gdipolefit} with $c=1$. For
comparison, the strangeness density distribution is shown with the
opposite sign.  The baryonic and strangeness density distributions
clearly indicate the dominance of the real parts over the imaginary
parts and spatially larger structure than the nucleon.  In addition,
the strangeness density distribution has a longer tail than the
baryonic one. Since the $\LamFOF$ at pole position $Z_{2}$ is dominated by 
the $\KbarN$ component, this means larger $\bar{K}$ 
distributions compared with $N$ inside the $\LamFOF$.

\begin{table}[!t]
  \caption{\label{tab:MSR_pole}
    Electromagnetic (upper), baryonic and strangeness (lower) 
    mean squared radii of the 
    $\LamFOF$, $\EMSR$, $\MMSR$, $\BMSR$, 
    and $\SMSR$, on the higher resonance pole position $Z_{2}$. 
  }
  \begin{ruledtabular}
    \begin{tabular}{rc}
      $\EMSR$ & 
      $-0.157 + 0.238 i \fm ^{2}$ \\
      $\MMSR$ & 
      $\phantom{-} 1.138 - 0.343 i \fm ^{2}$ \\
      $\BMSR$ & 
      $\phantom{-} 0.783 - 0.186 i \fm ^{2}$ \\ 
      $\SMSR$ & 
      $- 1.097 + 0.662 i \fm ^{2}$ 
    \end{tabular}
  \end{ruledtabular}
\end{table}

Finally we evaluate the electromagnetic, baryonic, and strangeness 
mean squared radii, which 
are calculated using Eq.~\eqref{eq:DefMSR} with the form factor.
For the magnetic mean squared radius, we use the calculated 
magnetic moment $\FM (Q^{2}=0) = (0.17 - 0.05 i)\mu _{\text{N}}$ 
as the normalization of the mean squared radius. 
The results are shown in Table~\ref{tab:MSR_pole}. 
We find that the absolute value of the electric (magnetic) mean 
squared radius is $|\EMSR|\simeq 0.29 \fm ^{2}$ 
($|\MMSR|\simeq 1.19 \fm ^{2}$), 
which is about two times larger than that of the neutron 
$\sim -0.12 \fm ^{2}$ ($\sim 0.66 \fm ^{2}$). 
Also the absolute values of the baryonic and strangeness mean squared 
radii are larger than the typical size of nucleon. 
We also observe in Table~\ref{tab:MSR_pole} larger radius of the 
strangeness distribution than the baryonic one.  This is due to 
the effect of the 
longer tail in strangeness density distribution compared with the baryonic 
one. Therefore, these results support that the resonant $\LamFOF$ 
state has a spatially-extended structure compared with the typical baryon size 
$\lesssim 1 \fm$. 


As a consequence, all of the results for the electromagnetic, baryonic, 
and strangeness structures show that the resonant $\LamFOF$ 
has a large size compared with typical hadrons.
Furthermore, it is interesting to observe that the strangeness 
density distribution of the $\LamFOF$ has longer tail than the 
baryonic one. Since the $\LamFOF$ ($Z_{2}$) is dominated by the $\KbarN$ 
component and $\bar{K}$ ($N$) carries the strangeness (baryon number), 
one can expect that such behaviors of the strangeness and baryonic 
distributions are understood by 
the widely spread $\bar{K}$ distribution around 
$N$ inside the $\LamFOF$. This expectation is supported by the charge 
distribution, since it has positive values in the inner part whereas 
negative values in the outer part, which will be caused by the 
$K^{-}p$ component inside the $\LamFOF$. 
In the next subsection, we will clarify these detailed structure of the 
$\LamFOF$ by decomposing the form factors and 
density distributions into the contribution from each meson-baryon 
component.

\subsubsection{Contribution from each meson-baryon component}
\label{subsubsec:decomposition}

\begin{figure}[!t]
  \centering
  \begin{tabular*}{\textwidth}{@{\extracolsep{\fill}}c}
    \includegraphics[width=8.6cm]{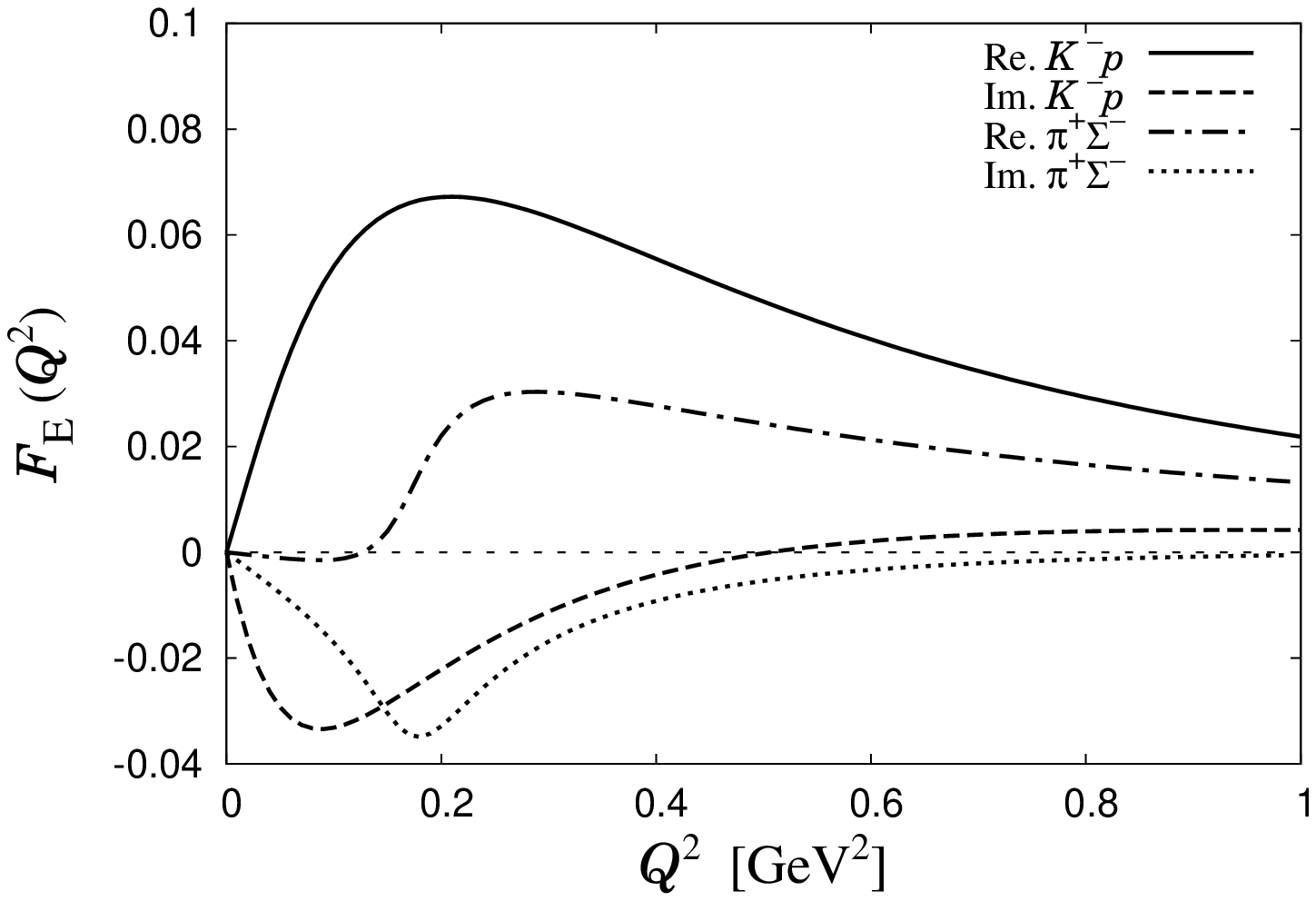} \\
    \includegraphics[width=8.6cm]{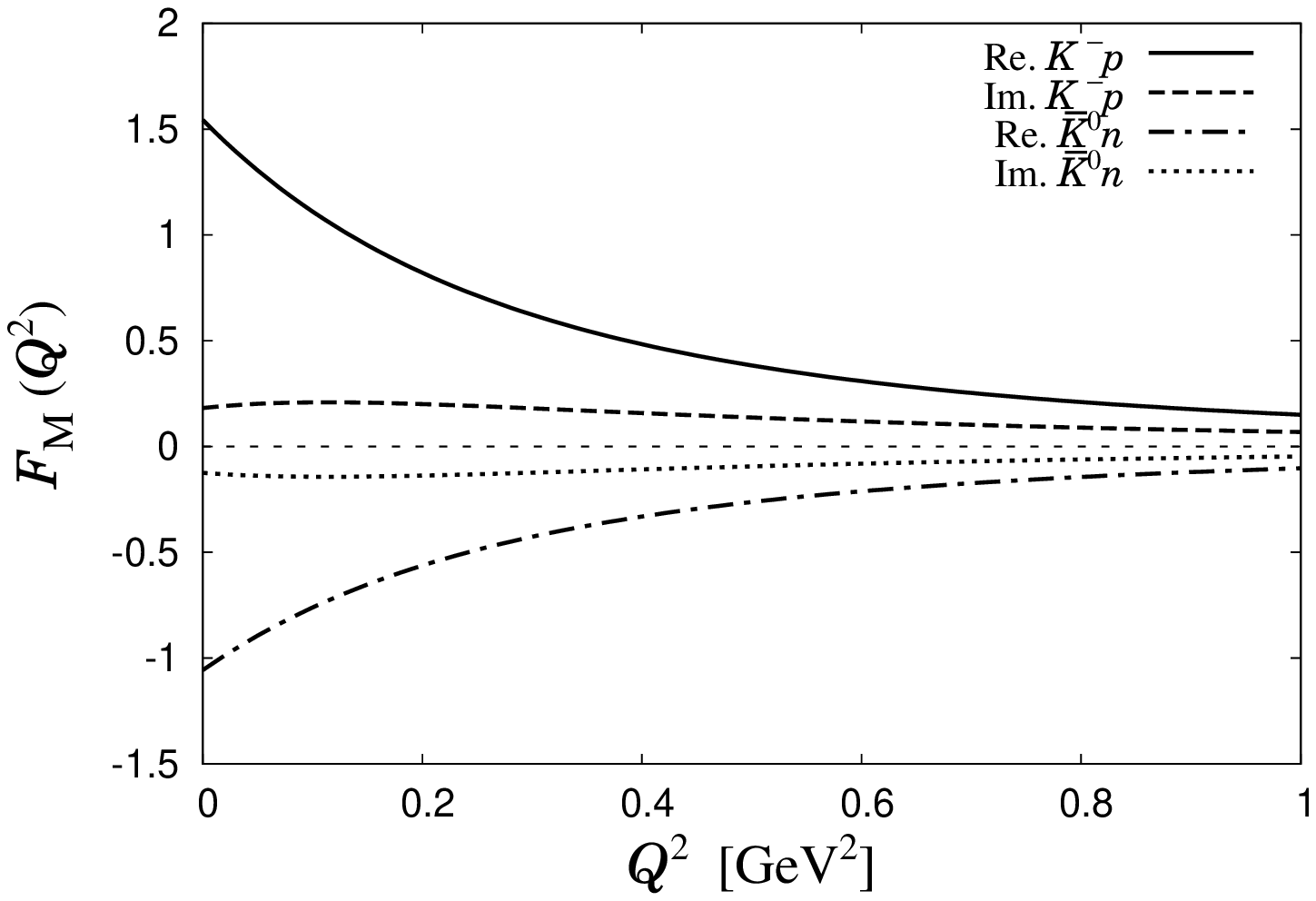} \\
    \includegraphics[width=8.6cm]{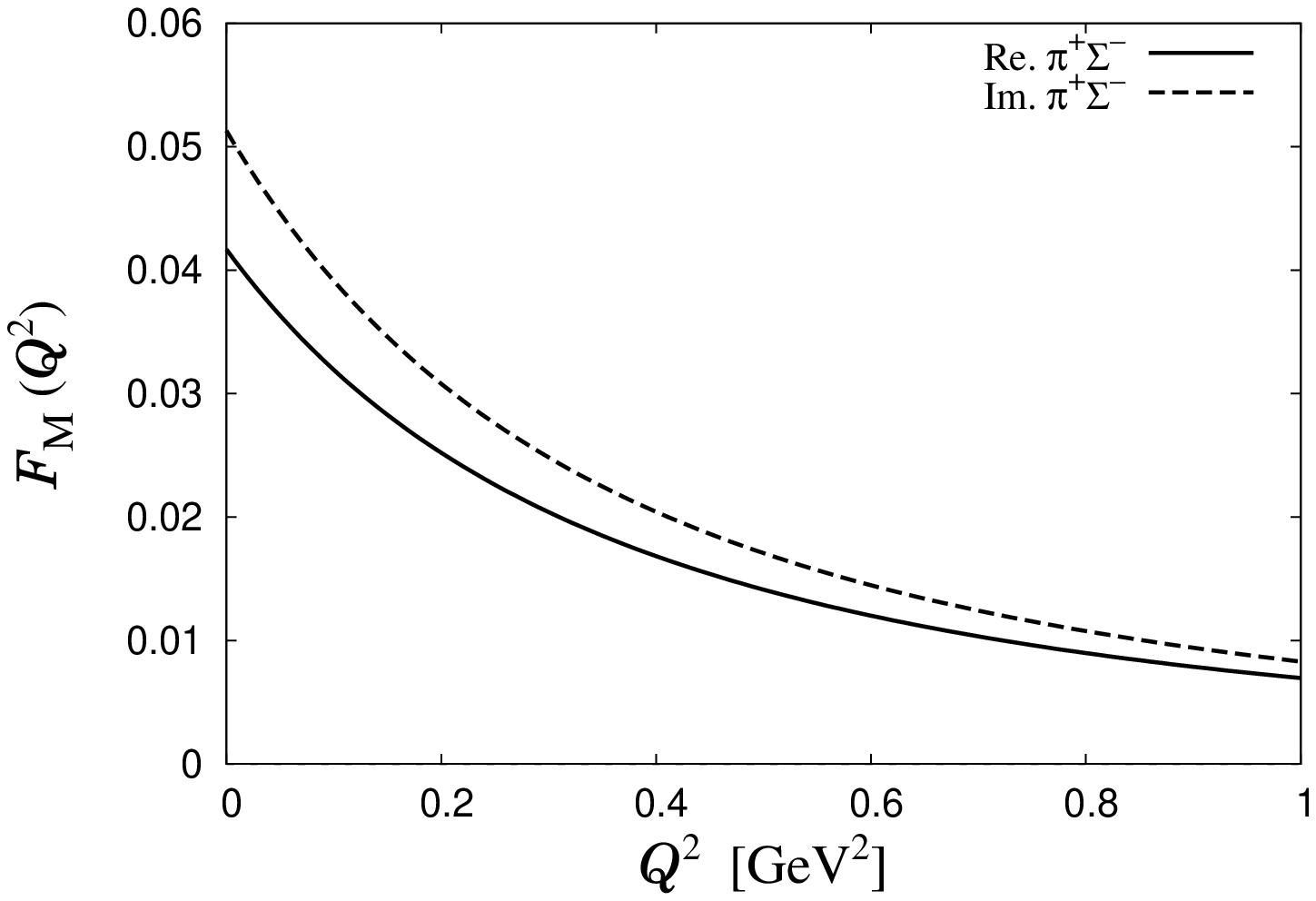} 
  \end{tabular*}
    \caption{
      \label{fig:FFEM_pole_comp} 
      Meson-baryon components of the 
      electromagnetic form factors of the $\LamFOF$ state on the higher pole 
      position $Z_{2}$. 
      Components of the elecric (magnetic) form factor 
      are shown in the upper panel (middle and lower panels). 
    }
\end{figure}%

In order to discuss the inner structure of the $\LamFOF$ resonance
from
the theoretical point of view, it is interesting to decompose the 
form factors into the contribution from each meson-baryon state
to which the external current couples. 
This decomposition can be done by calculating the amplitude 
$T_{\gamma (1)}^{\mu}+T_{\gamma (2)}^{\mu}$ in Eqs.~\eqref{eq:Tgamma1} and 
\eqref{eq:Tgamma2} without 
the summation of the intermediate channel $k$, so that the 
total form factor is obtained by summing up all the components. 
In the decomposition we do not include the contribution
from $T_{\gamma (3)}^{\mu}$, which is not important 
for the study of the spatial size of the resonance, 
since this is a contact interaction and gives an almost trivial 
momentum dependence from the form factor of the 
constituent hadrons\footnote{For the electric current, 
$T_{\gamma (3)}^{0}$ gives no contributions in case of 
neutral resonances. }. 
Here we mainly discuss the $\bar KN$ and $\pi\Sigma$ components 
in the form factors, 
since the $\bar KN$ and $\pi\Sigma$ intermediate states are
the dominant contributions in the $\LamFOF$ resonance
as we have seen in Table~\ref{tab:residua}, and we will check 
that the $\eta \Lambda$ and $K\Xi$ channels and the contact term 
coming from $T_{\gamma (3)}$ 
give negligibly small contributions to the form factors.

We first discuss the contributions to the electromagnetic form 
factors from the $\bar KN$ and $\pi \Sigma$ intermediate states. 
In the upper 
panel of Fig.~\ref{fig:FFEM_pole_comp}, we show
the electric form factor coming from the $K^{-}p$ and $\pi^{+} \Sigma^{-}$
states. As one can see from the figure, the 
electric form factor from the $K^{-}p$ state reproduces almost
the total form factor given in Fig.~\ref{fig:FFEM_pole}. 
The reason is as follows; 
the neutral hadrons do not contribute to the electric form
factor, and the $\LamFOF$ resonance ($Z_{2}$) has a tiny coupling to the
$K^{+}\Xi^{-}$ channel. Then, the sum of the contributions from $K^{-}p$ 
and $\pi ^{\pm} \Sigma ^{\mp}$ components dominate the electric form factor. 
However, since we are working in the isospin symmetric limit,
the $\pi^{\pm} \Sigma^{\mp}$ states give exactly same contributions
in magnitude for the isospin $I=0$ resonance with the opposite sign due 
to their electric charges. 
Thus, the sum of the $\pi^{\pm} \Sigma^{\mp}$ states 
does not contribute to the total electric form factor for the $\LamFOF$
resonance 
and the $\Kmp$ state reproduces almost the total electric form factor. 

In the $\pi^{+}\Sigma^{-}$ contribution, we observe the sudden
increase of the real part and the peak structure of the imaginary part
seen at $Q^{2}\simeq 0.2 \gev ^{2}$, which comes from the analytic
properties of the loop function $D_{\text{M}}^{0}$ in which the
external current attaches to the 
pion 
propagator.  We will discuss
the details in next subsection and Appendix~\ref{sec:Loop}.

The separated contributions to the magnetic form factor are shown in
the middle and lower 
panels of
Fig.~\ref{fig:FFEM_pole_comp}.  For the magnetic contribution, the
meson pole term $T_{\gamma (1)}^{\mu}$ shown in Fig.~\ref{fig:Tgamma}
does not contribute and we are left with the baryon pole term
$T_{\gamma (2)}^{\mu}$ (and the contact term $T_{\gamma (3)}^{\mu}$,
which contribution is not included in Fig.~\ref{fig:FFEM_pole_comp}.
Here we show the form factors with the $K^{-}p$, $\bar K^{0}n$, and
$\pi^{+}\Sigma^{-}$ intermediate channels. We find that the $\bar KN$
contribution is 
substantially 
larger than $\pi\Sigma$. Here we also find
that the $K^{-} p$ contribution has the opposite sign to the
$\bar{K}^{0} n$ and they largely cancel each other.  This is because
only the isosinglet component can couple to the $\LamFOF$ due to the
isospin symmetry and the isosinglet magnetic moment of nucleon is
known to be very small.  This result indicates that the magnetic form
factor of the $\LamFOF$ is composed mainly by nucleons in $\KbarN$
dynamics in which, however, large cancellation between $K^{-} p$ and
$\bar{K}^{0} n$ components takes place.

\begin{table}[!t]
  \caption{\label{tab:ChargeBS}
    Values of the baryonic and strangeness form factors at $Q^{2}=0$. The 
    separated contributions from each meson-baryon channel and contact 
    term are also listed. Each meson-baryon channel is given in the 
    isospin basis; for instance, $\KbarN$ represents the sum of the 
    contributions from $K^{-}p$ and $\bar{K}^{0}n$. }
  \begin{ruledtabular}
    \begin{tabular}{cc}
      Component & $\FB (0) = - \FS (0)$  \\ 
      \hline
      total & $1$  \\
      \hline
      \rule[0pt]{0pt}{10pt} $\KbarN$ & $\phantom{-} 0.994 + 0.048 i$  \\
      $\pi \Sigma $ & $-0.047 - 0.151 i$  \\
      $\eta \Lambda$ & $\phantom{-}0.052 + 0.012 i$  \\
      $K \Xi$ & $-0.002 + 0.002 i$ \\
      Contact & $\phantom{-}0.002 + 0.089 i$ 
    \end{tabular}
  \end{ruledtabular}
\end{table}

Next we discuss the contributions to the baryonic and strangeness form
factors from each meson-baryon channel. Here, instead of plotting the
form factors with respect to $Q^{2}$, 
we show in Table~\ref{tab:ChargeBS} the values of the baryonic and
strangeness form factors at $Q^{2}=0$ with individual contributions
from the meson-baryon channels 
to the $\LamFOF$, which corresponds to
the each channel contribution to the baryon number and strangeness of
the system.  Here we have the Gell-Mann-Nishijima relation $\FB =-
\FS$ at $Q^{2}=0$ for each meson-baryon channel. 
The decomposition to each
meson-baryon channel implies that the $\KbarN (I=0)$ channel gives
more than $90 \%$ of the total baryonic and strangeness charges,
whereas the $\pi\Sigma$, $\eta\Lambda$, $K\Xi$, and contact-term
components in $I=0$ channel are negligibly small.  The magnitude 
of the charge of each
component is determined by the coupling strength given in
Table~\ref{tab:residua} and the derivative of the loop function which
contributes to the form factor through Eqs.~\eqref{eq:DMGM} and
\eqref{eq:DBGB}.  Thus, due to the large coupling strength
$g_{\KbarN}$ the $\KbarN$ channel dominates the structure of the
$\LamFOF$.

\begin{figure}[!t]
  \centering
  \begin{tabular*}{\textwidth}{@{\extracolsep{\fill}}c}
    \includegraphics[width=8.6cm]{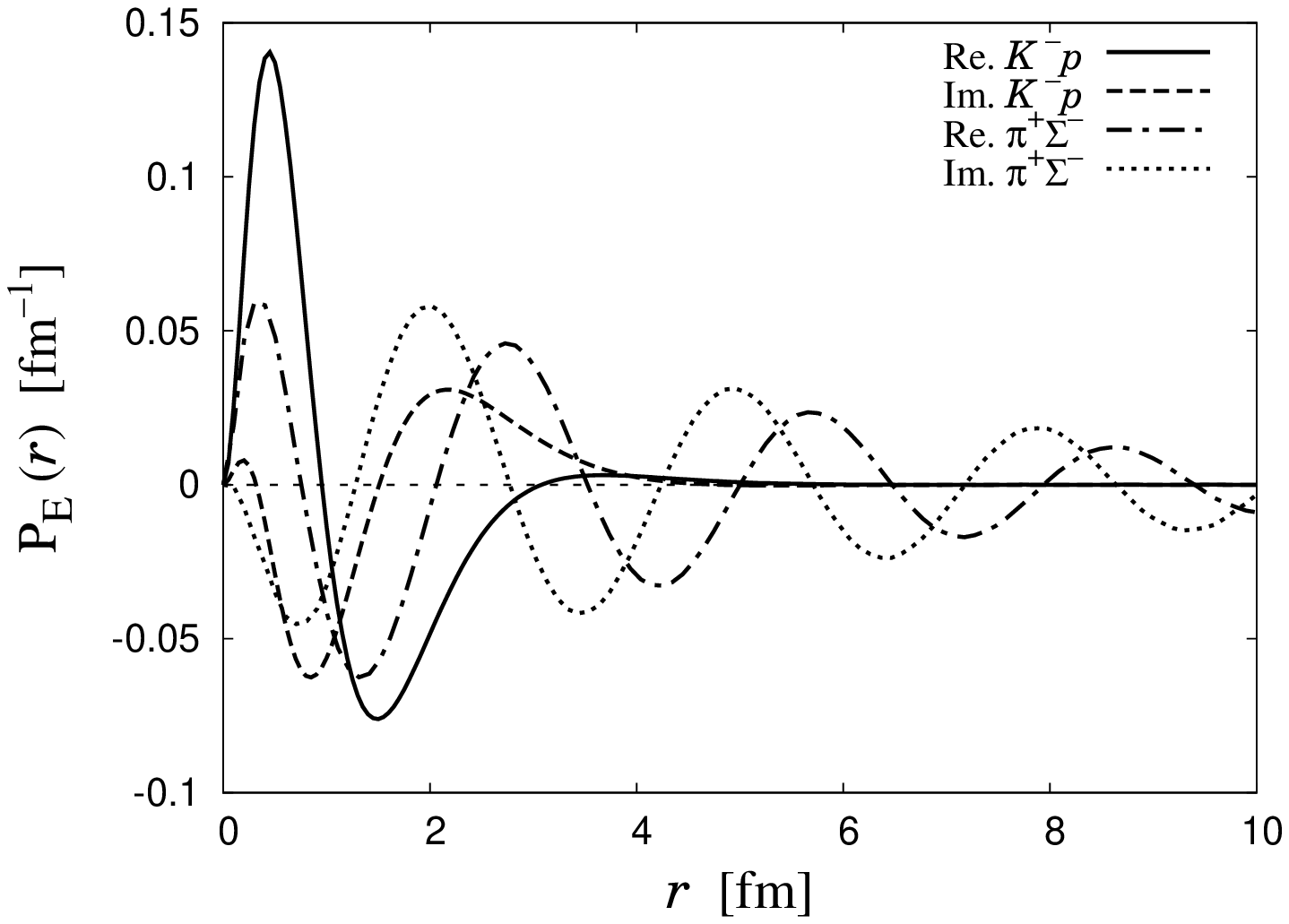} \\
    \includegraphics[width=8.6cm]{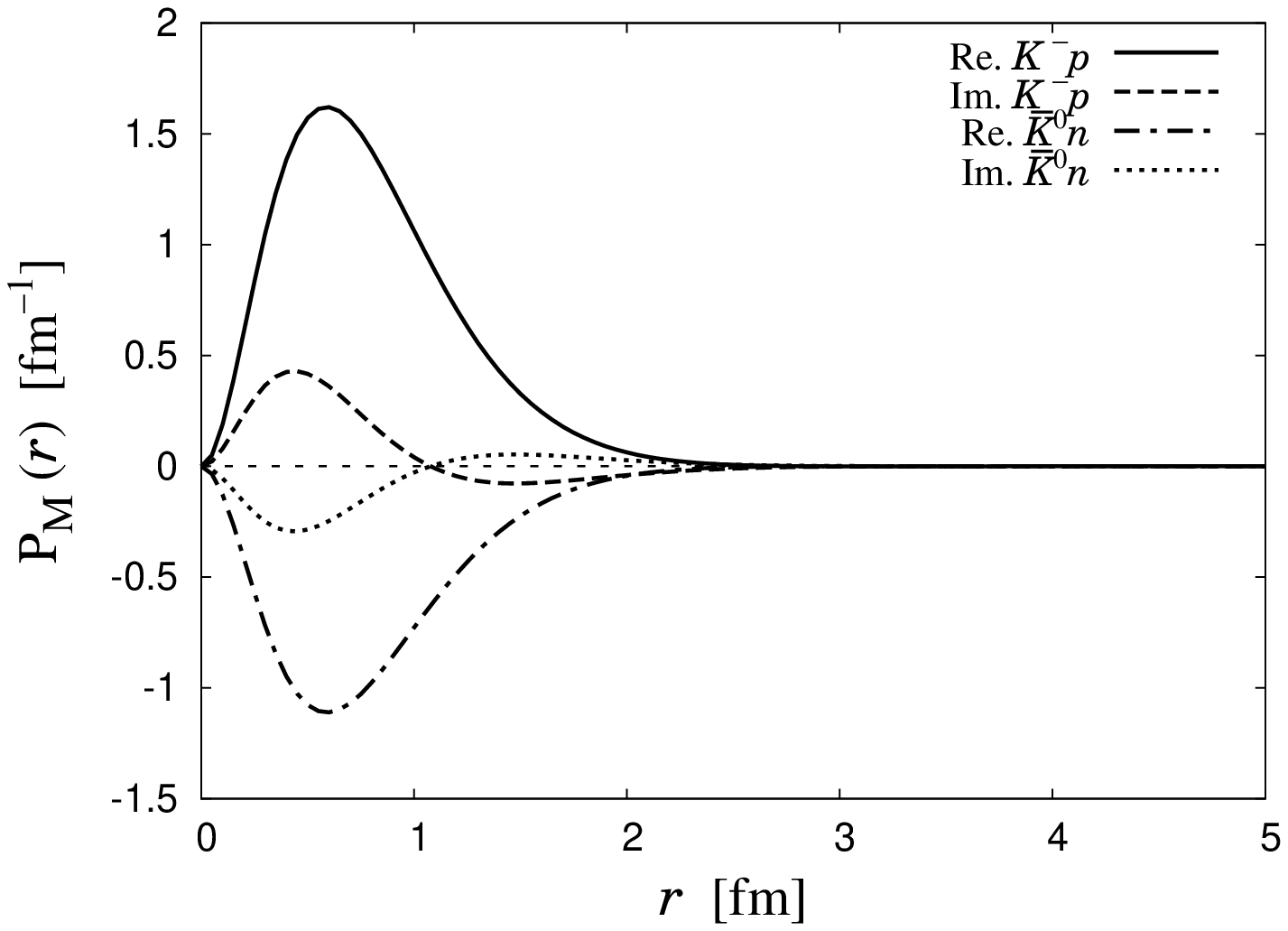} \\ 
    \includegraphics[width=8.6cm]{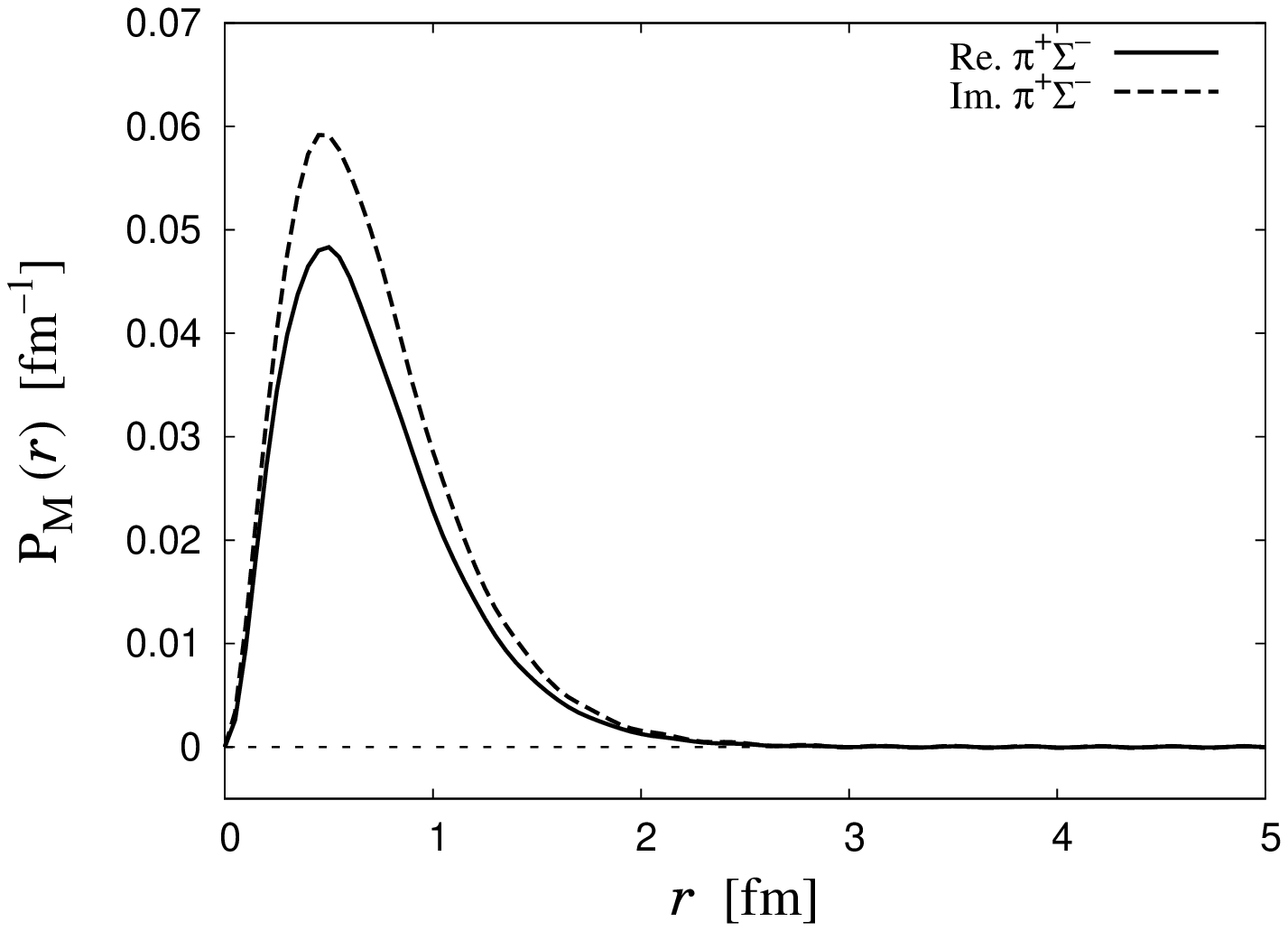} 
  \end{tabular*}
    \caption{
      \label{fig:RhoEM_pole_comp}
      Meson-baryon components of the 
      distributions $4\pi r^2\rho(r)$ of charge ($\PE$, upper) and 
      magnetic moment ($\PM$, middle and lower) 
      densities of the $\LamFOF$ state on the higher pole position $Z_{2}$. 
    }
\end{figure}%

Now we show the meson-baryon components of the electromagnetic
density distributions, $\PE$ and $\PM$, which are obtained by the
Fourier transformation of the corresponding meson-baryon components of
the form factors.  The results are shown in
Fig.~\ref{fig:RhoEM_pole_comp}. Note that $\PE$ is plotted up to $r=10
\fm$ instead of $5 \fm$.  We find again the dominance of the $\KbarN$
component in both $\PE$ and $\PM$.  
Hence, the negative (positive) charge distribution 
of the $\LamFOF$ 
in Fig.~\ref{fig:RhoEM_pole} is now understood as 
the lighter $K^{-}$ (heavier $p$) existence in outside (inside) region. 
It is a more interesting finding
that the $\pi ^{+} \Sigma ^{-}$ (equivalently $\pi^{-} \Sigma^{+}$
with the opposite sign) component of the electric density distribution
shows a characteristic behavior of dumping oscillation.  As we will
discuss in next subsection and 
Appendix~\ref{sec:Loop}, the oscillating behavior can be 
interpreted as the decay of the system into the $\pi\Sigma$ channels
through the photon coupling to the intermediate meson.  Although this
oscillation is interesting 
from the theoretical point of view, it does
not contribute to the total density distribution due to the
cancellation of $\pi ^{+} \Sigma ^{-}$ and $\pi ^{-} \Sigma ^{+}$
components.  
For the magnetic density distribution, we again observe large 
cancellation between $K^{-}p$ and $\bar{K}^{0}n$ components in 
the isosinglet $\LamFOF$.

\begin{figure}[!Ht]
  \centering
  \begin{tabular*}{8.6cm}{@{\extracolsep{\fill}}c}
    \includegraphics[width=8.6cm]{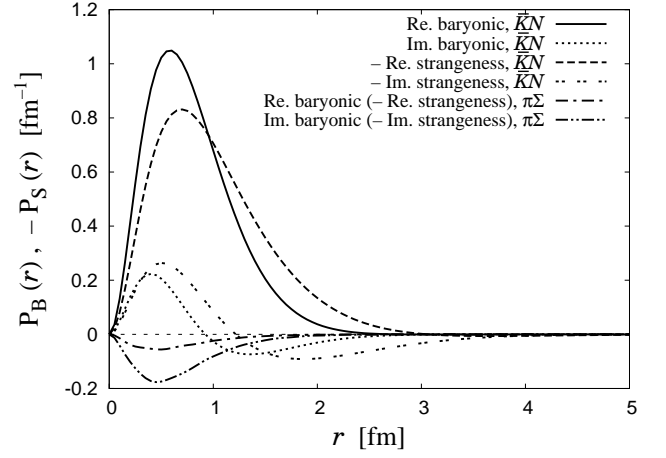} \\
  \end{tabular*}
  \caption{
    Meson-baryon components of 
    baryonic ($\PB$) and strangeness ($\PS$) density distributions of the 
    $\LamFOF$ state on the higher pole position $Z_{2}$. 
    The strangeness density distribution is presented with 
    opposite sign for comparison. 
    Here $\KbarN$ represents the sum of the contributions from $K^{-} p$ and 
    $\bar{K}^{0} n$, whereas $\pi \Sigma$ represents the sum of the 
    contributions from $\pi ^{0} \Sigma ^{0}$, $\pi ^{+} \Sigma ^{-}$ and 
    $\pi ^{-} \Sigma ^{+}$. 
  }
  \label{fig:RhoBS_pole_comp}
\end{figure}

Then, we decompose the baryonic and strangeness density distributions 
into the different meson-baryon contributions in
Fig.~\ref{fig:RhoBS_pole_comp}. In this figure, $\KbarN$ represents
the sum of the contributions from $K^{-} p$ and $\bar{K}^{0} n$,
whereas $\pi \Sigma$ represents the sum of the contributions from $\pi
^{0} \Sigma ^{0}$, $\pi ^{+} \Sigma ^{-}$, and $\pi ^{-} \Sigma ^{+}$.
For the $\pi\Sigma$ channel, we plot the baryonic density which is
equivalent to the strangeness $\pi \Sigma$ density with the opposite
sign. This is clear from the charges in Eqs.~\eqref{eq:BCh} and
\eqref{eq:SCh} where both baryonic and strangeness currents probe the
$\Sigma$ component with the same strength and the opposite sign 
in the $\pi \Sigma$ intermediate state. 
As one can see from the Fig.~\ref{fig:RhoBS_pole_comp}, the 
$\KbarN$ component where the baryonic (strangeness) current probes 
the $N$ ($\bar{K}$) distribution dominates both the baryonic and 
strangeness density distribution in Fig.~\ref{fig:RhoBS_pole}.  
Therefore, the longer tail of the 
strangeness distribution in Fig.~\ref{fig:RhoBS_pole} is now understood as the 
larger distribution of the $\bar{K}$ component than that of the 
baryon number distribution generated by the $N$.  This is consistent 
with the electric density distribution, that
the lighter $K^-$ locates outside the $p$, which should
also be the case for the $\bar{K}^0$ and $n$ through the isospin
symmetry. 

From the decomposition of the form factors and density distributions
into meson-baryon components, it is confirmed that the resonant
$\LamFOF$ is indeed dominated by the $\KbarN (I=0)$ component, giving
more than $90 \%$ of the total baryonic and strangeness charges, and it
is found that the $\LamFOF$ is dominantly composed of the $\bar{K}$ in the
outside region around the nucleon, with a large size compared with
typical hadrons.  We also find that the magnetic moment of the $\LamFOF$
is composed mainly by nucleons in $\KbarN$ dynamics in which large
cancellation between $K^{-} p$ and $\bar{K}^{0} n$ components takes
place.  In addition, we observe that the $\pi ^{+} \Sigma ^{-}$
component has the escaping oscillation in the charge distribution 
(see also next subsection),
which is although not observed in the total charge due to the $\pi
^{\pm} \Sigma ^{\mp}$ cancellation.  
In our results 
all of the
results for the electromagnetic, baryonic, and strangeness structure
of the $\LamFOF$ are consistent with each other.

\subsubsection{Escaping oscillation in decay channel}
\label{subsubsec:oscillation}

In the previous subsection we found an interesting behavior of the 
$\pi ^{+} \Sigma ^{-}$ escaping oscillation in the electric density 
distribution. Here we discuss the escaping oscillation 
found in the 
$\pi \Sigma$ channel, or decay channel in general. 

From a viewpoint of the Fourier transformation, 
what makes oscillation behavior in the $\pi ^{+} \Sigma ^{-}$ 
electric density distribution is the peak structure at 
$Q^{2}\simeq 0.2 \gev ^{2}$ in the electric form factor in 
the $\pi ^{+} \Sigma ^{-}$ channel. Namely, the Fourier 
transformation of the form factor picks up much contributions 
from the peak structure, making large magnitude in the corresponding 
wave number for the density distribution, which is seen as the 
oscillation component. 

Then, let us discuss the origin of the peak structure in the 
form factor. An important point is that the photon coupled loop 
integral of the intermediate channel $k$, 
$D_{\text{M}_{k}}$ ($D_{\text{B}_{k}}$), contains a divergent point at 
\be
Q^{2} = \frac{4 \tilde{q}_{k}^{2} s}{M_{k}^{2}} \quad 
\left ( Q^{2} = \frac{4 \tilde{q}_{k}^{2} s}{m_{k}^{2}} \right ) , 
\label{eq:Q2_singular}
\ee
with real energy $\sqrt{s}$, $\tilde{q}_{k}$ defined in
Eq.~\eqref{eq:q_k}, and the baryon (meson) mass $M_{k}$ ($m_{k}$) in the
intermediate state.  This divergence point corresponds to the
$t$-channel threshold.  The detailed discussion is given in
Appendix~\ref{sec:Loop}.  It should be emphasized that 
this singularity~\eqref{eq:Q2_singular} can be reached only in case
that the energy is above the threshold of the $k$ channel, 
$\sqrt{s}>M_{k}+m_{k}$, 
so that $4 \tilde{q}_{k}^{2} s>0$.  Such a singularity at
certain $Q^{2}$ generates peak structure in the form factor with
complex energy through the analytic continuation $\sqrt{s} \to z$.
Hence, that the resonance energy is above the threshold is essential
to the peak structure in the form factors, and the oscillation behavior 
of the density distributions can be interpeted as the decay of the 
system into the open channels through the photon coupling to the 
intermediate state, with kicked meson and baryon to the on-shell 
by the photon coupling.  In the present case, 
since only the $\pi \Sigma$ channel is open for the $\LamFOF$ decay, 
the peak structure appears only in the $\pi \Sigma$ channel. 

From the above discussion, it is obvious that the peak structures in
the form factors and the oscillation behaviors in the density
distributions should appear in decay channels for resonance states in
the meson-baryon picture. Such structures are, however, eventually not
observed in the total electric form factor and density distribution
for the resonant $\LamFOF$. This is because, as we have mentioned, the
$\pi ^{\pm} \Sigma ^{\mp}$ components largely cancel each other for
the electric structure due to the isospin symmetry.  Hence, if we
would observe a excited state for which its decaying channel 
contributes to the form factors without cancellation, 
we could observe the escaping oscillation in the total density distribution.

At last, we comment on that the magnetic, baryonic, and strangeness
density distributions do not show the (visible) oscillation behavior
even in the decay channel, $\pi \Sigma$.  This is because the
magnetic, baryonic, and strangeness currents couple to $\Sigma$ rather
than pion in the $\pi \Sigma$ channel; the current coupling to the 
$\Sigma$ propagator also provides the oscillation behavior in the
electric as well as the magnetic, baryonic, and strangeness density
distributions.
However, Eq.~\eqref{eq:Q2_singular} indicates that, 
due to the small pion mass $m_{\pi}$, 
the $\Sigma$-current
coupling makes the peak structure in the form factor with very high
$Q^{2}$ value [$Q^{2}\sim 10 \gev ^{2}$ for the $\pi
\Sigma$ channel in the $\LamFOF$ ($Z_{2}$)], thus such a high $Q^{2}$
coupling should be strongly suppressed by form factors of the
constituent hadrons. As a consequence, the oscillation contributions
from the $\Sigma$-current coupling are numerically small and not
visible in Figs.~\ref{fig:RhoEM_pole_comp} and
\ref{fig:RhoBS_pole_comp}.

\subsection{Effective form factors on the real energy axis}
\label{subsec:on_real}

Now we evaluate the effective form factors of the $\LamFOF$ on the
real energy axis defined in Eq.~\eqref{eq:Feffective}, in order to
study how the form factors obtained at the resonance position are seen
on the real energy axis.  These effective form factors on the real
axis will provide the quantities which can be compared with the
experimental observations, in contrast to the form factors obtained in
the complex energy plane.  Here we comment that the mean
  squared radii are not well defined for resonances on the real energy
  axis due to insufficient fall-off of the densities at large $r$, in
  contract to the case on the resonance pole position.

\begin{figure}[t]
  \centering
  \begin{tabular*}{8.6cm}{@{\extracolsep{\fill}}c}
    \includegraphics[width=8.6cm]{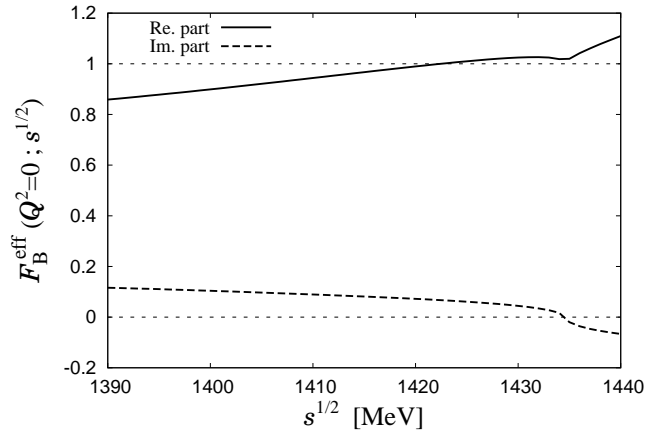} 
  \end{tabular*}
  \caption{Effective baryon number of the resonance, 
    $\Feff _{\text{B}} (Q^{2}=0)$, as a
    function of meson-baryon center-of-mass energy $\sqrt{s}$. 
    Calculations are performed in the $\KbarN (I=0) \gamma ^{\ast} 
    \rightarrow \KbarN (I=0)$ process on the real energy axis.}
  \label{fig:QE_eff}
\end{figure}

As discussed in Sec.~\ref{subsec:photon-coupled}, the form factors 
calculated with three diagrams shown in Fig.~\ref{fig:Tgamma}
keep correct normalizations 
only on the resonance pole position. Off the resonance pole position
the additional diagrams shown in Fig.~\ref{fig:Tgamma-other} are necessary
to keep the correct normalizations. Nevertheless, since the double-pole diagrams
dominate the photon-coupled amplitude at the resonance energies 
if the pole position
is not far from the real axis, the contributions from the supplemental
diagrams in Fig.~\ref{fig:Tgamma-other} to the form factors may be
negligible.

This can be checked by the calculation of 
the effective baryon number $\Feff _{\text{B}} (Q^{2}=0)$ 
for the $\LamFOF$ state with the diagrams in Fig.~\ref{fig:Tgamma} on the 
real axis. 
Here we choose the $\KbarN \to \KbarN$
channel in $I=0$ with the baryonic current, so that the
less singular contributions are suppressed
through the large coupling constant 
$g_{\KbarN}$ [see Eq.~\eqref{eq:effFFapprox}].
In Fig.~\ref{fig:QE_eff}, we plot 
$\Feff _{\text{B}} (Q^{2}=0)$ 
for the $\LamFOF$ on the real energy axis as a function of $\sqrt{s}$, 
which is evaluated with the three diagrams shown in Fig.~\ref{fig:Tgamma}. 
If the amplitude maintains 
the conservation law by taking account of 
all the diagrams, $\Feff _{\text{B}} (Q^{2}=0)$ should be 
pure real and unity 
independently 
of $\sqrt{s}$. Figure~\ref{fig:QE_eff} indicates 
that the real part of $\Feff _{\text{B}} (Q^{2}=0)$ calculated 
with the three diagrams is close to one in the energy $\LamFOF$. 
This means that the three diagrams dominate the amplitude of the 
$\KbarN (I=0) \gamma ^{\ast} \to \KbarN (I=0)$ process, and that 
contributions coming from the neglected diagrams are less than $10 \%$. 
To study the energy dependence, we 
evaluate the form factors at three meson-baryon center-of-mass energies 
$\sqrt{s}=1410$, $1420$, and $1430 \mev$. 

\begin{figure*}[t]
  \centering
  \begin{tabular*}{\textwidth}{@{\extracolsep{\fill}}cc}
    \includegraphics[width=8.6cm]{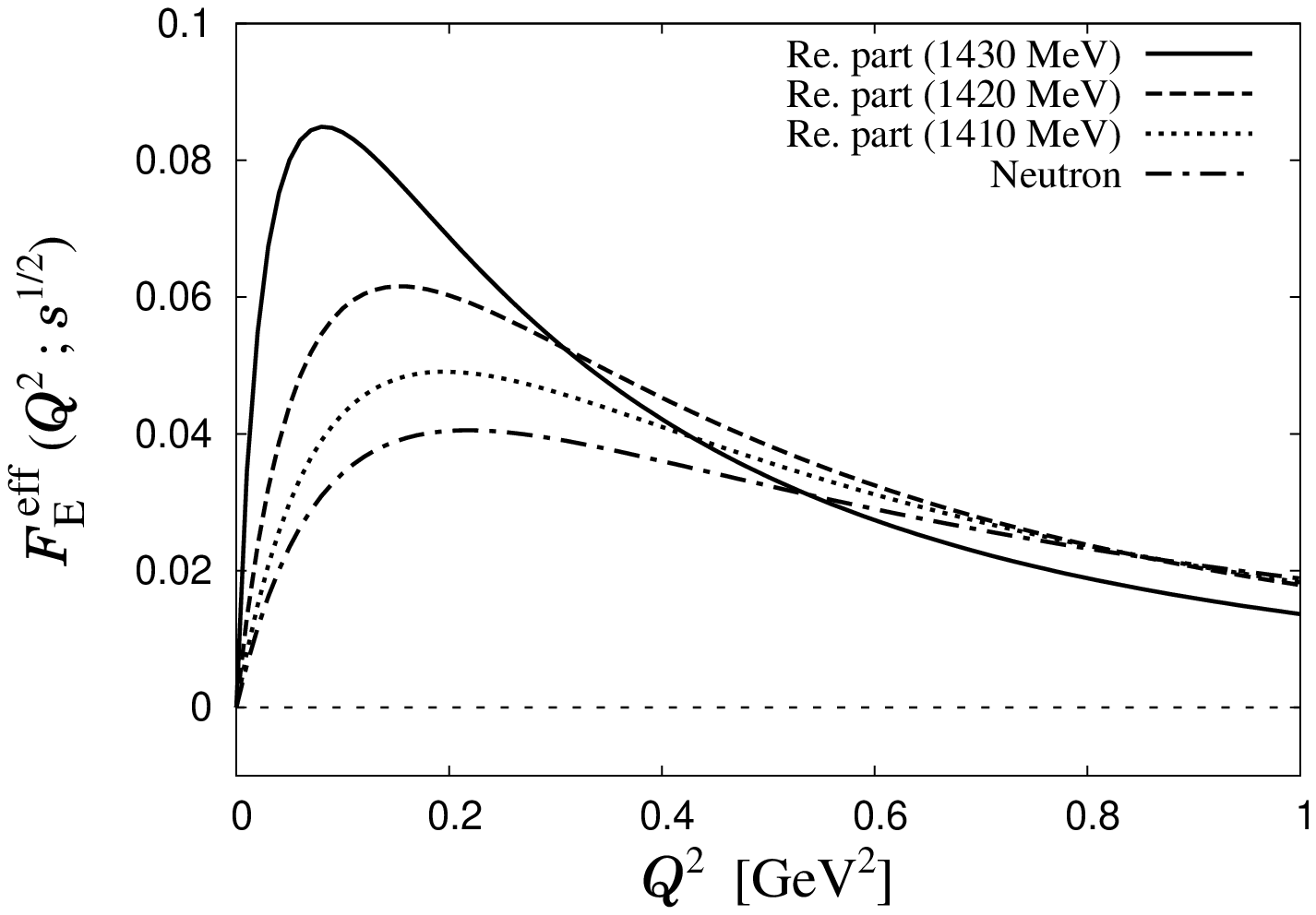} &
    \includegraphics[width=8.6cm]{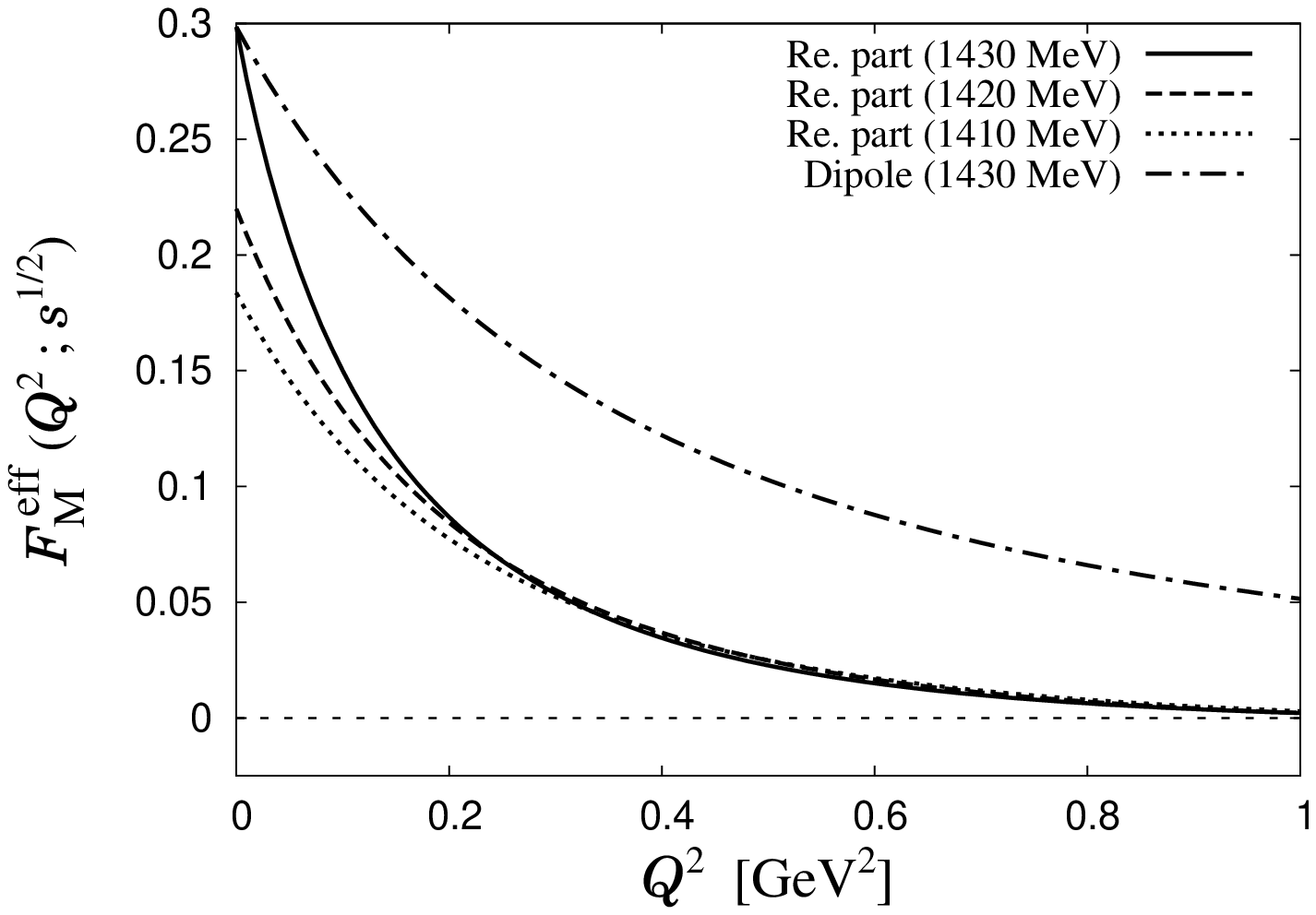} 
  \end{tabular*}
  \caption{
    Real part of the 
    effective electromagnetic form factors ($\FEeff$ and $\FMeff$) 
    on the real energy axis, together with the 
    empirical form factors of the neutron. 
    Calculations are performed with the center-of-mass energy 
    $\sqrt{s}=1410$, $1420$, and $1430 \mev$. 
    The parameter $c$ in the dipole form factor is 
    chosen to be $c=\text{Re} [\FMeff (Q^2=0; \sqrt{s}=1430 \mev)]$, 
    the real part of the magnetic moment at $\sqrt{s}=1430 \mev$. }
  \label{fig:FFEM_eff}
\end{figure*}

The results of the real part of the effective electromagnetic form factors 
on the real energy axis are shown in Fig.~\ref{fig:FFEM_eff}. 
We find that the effective electromagnetic form factors for 
$\sqrt{s}=1420 \mev$ are 
qualitatively very similar with the form factors evaluated at the pole 
position in Fig.~\ref{fig:FFEM_pole}. 
We also find in Fig.~\ref{fig:FFEM_eff} that the effective electromagnetic 
form factors have mild energy dependence even 
in energies close to the resonance position. 
This is because, off the resonance pole position, less singular terms also 
contribute to the effective form factors and brings energy 
dependence to them. 
We also see that, in the energies closer to the $\KbarN$ threshold,
both the $\FEeff$ and $\FMeff$ have 
steeper change at low $Q^{2}$.

\begin{figure*}[!t]
  \centering
  \begin{tabular*}{\textwidth}{@{\extracolsep{\fill}}cc}
    \includegraphics[width=8.6cm]{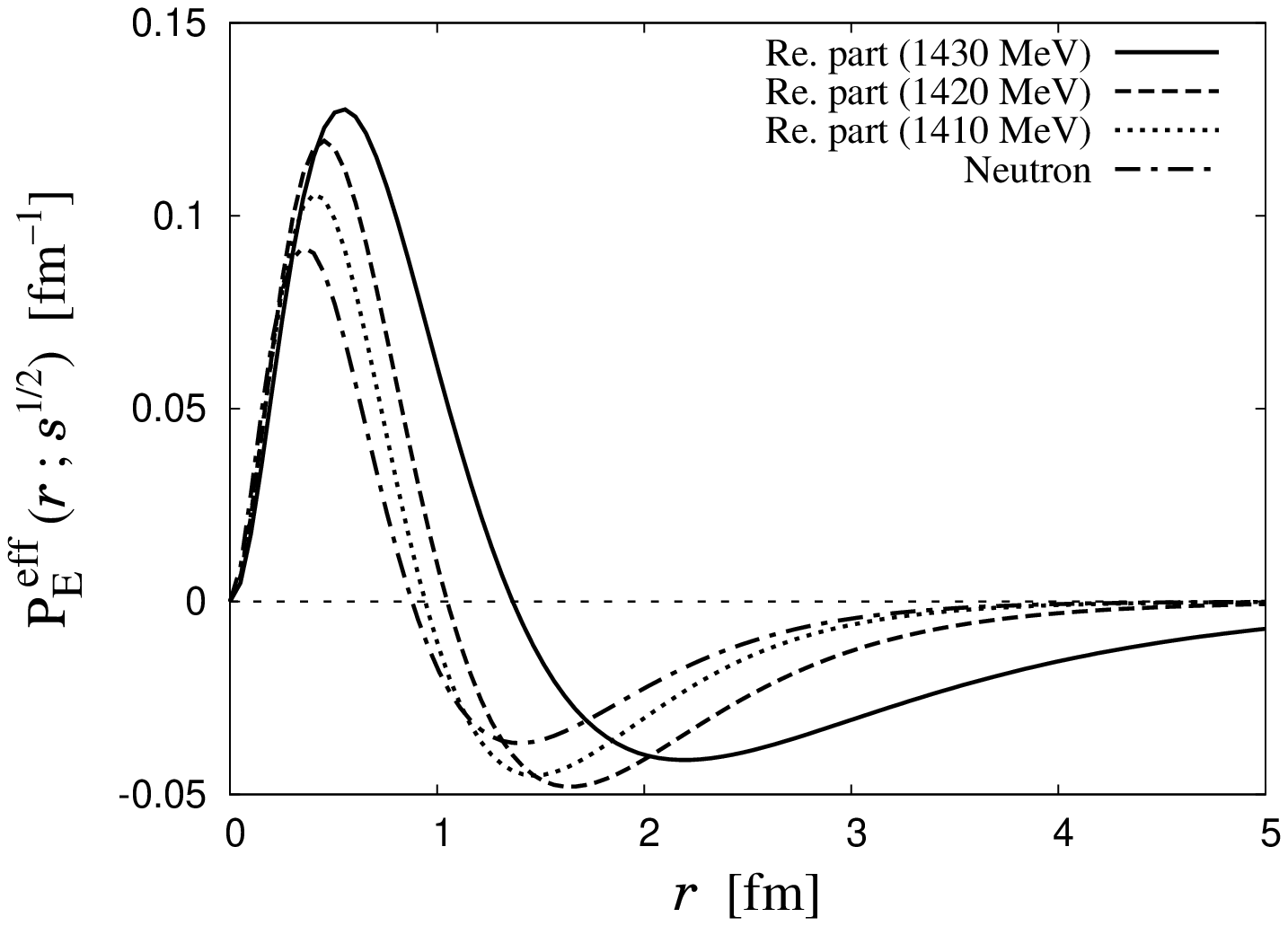} &
    \includegraphics[width=8.6cm]{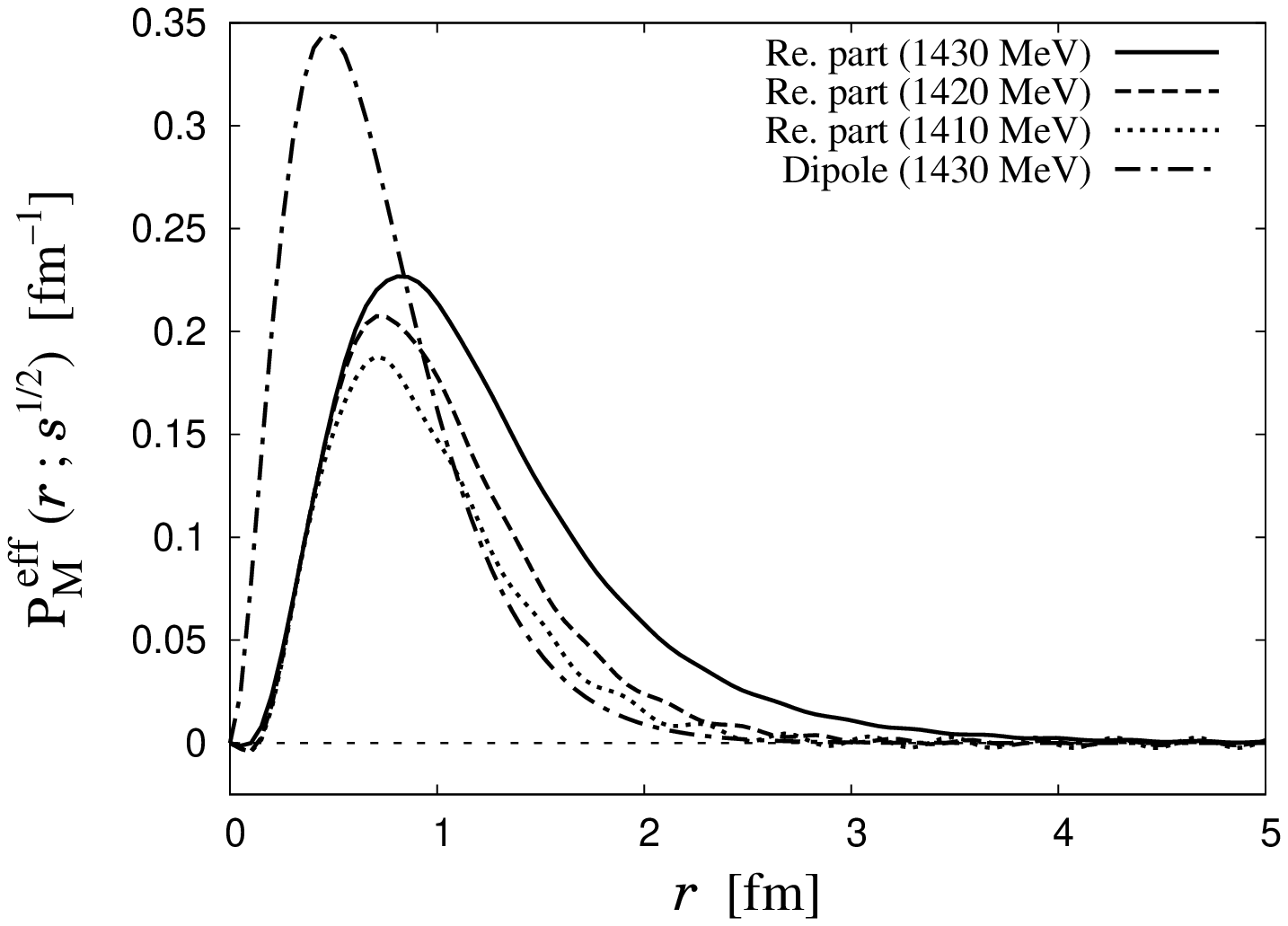} 
  \end{tabular*}
  \caption{
    Real part of the 
    effective charge ($\PEeff$, left) and magnetic moment ($\PMeff$, right) 
    density distributions on the real energy 
    axis. Calculations are performed with 
    the center-of-mass energy $\sqrt{s}=1410$, $1420$, and $1430 \mev$. 
    Empirical charge distribution in the neutron 
    is evaluated by Eq.~\eqref{eq:FEneutron}. 
    Line denoted as ``Dipole'' in magnetic moment density 
    is evaluated by Eq.~\eqref{eq:Gdipolefit} with 
    $c=\text{Re} [\FMeff (Q^2=0 \gev ^2;
      \sqrt{s}=1430 \mev )]$. }
  \label{fig:RhoEM_eff}
\end{figure*}

Now let us discuss the electromagnetic density distributions $\PEeff$
and $\PMeff$ obtained from $\FEeff$ and $\FMeff$.  We plot the real
parts of $\PEeff$ and $\PMeff$ in Fig.~\ref{fig:RhoEM_eff} with
$\sqrt{s}= 1410$, $1420$, and $1430 \mev$, in comparison with the
empirical neutron density distribution. 
The important point is that the effective distribution in the real 
axis also catches the properties found in the density 
distribution on the pole position, that is, 
spatially larger structure than the neutron with
the outward negative charge of $K^-$ and the inward positive charge of
$p$.
As the energy $\sqrt{s}$ increases, the observed distribution becomes effectively wider. 

Finally we summarize the numerical analyses of the internal
structure of the resonant $\LamFOF$ state, which is done by evaluating 
two observables; 
one is the resonance form factor~\eqref{eq:Res_scheme}  
on the resonance pole position (see Sec.~\ref{subsec:on_pole}, and the 
other is the effective form factor~\eqref{eq:Feffective} on the real energy
axis (Sec.~\ref{subsec:on_real}.  Although there is a quantitative 
difference between these two analyses 
such as
the energy $\sqrt{s}$ dependence of the effective form
factors, it is found that the several peculiar
features of the resonance form factor in the complex plane are mostly
maintained in the effective form factor on the real axis. This means
that the properties of the $\LamFOF$ defined in the complex plane may be
within our reach of the experimental searches achieved by the real
energies.  This consequence comes from the small imaginary part of the
pole $Z_2$ of the $\LamFOF$.

From these analyses we have drawn the following conclusions. 
The $\LamFOF$ has softer form factors than those of neutron.
Consequently the mean squared radii are larger than the neutron radii
which can be regarded as a typical baryon, as found in
Ref.~\cite{Sekihara:2008qk}.  Through the decomposition into
meson-baryon channels and the analysis of the form factors with
different probe currents, the internal structure of
the resonant $\LamFOF$ is found to be dominated by the $\bar{K}N$
component, with the nucleon in the center being surrounded by the
antikaon.

\section{Discussions}
\label{sec:Discussion}

We have considered the form factors and density distributions of the
resonant $\LamFOF$ state.  Because the $\LamFOF$ resonance has the
finite decay width, the obtained form factors are complex numbers. Although
we have deduced the structure of the $\LamFOF$ mainly from the
behavior of the real part of the form factors, the interpretation may 
not be as straightforward as the stable particle.

The $\LamFOF$ has been considered as a quasi-bound state of the $\KbarN$
system\footnote{In this section 
$\KbarN$ represents $\KbarN (I=0)$ channel in the isospin basis. } 
having the $\pi \Sigma$ decay 
channel~\cite{Dalitz:1960du,Dalitz:1967fp}. This 
picture is also supported by 
the chiral unitary approach, where the $\KbarN$ bound state
is generated only by the attractive $\KbarN$ interaction and 
channel coupling to $\pi \Sigma$ provides the $\LamFOF$ with the decay 
width~\cite{Hyodo:2007jq}. In the present study, we confirmed
this picture by the decomposition of the baryonic charge in 
the previous section. Therefore,
it is instructive to understand the structure of the $\KbarN$ bound
state, by switching off the couplings of $\bar{K} N$ to other
channels.

In the case of the bound sate, the form factors are obtained 
as the real numbers for all $Q^{2}$, hence we can interpret the 
form factors as physical quantities.  Furthermore, 
the single-channel model allows us to investigate the
internal structure of a dynamically generated bound state. For
instance, we expect that the spatial size will be larger for 
the system with smaller binding energy. 
This does not trivially follow from the 
calculation of the form factor and is worth examining in the present
framework. Thus, we examine the structure of the $\bar{K}N$ bound
state by changing the model parameters.

In Sec.~\ref{subsec:Bound}, we will study the structure of the
$\KbarN$ bound state with the natural subtraction constant
$a_{\KbarN}=-1.95$, which is obtained so as to exclude explicit
pole contributions from the loop function~\cite{Hyodo:2008xr}.  This
ensures that the bound state has the pure molecule structure.  Then we will
study the structure of the dynamically generated $\KbarN$ bound states
with different binding energies in Sec.~\ref{subsec:Binding}.  The 
interaction 
strength is adjusted to fix the binding energy within the natural
condition.  We also discuss the meson-nucleon bound state with different
meson masses instead of the physical kaon mass, in order to see the 
effect of meson masses to the bound state in
Sec.~\ref{subsec:Kmass}, where the spatial size of the bound state is
kept fixed.

\subsection{Structure of the $\bm{\KbarN}$ bound state}
\label{subsec:Bound} 

In this subsection, we consider the $\KbarN$ single channel 
model with a
bound state.  We use the Weinberg-Tomozawa interaction~\eqref{eq:Vij}
for the $\KbarN$ channel and the subtraction constant
$a_{\KbarN}=-1.95$ with the regularization scale $\mu _{\text{reg}}=
630 \mev$, which is obtained by the condition
$G_{\KbarN}(M_{N})=0$~\cite{Hyodo:2008xr}.  With this subtraction
constant, solving the $\KbarN$ single-channel scattering
equation~\eqref{eq:BSEq} we obtain the $\KbarN$ bound state at $1424
\mev$ with the coupling strength to the $\KbarN$ [see
Eq.~\eqref{eq:T_mat}] as $g_{\KbarN}=2.17$. This bound state is
generated only by the attractive $\KbarN$ interaction 
and has a binding energy of $10 \mev$.


\begin{figure*}[!ht]
  \centering
  \begin{tabular*}{\textwidth}{@{\extracolsep{\fill}}cc}
    \includegraphics[width=8.6cm]{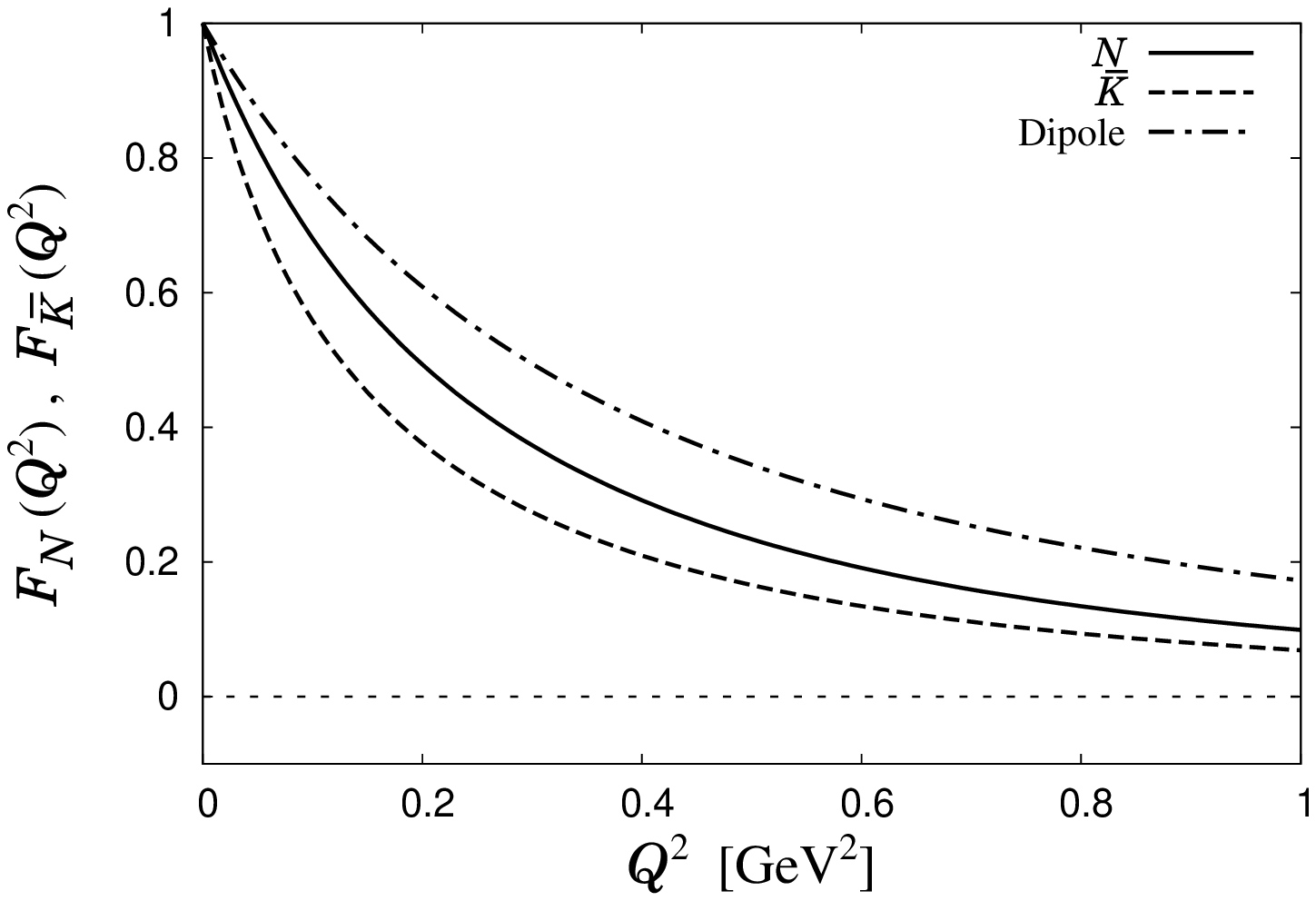} &
    \includegraphics[width=8.6cm]{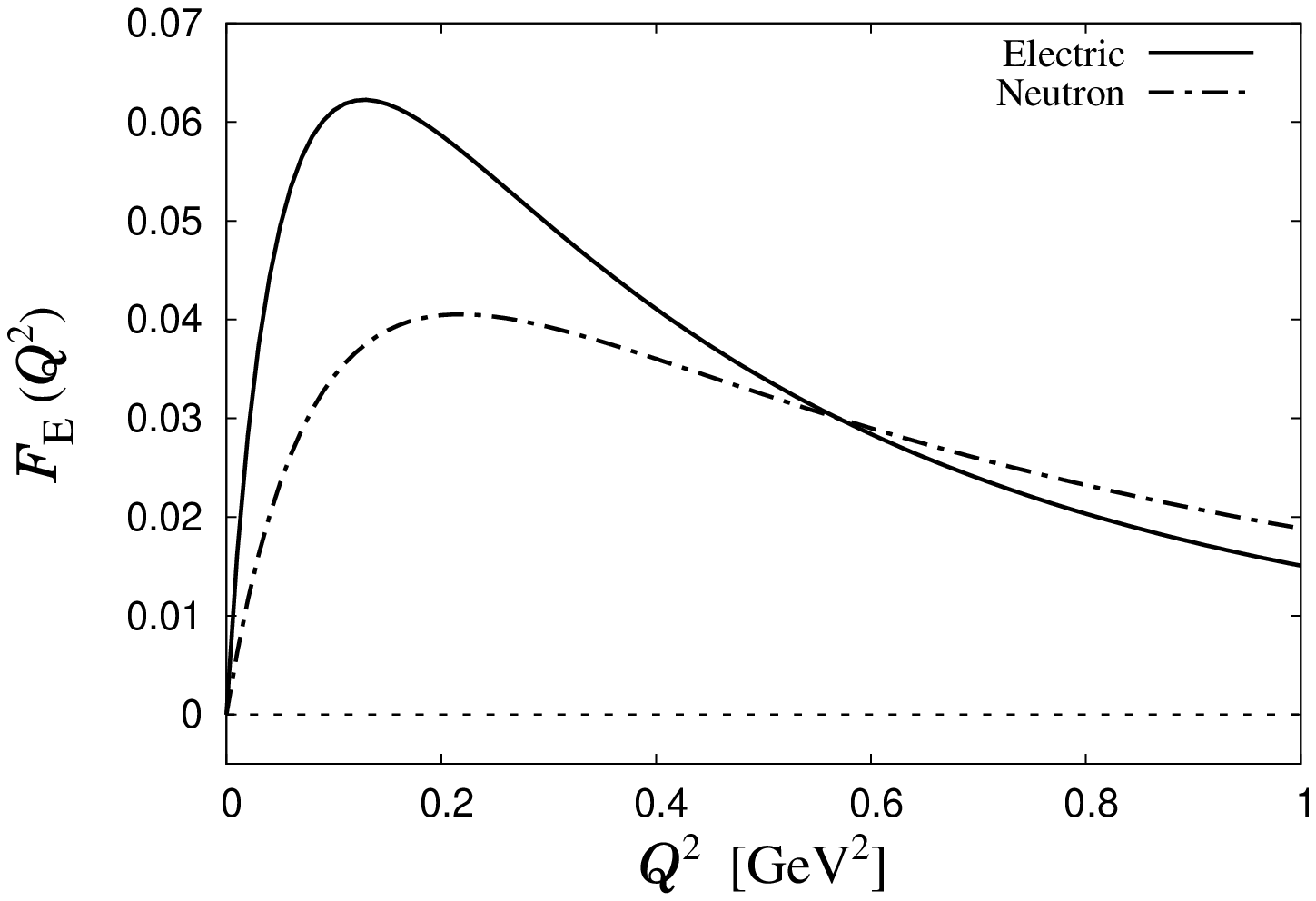} 
  \end{tabular*}
  \caption{The $\bar{K}$, $N$ (left) and electric (right) 
    form factors of the $\KbarN$ bound state 
    with mass $1424 \mev$, together with the empirical form factors 
    of the neutron. 
    The parameter $c$ in the dipole form factor is chosen to be $c=1$. }
  \label{fig:FF_KbarN}
\end{figure*}

Here we first study the structures of the $\bar{K}$ and $N$ components 
in the $\KbarN$ bound state.  Since the $\bar{K}$ ($N$) has strangeness 
$-1$ ($0$) and baryon number $0$ ($1$), we can observe $\bar{K}$ and $N$ 
distributions in the bound state by using the strangeness probe current 
with the opposite sign and the baryonic current, respectively, which 
leads to the relations for the form factors as, 
\be
F_{\bar{K}} (Q^{2}) = - \FS (Q^{2}) , \quad F_{N} (Q^{2}) = \FB (Q^{2}) , 
\ee
where $F_{\bar{K} (N)}$ represents the $\bar{K}$ ($N$) form factors 
probing the $\bar{K}$ ($N$) component in the bound state.  We also study 
the electric 
structure of the bound state, which is related to the $\bar{K}$ and 
$N$ form factors {\it via} the generalized Gell-Mann-Nishijima 
relation~\eqref{eq:genGN}.  Here we do
not consider magnetic component of the bound state.  
In Fig.~\ref{fig:FF_KbarN}, the $\bar{K}$, $N$, and electric form 
factors are shown together with the empirical form factors 
of the neutron~\eqref{eq:FEneutron} and~\eqref{eq:Gdipolefit}.  
Comparing with
the previous results for the resonant $\LamFOF$, 
we find that all of the electric, $N$, and $\bar{K}$ 
form factors of the $\KbarN$ bound state show
similar characteristic behaviors with the real part of the electric, 
baryonic, and opposite-sign-strangeness form factors 
of the resonant $\LamFOF$, respectively.  For the $\bar{K}$ and $N$ 
form factors $F_{\bar{K}, N}$, we observe the
steeper derivative at $Q^2 = 0$ compared with the dipole 
form factor and the faster decreasing $\bar{K}$ form factor 
than the $N$ one, which reflect a salient structure of
the bound state, as the resonant $\LamFOF$ discussed before.
Also the electric form factor $\FE$ exhibits large enhancement at 
smaller $Q^{2}$ region
($\lesssim 0.1 \gev ^{2}$) and slowly decrease above $Q^{2} \gtrsim
0.2 \gev ^{2}$, as the real part of the form factor of the resonant
$\LamFOF$.  

\begin{figure*}[!ht]
  \centering
  \begin{tabular*}{\textwidth}{@{\extracolsep{\fill}}cc}
    \includegraphics[width=8.6cm]{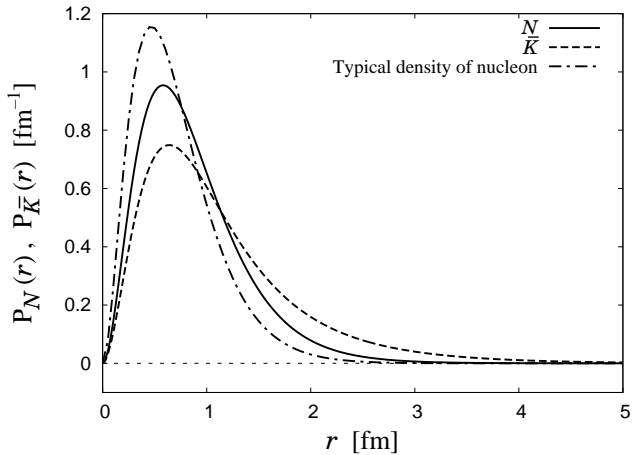} &
    \includegraphics[width=8.6cm]{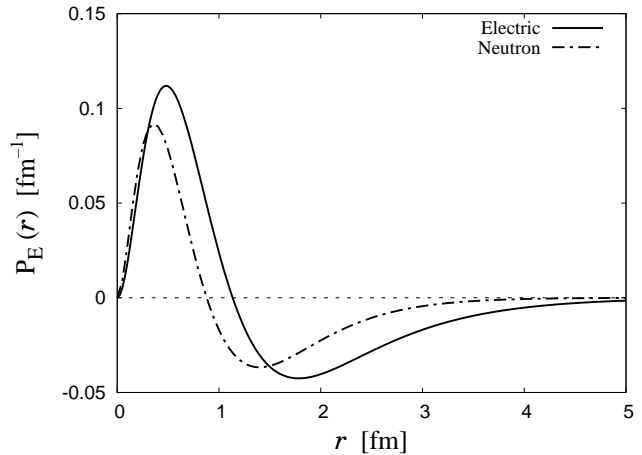} 
  \end{tabular*}
  \caption{The $\bar{K}$, $N$ (left) and charge (right) density 
    distributions 
    of the $\KbarN$ bound state with mass $1424 \mev$. 
    Typical density of nucleon is evaluated by 
    Eq.~\eqref{eq:Gdipolefit} with $c=1$. 
    Empirical charge distribution in the neutron 
    is evaluated by Eq.~\eqref{eq:FEneutron}. 
  }
  \label{fig:Rho_KbarN}
\end{figure*}


Using the form factors we calculate the density distributions through
the Fourier transformation.  In Fig.~\ref{fig:Rho_KbarN}, the $\bar{K}$, 
$N$, and charge density distributions from the form factors are
shown with the normalization $\text{P} (r) =4 \pi r^{2} \rho (r)$. 
As we can see from the figure, the behaviors of the density distributions are
similar to those for the resonant $\LamFOF$ in 
Figs.~\ref{fig:RhoEM_pole} and \ref{fig:RhoBS_pole}, 
thanks to the large $\KbarN$ coupling $g_{\KbarN}$ of the $\LamFOF$; 
the charge distribution has positive values in the inner part whereas the 
negative values in the outer part, and the $\bar{K}$ distribution has
longer tail than the $N$ one.  

\begin{table}[!t]
  \caption{\label{tab:MSR_KbarN}
    The $\bar{K}$, $N$, and electric mean squared radii, 
    $\MSR _{\bar{K}}$, $\MSR _{N}$, and $\EMSR$, and mean squared distance 
    between $\bar{K}$ and $N$, $\MSD _{\KbarN}$, of 
    $\KbarN$ bound state with mass $1424 \mev$. }
  \begin{ruledtabular}
    \begin{tabular}{cccc}
      $\MSR _{\bar{K}}$ [fm$^{2}$] & 
      $\MSR _{N}$ [fm$^{2}$] & 
      $\EMSR$ [fm$^{2}$]
      & $\MSD _{\KbarN}$ [fm$^{2}$] \\ 
      \hline
      $1.878$ & 
      $0.998$ & 
      $-0.440$ &
      $2.848$
    \end{tabular}
  \end{ruledtabular}
\end{table}

The results of the $\bar{K}$, $N$, and electric mean squared
radii evaluated from the form factors are displayed in 
Table~\ref{tab:MSR_KbarN}.  The results of the $\bar{K}$ and $N$ mean squared
radii, $\MSR _{\bar{K}} = 1.878 \fm ^{2}$ and $\MSR _{N}  = 0.998 \fm ^{2}$, 
respectively, 
indicate that both the $\bar{K}$ and $N$ distributions spread
compared with the typical nucleon size in the $\KbarN$ bound state,
and the $\bar{K}$ has larger distribution than the $N$ inside the
bound state.  The electric mean
squared radius $\EMSR$ is $-0.440 \fm ^{2}$, whose absolute value is
four times larger than that of the neutron $\sim -0.12 \fm ^{2}$.
Indeed, the charge density distribution $\PE$ clearly shows the larger
structure of the $\bar{K}N$ bound state than the typical neutral
hadron.

Next, it is also interesting to evaluate the mean squared distance between
$\bar{K}$ and $N$ from the form factors in our approach.  
For this purpose we take the nonrelativistic limit and 
treat the constituent hadrons as point particles, 
neglecting the common form factor effects [see Eq.~\eqref{eq:dipole}].  
In the nonrelativistic limit, both the
$\bar{K}$ and $N$ form factors $F_{\bar{K}}$ and $F_{N}$ are
determined from the form factor for the relative motion in the
two-body system $F_{\text{rel}}$ with appropriate scale
transformations as,
\be
F_{\bar{K}} ( Q^{2} ) = 
F_{\text{rel}} 
\left ( \left ( \frac{M_{N}}{m_{\bar{K}} + M_{N}} \right )^{2} 
Q^{2} \right ) , 
\ee
\be
F_{N} ( Q^{2} ) = 
F_{\text{rel}} 
\left ( \left ( \frac{m_{\bar{K}}}{m_{\bar{K}} + M_{N}} \right )^{2} 
Q^{2} \right ) , 
\ee
due to the kinematics of the system. 
Therefore, the mean squared distance between $\bar{K}$ and $N$, 
$\MSD _{\KbarN}= 6 \, dF_{\text{rel}}/dQ^{2}|_{Q^{2}=0}$, can be determined 
from both the $\bar{K}$ and $N$ form factors $F_{\bar{K}} (Q^{2})$ and 
$F_{N} (Q^{2})$ with appropriate coefficients.  The results are given by, 
\be
\left ( \frac{m_{\bar{K}} + M_{N}}{M_{N}} \right )^{2} \times \MSR _{\bar{K}} 
= 2.849 \fm ^{2} , 
\label{eq:MSD_K}
\ee
\be
\left ( \frac{m_{\bar{K}} + M_{N}}{m_{\bar{K}}} \right )^{2} \times \MSR _{N} 
= 2.846 \fm ^{2} , 
\label{eq:MSD_N}
\ee
for the mean squared distance between $\bar{K}$ and $N$.  The two
values in Eqs.~\eqref{eq:MSD_K} and \eqref{eq:MSD_N} show very good
agreement with each other.  A small difference between two values in
Eqs.~\eqref{eq:MSD_K} and \eqref{eq:MSD_N} is expected to come from
the field theoretical evaluation of the form factor and the
relativistic correction with respect to the binding energy.  In this
study we define the mean squared distance as the average of the values
evaluated from the $\bar{K}$ and $N$ radii, and we obtain the mean
squared distance between $\bar{K}$ and $N$ as 
\be \MSD _{\KbarN} =
2.848 \fm ^{2} , 
\ee 
for the $\KbarN$ bound state with binding energy $10 \mev$.

\subsection{$\bm{\KbarN}$ system with different binding energies}
\label{subsec:Binding} 

\begin{figure}[t]
  \includegraphics[width=8.6cm]{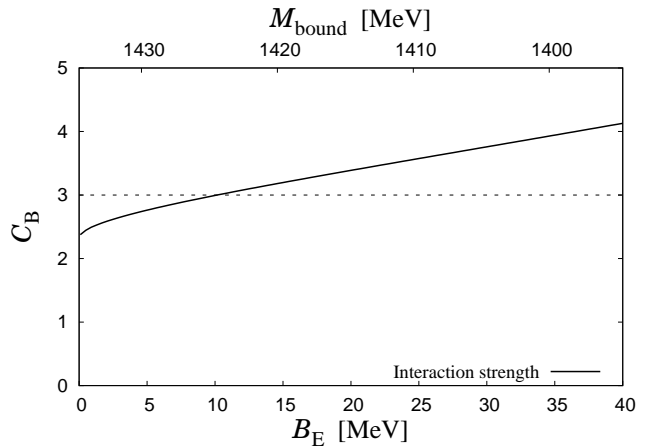}
  \caption{Interaction strength needed to generate a bound state 
    with different binding energies, $C_{\text{B}}$. 
    In the figure, $M_{\text{bound}}$ represents the mass of the bound state. }
  \label{fig:coupBE}
\end{figure}

Here we discuss the $\KbarN$ bound state with different binding 
energies, in order to see the dependence of the structure on the 
binding energy. This can be achieved by replacing the interaction 
strength $C=3$ for $\KbarN (I=0)$ in Eq.~\eqref{eq:Vij} with 
$C_{\text{B}}$ representing an interaction strength for $\KbarN$ to 
generate a bound state with a specified binding energy $B_{\text{E}}$, 
and we keep the subtraction constant $a_{\KbarN}=-1.95$ in order to 
exclude the explicit pole contribution. In Fig.~\ref{fig:coupBE} we plot 
interaction strength $C_{\text{B}}$ as a function of the 
binding energy $B_{\text{E}}$. 

\begin{figure}[!Ht]
  \includegraphics[width=8.6cm]{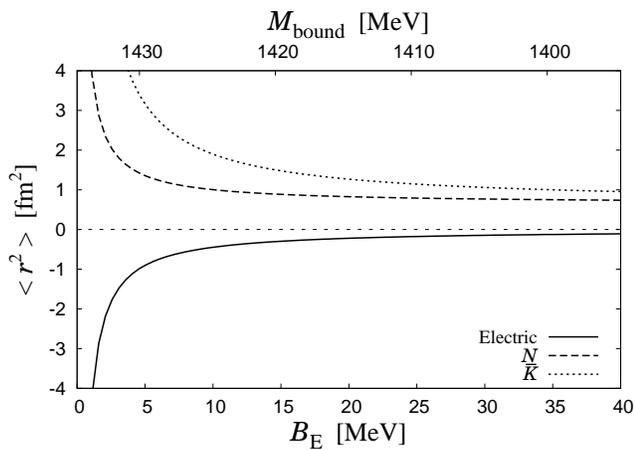}
  \caption{Mean squared radii of $\KbarN$ bound state with 
    different binding energies. 
    In the figure, $M_{\text{bound}}$ represents the mass of the bound state. }
  \label{fig:MSR_BE}
\end{figure}

We 
dynamically generate a $\KbarN$ bound system with different 
binding energies $B_{\text{E}}$
and show the
electric, $\bar{K}$, and $N$ mean squared radii 
as functions of the binding energy 
in
Fig.~\ref{fig:MSR_BE}.  It is obvious that the distribution of the
constituent hadrons in the $\KbarN$ bound state spreads if the binding
energy decreases, in accordance with our expectation from quantum
mechanics.  In addition, the mean squared radii are much sensitive to
the binding energy in the near-threshold region ($B_{\text{E}}
\lesssim 10 \mev$), which indicates that for the shallow bound state 
the binding energy is a key quantity for the spatial structure of the 
bound state.  We also 
find 
that the $\bar{K}$ distribution in the 
$\KbarN$ bound system is 
more 
sensitive to the binding energy in this 
region, since the $\bar{K}$ is lighter than the $N$.  
In the large binding energy region, the $\bar{K}$ and $N$ mean squared
radii asymptotically goes to finite values. 
This behavior in the large
binding energy region is caused by that the distributions of both 
$\bar{K}$ and $N$ shrink to the finite $\bar{K}$ and $N$ radii. 
In contrast, the electric radius goes to zero, because 
$\bar{K}$ and $N$
get close to each other, which leads to almost zero
electric mean squared radius.

\begin{figure}[!Ht]
  \includegraphics[width=8.6cm]{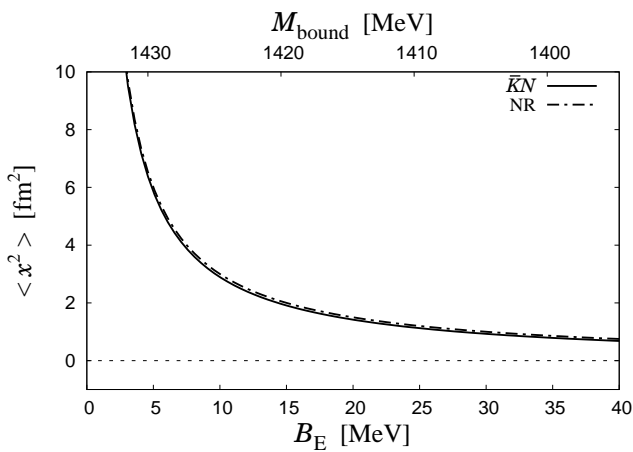}
  \caption{Mean squared distance between $\bar{K}$ and $N$ in the
    bound system in our approach (solid line), together with that
    obtained from the nonrelativistic wave function $\psi (x)$
    (dashed-dotted line, denoted as ``NR'').  In the figure,
    $M_{\text{bound}}$ represents the mass of the bound state. }
  \label{fig:MSD_BE}
\end{figure}

Next let us make a simple comparison of our results of mean squared
radii, which is based on the field theory, with that of the
nonrelativistic wave function of the two-body bound state, 
in order to check the consistency of these methods. Outside the
interaction range $R_{\text{int}}$ the nonrelativistic wave function
for the bound state takes an asymptotic form $\sim \exp \left(
  -\sqrt{2 \mu B_{\text{E}}} x \right)/x$ in the relative coordinate 
with the distance $x$ and the reduced mass $\mu = M_{N}
m_{\bar{K}} / (M_{N} + m_{\bar{K}})=324 \mev$.  Here we adopt a simple
form having the correct asymptotic behavior,
\be
\psi (x) = (\text{const.}) \times 
\frac{\exp \left( -\sqrt{2 \mu B_{\text{E}}} x \right)}{x} , 
\label{eq:wavefunction}
\ee
as a typical wave function for comparison. Using this wave function 
$\psi (x)$, one can evaluate the mean squared distance of 
the two particles in the bound state as, 
\be
\MSD _{\text{NR}} = 
\frac{\dsp \int _{0}^{\infty} d x \, 4 \pi x^{4} 
\left | \psi (x) \right |^{2}}
{\dsp \int _{0}^{\infty} d x \, 4 \pi x^{2}
\left | \psi (x) \right |^{2}}
= \frac{1}{4 \mu B_{\text{E}}} . 
\label{eq:NR-MSdistance}
\ee
Although it is derived from the specific wave
function~\eqref{eq:wavefunction}, this relation holds
model-independently for the state with small binding energy, where the
inner part ($x<R_{\text{int}}$) of the wave function is irrelevant and
the asymptotic form of the wave function 
dominates the integration in Eq.~\eqref{eq:NR-MSdistance}.  Now in
Fig.~\ref{fig:MSD_BE} we compare our result of the mean squared
distance between $\bar{K}$ and $N$, $\MSD _{\KbarN}$, 
which is evaluated by using method developed in the previous 
subsection for a bound state of point particles,
with that
obtained from the nonrelativistic wave
function~\eqref{eq:NR-MSdistance}, $\MSD _{\text{NR}}$.  As we can see
from the figure, the difference between $\MSD _{\KbarN}$ and $\MSD
_{\text{NR}}$ is quite small for every binding energy.  This is an
interesting result, because it is nontrivial that the mean squared
distance between $\bar{K}$ and $N$ in our approach, where the bound
state is dynamically generated and the size is probed by external
currents within the framework of the field theory, shows very similar
value to that from the nonrelativistic wave function $\psi$.

\subsection{Meson-nucleon bound system with various meson masses}
\label{subsec:Kmass} 

In this subsection, we vary the meson mass from the physical kaon mass, 
in order to see the effect of the meson 
mass to the meson-nucleon bound state. 
Here we consider the 
meson ($m^{-}$, $m^{0}$) 
having same quantum numbers 
as the antikaon ($K^{-}$, $\bar{K}^{0}$) except for the mass. 

\begin{figure}[t]
  \centering
  \includegraphics[width=8.6cm]{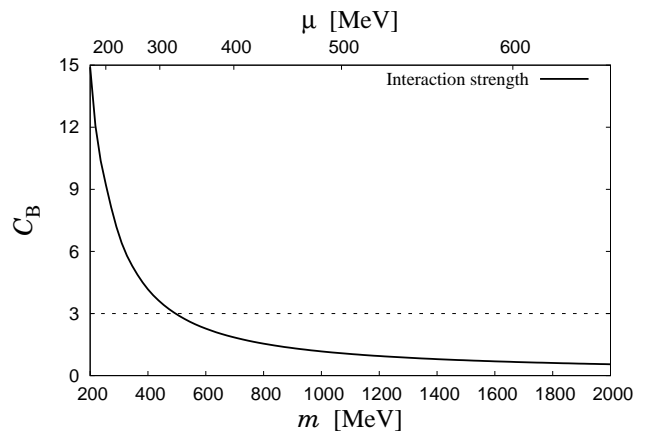}
  \caption{Interaction strength $C_{\text{B}}$ as a function of the 
    meson mass $m$, in order to generate a bound state 
    with the condition~\eqref{eq:condition}. 
  }
  \label{fig:coupMK}
\end{figure}

During the calculation, we fix the spatial structure of the bound system 
so that the meson-nucleon distance is to be in hadronic interaction range. 
For this purpose, we impose the condition, 
\be
\mu B_{\text{E}} = \text{const.} , 
\label{eq:condition} 
\ee
where $B_{\text{E}}$ is the binding energy of the system
and $\mu$ is the reduced mass, $\mu = m M_{N} 
/ (m + M_{N})$\footnote{For the study of the different meson
  masses, we denote the meson mass by $m$. }.  
The interaction strenght $C_{\text{B}}$ is adjusted by the binding energy 
through~\eqref{eq:condition} and the natural subtraction constant 
determined by $G_{\KbarN}(\sqrt{s}=M_{N}; \, 
m_{\bar{K}} \to m)=0$~\cite{Hyodo:2008xr}. 
We start with the bound state with the physical kaon mass
$m=m_{\bar{K}}=495.67 \mev$ which leads $B_{\text{E}}=10 \mev$ as
studied in section~\ref{subsec:Bound}.  
In Fig.~\ref{fig:coupMK} we plot interaction strength $C_{\text{B}}$ as a
function of the meson mass.

\begin{figure}[t]
  \includegraphics[width=8.6cm]{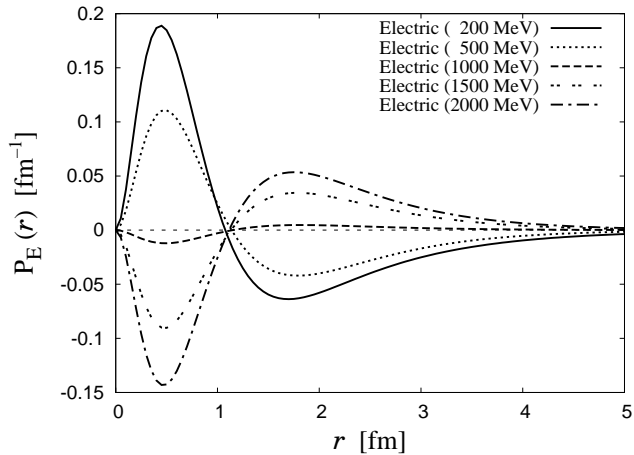}
  \caption{Charge density distribution of meson-nucleon bound state with 
    different meson masses, $200$, $500$, $1000$, $1500$, and $2000 \mev$. }
  \label{fig:Rho_MK}
\end{figure}

For the meson-nucleon bound state, the electric probe is useful to 
study the structure of the system, since the electric current probes 
the negatively charged $m^{-}$ and positively charged $p$ in the bound 
state, hence we can observe relative distribution of the meson and nucleon 
in the system as well as the spatial size of the system. 
In Fig.~\ref{fig:Rho_MK}, we show the charge density distribution of
the meson-nucleon bound state for different meson masses. As we have
discussed for the $\KbarN$ bound state, in the meson-nucleon bound 
state, concentration of positive
(negative) charge density corresponds to the region dominated by the
$p$ ($m^-$). For the small meson mass region ($m = 200$, $500
\mev$), the meson distributes at larger distance than the $N$, as 
the $\KbarN$ bound state in the preceding subsections.  If
the meson mass is compatible with the nucleon mass ($m = 1000
\mev$) the charge distribution is almost flat around zero, which
indicates that meson and $N$ have almost same distribution, and
cancel each other in the charge distribution. For the meson mass larger
than the nucleon ($m = 1500$, $2000 \mev$), the meson 
and $N$ distributions interchange their role, where the lighter
nucleon goes outside and the heavier meson stays inside.  Such a behavior is
consistent with our expectation from the classical picture and
indicates that we correctly observe internal structure of the bound
state with respect to the mass ratio of the constituent hadrons.

Here we note that the difference of the structures between the bound states 
of $\KbarN$ and heavy meson-nucleon, if existed, might be realized as a
difference of the behaviors of the bound states in nuclear
medium. For the $\KbarN$ (heavy meson-nucleon) bound state, $N$ has
smaller (larger) density distribution in the system, which implies
that $N$ has larger (smaller) momentum distribution due to the
uncertainty principle.  Therefore, in nuclear medium, where the
momentum of $N$ is filled up to the Fermi momentum, the existence of
heavy meson-nucleon bound state might be suppressed even in the
lower density compared with $\KbarN$.  

\begin{figure}[t]
  \includegraphics[width=8.6cm]{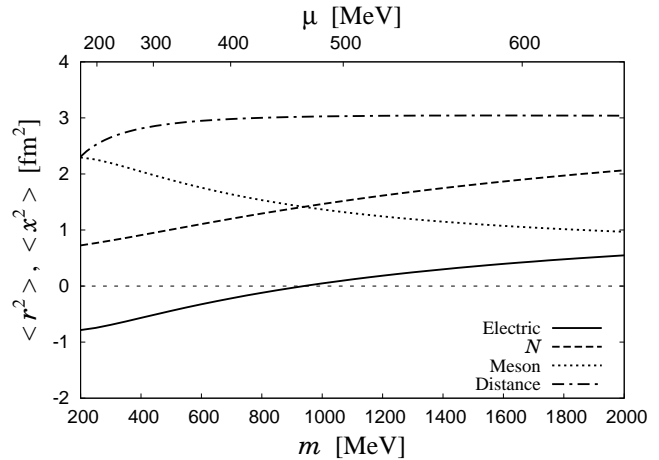}
  \caption{Electric, baryonic and strangeness mean squared radii of 
    meson-nucleon bound state with different 
    meson masses.  We also plot mean squared distance between meson and 
    nucleon, $\MSD$. }
  \label{fig:MSR_MK}
\end{figure}

It is also illustrative to plot the various mean squared radii and the
mean squared distance between meson and nucleon as functions of the 
meson mass (Fig.~\ref{fig:MSR_MK}. The mean squared distance is almost
independent of the meson mass, as a consequence of the fixed $\mu
B_{\text{E}}$. This suggests that the spatial size of the bound state
is indeed constrained by this condition. 
The suppression of the mean squared distance in the smaller 
meson mass region ($m \simeq 200 \mev$) is caused by the condition that 
the binding energy, which is restricted by Eq.~\eqref{eq:condition}, 
is not negligible compared with the meson mass. 
In addition, the electric, mesonic, and
nucleon mean squared radii behave consistently with our
interpretation discussed above.  The sign change of the electric radius
at $m = M_{N}$ reflects the inversion of the
spatial distribution of the meson and $N$.

At last we comment on the relation between the meson mass $m$ 
and the interaction strength $C_{\text{B}}$.  
The result in Fig.~\ref{fig:coupMK} shows that in 
the small meson mass region a strongly attractive
interaction is required to keep the light meson binding with the
nucleon, whereas the heavier meson can be bound as the molecular type
(with $\MSD \simeq 3 \fm ^{2}$) by the moderate attraction, in
accordance with the classical expectation.  From the group theoretical
point of view, the coupling strength of the scattering of the octet
baryon and the octet meson is at most
$C=6$~\cite{Hyodo:2006yk,Hyodo:2006kg}. The result in
Fig.~\ref{fig:coupMK} suggests that if the meson were as light as $200$
MeV, we would need the coupling strength $C\sim 15$ to generate a
bound state with the size $\MSD \simeq 3 \fm ^{2}$. This means that
the pion is too light to generate a bound state of the molecular
type\footnote{It is possible to generate a resonance, such as the
  $\sigma$ meson in $\pi\pi$ scattering, the lower pole of the
  $\Lambda(1405)$ resonance in $\pi\Sigma$ scattering, and the $a_1$
  meson in $\pi\rho$ scattering.}. On the other hand, the sufficiently
heavy meson mass 
($m \gtrsim 400 \mev$) 
is essential to generate a 
meson-nucleon bound state with
spatial size of a hadronic molecular state by using the chiral
interaction.

\section{Conclusion}
\label{sec:conclusion}

We have studied the internal structure of the resonant $\LamFOF$ state
with electromagnetic, baryonic, and strangeness probes. The resonant
$\LamFOF$ state has been described by the meson-baryon interaction picture
based on chiral dynamics. The form factors and the density
distributions of the $\LamFOF$ have been defined carefully as an extension
of ordinary stable particles and been directly evaluated from the
meson-baryon scattering amplitude, paying attention to the gauge
invariance of the probe interaction.

The internal structure of the resonant $\LamFOF$ with full
coupled-channels has been studied in two ways, one on the resonance
pole position in the complex energy plane and the other on the real
energy axis around the $\LamFOF$ resonance region. The first method has 
produced 
the resonant $\LamFOF$ structure as the matrix element of
the probe current by the resonance vector without contamination from
nonresonant background, and exactly kept the gauge invariance. The
second one, on the other hand, has given values which may be observed in
experiments while the nonresonant contamination has been involved.  We 
have found that the resonant $\LamFOF$ state has softer form factors
and larger spatial radii than those of the typical baryon, and
the structure is largely dominated by the $\bar{K}N$ component. The
charge distribution has indicated that the light $\bar{K}$ distributes
around the proton which sits in the central region. The characteristic 
structure of the $\LamFOF$ shown in the resonance form factor on the
resonance pole position can be kept in the effective form factor on
the real axis. We have also found
that the decay of the $\LamFOF$ into the $\pi \Sigma$ channel through
the photon coupling causes an escaping oscillation pattern in the
density distribution in the coordinate space.

We have also discussed internal structure of a $\KbarN$ bound state
without the decay width to extract an intuitive picture of the bound
state. We have found the similar structure with the resonant
$\LamFOF$, thanks to the dominance of the $\bar{K}N$ component. We
also have observed that such a structure shrinks (stretches) as the
binding energy increases (decreases), which is consistent with the
behavior of the bound state in the quantum mechanics.  We have seen
the meson mass dependence of the bound state which indicates that the
behavior of the relative distribution of the meson and nucleon inside
the bound state is consistent with the kinematics of meson-nucleon
system.  These results have verified that the form factor defined
through the scattering amplitude serves as a natural generalization of
the form factor for the resonance state.  Also the relation between the
coupling strength and the meson mass shows that the physical kaon mass
appears to be within the suitable range to form a molecular bound
state with the nucleon through the chiral SU(3) interaction.

\begin{acknowledgments}
We acknowledge 
Y.~Kanada-En'yo, 
T.~Myo, 
A.~Hosaka, 
A.~Ohnishi, 
H.~Suganuma, and 
K.~Yazaki 
for useful discussions. This work is partly supported by the 
Grand-in-Aid for Scientific Research from MEXT and JSPS 
(Nos. 
21840026, 
22105507, 
22740161, 
and 22-3389
), and 
the Grant-in-Aid for the Global COE Program ``The Next Generation of Physics, 
Spun from Universality and Emergence'' from MEXT of Japan. 
One of the authors, T.S. acknowledges the support by the Grand-in-Aid 
for JSPS fellows. 
T.H. thanks the support from the Global Center of Excellence Program by MEXT, Japan through the Nanoscience and Quantum Physics Project of the Tokyo Institute of Technology.
This work is part of the Yukawa International Program for Quark-Hadron 
Sciences (YIPQS).
\end{acknowledgments}

\appendix

\section{Normalization of form factors}
\label{sec:normalization}

In this Appendix we discuss the normalization of the $MB \gamma ^{\ast} \to
M^{\prime} B^{\prime}$ amplitude $T_{\gamma}^{\mu}(\sqrt{s};\, Q^2)$ induced 
by a conserved current $J^{\mu}$ 
in the soft limit $Q^{2}\to 0$, 
where $\sqrt{s}$ and $Q^{2}$ are the meson-baryon
center-of-mass energy and the squared momentum transfer with
opposite sign, respectively.  
Because we are considering the matrix element with the conserved current 
$J^{\mu}$, the amplitude $MB \gamma ^{\ast} \to
M^{\prime} B^{\prime}$ should automatically be 
normalized by Ward-Takahashi identity: 
\be
\hat{Q} \frac{d T_{ij} (\sqrt{s} )}{d \sqrt{s}} 
= T_{\gamma ij}^{\mu =0} (\sqrt{s} ; \, Q^{2}=0) , 
\label{eq:app-norm}
\ee
where $T(\sqrt{s})$ is the $MB \to M^{\prime} B^{\prime}$ amplitude, 
$\hat{Q}$ the ``charge'' of the meson-baryon system with respect to 
the probe current, and 
the indices $i$ and $j$ 
stand for the final and initial
meson-baryon channels, respectively\footnote{The Ward identity of the 
general unitarized amplitude was discussed in Ref.~\cite{Borasoy:2005zg}. 
Here we confirm the Ward-Takahashi identity in a derivative 
form~\eqref{eq:app-norm} within our formulation. }. Thus, we
prove that the amplitudes discussed in 
sections~\ref{subsec:dynamics} and \ref{subsec:EMint} 
satisfy Eq.~\eqref{eq:app-norm}, concentrating on the $Q^{2}=0$ case.

In this study the meson-baryon amplitude $T$ is obtained 
as the solution of the Bethe-Salpeter (BS) equation 
\be
T_{ij} = V_{ij} + \sum _{k} V_{ik} G_{k} T_{kj} 
= V_{ij} + \sum _{k} T_{ik} G_{k} V_{kj} , 
\label{eq:app-BS}
\ee
with the interaction kernel $V$ and the meson-baryon loop integral $G$
[see Eq.~\eqref{eq:BSEq}]. The $T_{\gamma}^{\mu}$ amplitude is 
obtained by evaluating ten diagrams given in
Figs.~\ref{fig:Tgamma} and \ref{fig:Tgamma-other} in which the probe
current is attached to the meson-baryon scattering amplitude. Among
them, the amplitudes $T_{\gamma (7)}^{\mu}$, $T_{\gamma (8)}^{\mu}$,
$T_{\gamma (9)}^{\mu}$, and $T_{\gamma (10)}^{\mu}$ diverges in the
soft limit $Q^{2}\to 0$. These terms correspond to the infrared divergences
coming from the bremsstrahlung processes of the external hadrons and
are canceled by the vertex corrections, leaving no contributions to
the amplitude $T_{\gamma}^{\mu}$ at $Q^{2}=0$.  
Hence, in the soft limit, only the
six diagrams are relevant for $T_{\gamma}^{\mu}$,
\be
T_{\gamma ij}^{\mu} = \sum _{a=1}^{6} T_{\gamma (a) ij}^{\mu} , 
\label{eq:app-Tgamma}
\ee
with each contribution calculated in our formulation as, 
\be
T_{\gamma (1) ij}^{\mu} 
= \sum _{k} T_{ik} D_{\text{M}_{k}}^{\mu} T_{kj} , 
\label{eq:app-T1}
\ee
\be
T_{\gamma (2) ij}^{\mu} 
= \sum _{k} T_{ik} D_{\text{B}_{k}}^{\mu} T_{kj} , 
\label{eq:app-T2}
\ee
\be
T_{\gamma (3) ij}^{\mu} 
= \sum _{k, l} T_{ik} G_{k} \Gamma _{kl}^{\mu} G_{l} T_{lj} , 
\label{eq:app-T3}
\ee
\be
T_{\gamma (4) ij}^{\mu} 
= \Gamma _{ij}^{\mu} , 
\label{eq:app-T4}
\ee
\be
T_{\gamma (5) ij}^{\mu} 
= \sum _{k} \Gamma _{ik}^{\mu} G_{k} T_{kj} , 
\label{eq:app-T5}
\ee
\be
T_{\gamma (6) ij}^{\mu} 
= \sum _{k} T_{ik} G_{k} \Gamma _{kj}^{\mu} , 
\label{eq:app-T6}
\ee
where the loop integrals with photon couplings 
$D_{\text{M}}^{\mu}$, $D_{\text{B}}^{\mu}$, and the vertex function 
$\Gamma ^{\mu}$ are defined in Sec.~\ref{sec:formulation}. 

The key relations for proving Eq.~\eqref{eq:app-norm} are 
the derivative of the vertex and the loop integrals at $Q^{2}=0$, 
\be
\hat{Q} \frac{d V_{ij}}{d \sqrt{s}} = \Gamma _{ij}^0 |_{Q^{2}=0}, 
\label{eq:app-dVds}
\ee
\be
\hat{Q} \frac{d G_{k}}{d \sqrt{s}} = 
( D_{\text{M}_{k}}^0 + D_{\text{B}_{k}}^0 )|_{Q^{2}=0} . 
\label{eq:app-dGds}
\ee
Equations~\eqref{eq:app-dVds} and \eqref{eq:app-dGds} are the 
Ward-Takahashi identity for the elementary vertex $V$ 
and two-body free propagator $G$. Thus, if the vertex 
$\Gamma$ and the loop integrals $D_{\text{M}}$ and $D_{\text{B}}$ 
conserve the charge associated with the current, 
Eqs.~\eqref{eq:app-dVds} and \eqref{eq:app-dGds} should be satisfied. 
Then, using
these relations and the BS equation~\eqref{eq:app-BS}, we can prove
Eq.~\eqref{eq:app-norm} by an algebraic calculation. Omitting the
superscript $\mu=0$ and evaluating all the terms at $Q^{2}=0$, 
we can write the left-hand side of Eq.~\eqref{eq:app-norm} as, 
\begin{align}
& [\text{Left-hand side of \eqref{eq:app-norm}}]
\nonumber \\ 
& = \hat{Q} \frac{d T}{d \sqrt{s}} 
= \hat{Q} \frac{d}{d \sqrt{s}} (V + V G T) 
\nonumber \\ 
& = \Gamma + \Gamma G T 
+ \hat{Q} V \left ( \frac{d G}{d \sqrt{s}} T + G \frac{d T}{d \sqrt{s}} 
\right ) , 
\label{eq:app-prove1}
\end{align}
where we have use Eqs.~\eqref{eq:app-BS} and \eqref{eq:app-dVds}. 
From Eqs.~\eqref{eq:app-T4} and \eqref{eq:app-T5}, the first 
and second terms in the right-hand side of~\eqref{eq:app-prove1} are 
$T_{\gamma (4)}$ and $T_{\gamma (5)}$, respectively.  Replacing 
$V$ in the third term of Eq.~\eqref{eq:app-prove1} with $T-TGV$, we have, 
\begin{align}
& [\text{Left-hand side of \eqref{eq:app-norm}}] 
\nonumber \\ & 
= \sum _{a=4,5} T_{\gamma (a)} 
+ \hat{Q} (T - T G V) \left ( \frac{d G}{d \sqrt{s}} T + 
G \frac{d T}{d \sqrt{s}} \right ) 
\nonumber \\ 
& 
= \sum _{a=4,5} T_{\gamma (a)} 
+ \hat{Q} T \frac{d G}{d \sqrt{s}} T 
- \hat{Q} T G V \frac{d G}{d \sqrt{s}} T 
\nonumber \\ 
& \phantom{=} 
+ \hat{Q} T G \frac{d T}{d \sqrt{s}} 
- \hat{Q} T G V G \frac{d T}{d \sqrt{s}} . 
\label{eq:app-prove2}
\end{align}
From Eqs.~\eqref{eq:app-T1} and \eqref{eq:app-T2} with \eqref{eq:app-dGds}, 
the second term of the right-hand side in~\eqref{eq:app-prove2} is equal 
to $T_{\gamma (1)}+T_{\gamma (2)}$. Then, using the product rule of the 
differentiation for the third and fifth terms of Eq.~\eqref{eq:app-prove2}, 
we have, 
\begin{align}
& [\text{Left-hand side of \eqref{eq:app-norm}}] 
\nonumber \\ 
& 
= \sum _{a=1,2,4,5} T_{\gamma (a)} 
- \hat{Q} T G \left ( \frac{d}{d \sqrt{s}} (V G T) - 
\frac{d V}{d \sqrt{s}} G T \right ) 
\nonumber \\ 
& \phantom{=} 
+ \hat{Q} T G \frac{d T}{d \sqrt{s}} .
\label{eq:app-prove3}
\end{align}
From Eqs.\eqref{eq:app-T3} and \eqref{eq:app-dVds}, 
the second term in the parenthesis in Eq.~\eqref{eq:app-prove3} 
is $T_{\gamma (3)}$. Therefore, collecting remained terms, 
\begin{align}
& [\text{Left-hand side of \eqref{eq:app-norm}}] 
\nonumber \\ 
& 
= \sum _{a=1,2,3,4,5} T_{\gamma (a)} 
+ \hat{Q} T G \frac{d}{d \sqrt{s}} (T - V G T) 
\end{align}
Then, using Eqs.~\eqref{eq:app-BS}, \eqref{eq:app-T6}, 
and \eqref{eq:app-dVds}, 
we finaly achieve from the left-hand side of Eq.~\eqref{eq:app-norm} 
to its right-hand side, 
\begin{align}
& [\text{Left-hand side of \eqref{eq:app-norm}}] 
\nonumber \\
& = \sum _{a=1}^{6} 
T_{\gamma (a)} = [\text{Right-hand side of \eqref{eq:app-norm}}]. 
\end{align}
Thus, with~\eqref{eq:app-Tgamma}, we obtain Eq.~\eqref{eq:app-norm}.
We note that this relation is valid at any value of the meson-baryon 
center-of-mass energy $\sqrt{s}$, even in complex energy plane. 
Since Eq.~\eqref{eq:app-norm} is the relation between meson-baryon
scattering amplitude and that with photon attached, the existence of
the resonance pole is not necessary for the argument above.  

The relation~\eqref{eq:app-norm} immediately leads to correct 
normalization of the
effective form factor defined in Eq.~\eqref{eq:Feffective}, as,
\be
\left . F^{\text{eff}\, \mu =0} (Q^{2}=0 ; \, \sqrt{s}) 
= \frac{T^{\mu=0}_{\gamma ij}}{d T_{ij} / d \sqrt{s}} \right | _{Q^{2}=0}
= \hat{Q} . 
\ee
This normalization is achieved 
at any energies $\sqrt{s}$, regardless 
of existence of the resonance, as long as 
one includes all the relevant 
contributions in Eq.~\eqref{eq:app-Tgamma}. 
If there exists the resonance pole at the complex energy 
$Z_{\text{R}}$ in the meson-baryon scattering, 
taking limit $\sqrt{s}\to Z_{\text{R}}$ we see that 
$\Feff =T_{\gamma ij}^{\mu =0}/(dT_{ij}/d\sqrt{s})$ coincides with the resonance 
form factor $F$ evaluated in Eq.~\eqref{eq:Res_scheme}.  Hence 
the relation~\eqref{eq:app-norm} also guarantees the correct 
normalization of the resonance form factor $F$.  
In the case that the energy $\sqrt{s}$ is close to the 
resonance pole position ($\sqrt{s} \simeq 1420 \mev$ in our  
case) on the real axis, 
the charge $\hat{Q}$ will be dominated by the contributions 
from the three diagrams in Fig.~\ref{fig:Tgamma}, 
as discussed in Sec.~\ref{subsec:photon-coupled}. 

We also derive, 
from Ward-Takahashi identity~\eqref{eq:app-norm}, 
a relation between the 
coupling strengths of the resonance state to the meson-baryon channel $g$ and 
the derivative of the vertex $V$ and the loop function $G$: 
\be
\sum _{i, j} g_{i} g_{j} \left . 
\left ( \frac{d G_{i}}{d \sqrt{s}} \delta _{ij} 
+ G_{i} \frac{d V_{ij}}{d \sqrt{s}} G_{j} \right ) 
\right |_{\sqrt{s} \to Z_{\text{R}}} 
= - 1 . 
\label{eq:app-g_dVds}
\ee
To prove Eq.~\eqref{eq:app-g_dVds}, we consider explicit forms of 
$T_{\gamma ij}^{\mu =0}$ and $dT_{ij}/d \sqrt{s}$ for the resonance 
contribution. 
The derivative of the $MB \to M^{\prime} B^{\prime}$ amplitude 
$dT_{ij}/d \sqrt{s}$ 
is shown in Eq.~\eqref{eq:delT}.  
The $MB \gamma ^{\ast} \to M^{\prime} B^{\prime}$ amplitude 
$T_{\gamma ij}^{\mu =0}$ has double-pole terms for the 
resonance as well as the less singular terms.  
The former corresponds to Fig.~\ref{fig:Tgamma} and is expressed in 
Eqs.~\eqref{eq:app-T1}--\eqref{eq:app-T3}, while the latter 
does to Fig.~\ref{fig:Tgamma-other} and is in 
Eqs.~\eqref{eq:app-T4}--\eqref{eq:app-T6}.  Using the explicit 
form of $T_{ij}$ given in Eq.~\eqref{eq:T_matreal} and the 
Ward-Takahashi identity~\eqref{eq:app-dVds} and \eqref{eq:app-dGds} 
for the double-pole terms, $T_{\gamma ij}^{\mu =0}$ can be expressed as, 
\begin{align}
& T_{\gamma ij}^{\mu =0}(\sqrt{s}; \, Q^{2}=0) 
\nonumber \\
& = \sum _{k. l} \frac{g_{i} g_{k}}{\sqrt{s} - Z_{\text{R}}} 
\hat{Q} \left ( \frac{d G_{k}}{d \sqrt{s}} \delta _{kl} 
+ G_{k} \frac{d V_{kl}}{d \sqrt{s}} G_{l} \right ) 
\frac{g_{l} g_{j}}{\sqrt{s} - Z_{\text{R}}} 
\nonumber \\ 
& \phantom{=} + T_{\gamma ij}^{\text{less}} , 
\end{align}
where $T_{\gamma ij}^{\text{less}}$ represents the less singular terms 
than the double-pole contribution.  Thus, calculating 
$\hat{Q} = T_{\gamma ij}^{\mu =0}/(dT_{ij}/d\sqrt{s})$ and taking limit 
$\sqrt{s}\to Z_{\text{R}}$, 
both the nonresonant contribution 
in $dT_{ij}/d\sqrt{s}$ and less singular terms $T_{\gamma ij}^{\text{less}}$ 
in $T_{\gamma ij}^{\mu =0}$ automatically drop and 
we have, 
\begin{align}
  \hat{Q} & = \left . \frac{T_{\gamma ij}^{\mu =0}}{dT_{ij}/d\sqrt{s}} 
  \right |_{\sqrt{s}\to Z_{\text{R}}, \, Q^{2}=0} 
  \nonumber \\
  & = - \hat{Q} 
  \sum _{k, l} g_{k} g_{l} \left . 
    \left ( \frac{d G_{k}}{d \sqrt{s}} \delta _{kl} 
      + G_{k} \frac{d V_{kl}}{d \sqrt{s}} G_{l} \right ) 
  \right |_{\sqrt{s} \to Z_{\text{R}}} , 
\end{align}
hence, we obtain Eq.~\eqref{eq:app-g_dVds}. 
This is the exact form of the relation between the coupling 
strength $g_{i}$ and the derivative of the loop integral $dG_{i}/d\sqrt{s}$, 
discussed in~\cite{Gamermann:2009uq,YamagataSekihara:2010pj}. 
Equation~\eqref{eq:app-g_dVds} exactly corresponds to the 
relation known as the Ward identity: 
\be
Z_{(1)}^{-1} Z_{(2)} = 1 , 
\ee
with the vertex-renormalization factor 
$Z_{(1)}^{-1}$ and the wave-function one $Z_{(2)}$, respectively. 
In our case with the resonance state, 
$Z_{(2)}$ is the residue of the resonance contribution in the meson-baryon 
scattering amplitude, $Z_{(2) ij} = -g_{i}g_{j}$, and $Z_{(1)}^{-1}$ 
is the current-coupling term, 
\be
Z_{(1) ij}^{-1} = 
\left ( \frac{d G_{i}}{d \sqrt{s}} \delta _{ij} 
  + G_{i} \frac{d V_{ij}}{d \sqrt{s}} G_{j} \right ) , 
\ee
through Eqs.~\eqref{eq:app-dVds} and \eqref{eq:app-dGds}.  
Using $Z_{(1) ij}^{-1}$ and $Z_{(2) ij}$, Eq.~\eqref{eq:app-g_dVds} 
can be rewritten as, 
\be
\trace \left [ Z_{(1)}^{-1} Z_{(2)} \right ]
= \sum _{i, j} Z_{(1) ij}^{-1} Z_{(2) ji} = 1 . 
\ee

\section{Loop integrals $\bm{D_{\text{M}}^{\mu}}$ and $\bm{D_{\text{B}}^{\mu}}$}
\label{sec:Loop}

In this Appendix we discuss the properties of the loop integral
$D_{\text{M}, \text{B}}^{\mu} ( \sqrt{s}; \, Q^{2} )$ [given in 
Eqs.~\eqref{eq:DloopM} and 
\eqref{eq:DloopB}], which describes that one photon is attached to the 
one of the internal propagators. For concreteness, we consider the 
electromagnetic probe interaction for the $\LamFOF$.

\begin{figure*}[!Ht]
  \centering
  \begin{tabular*}{\textwidth}{@{\extracolsep{\fill}}cc}
    \includegraphics[width=8.6cm]{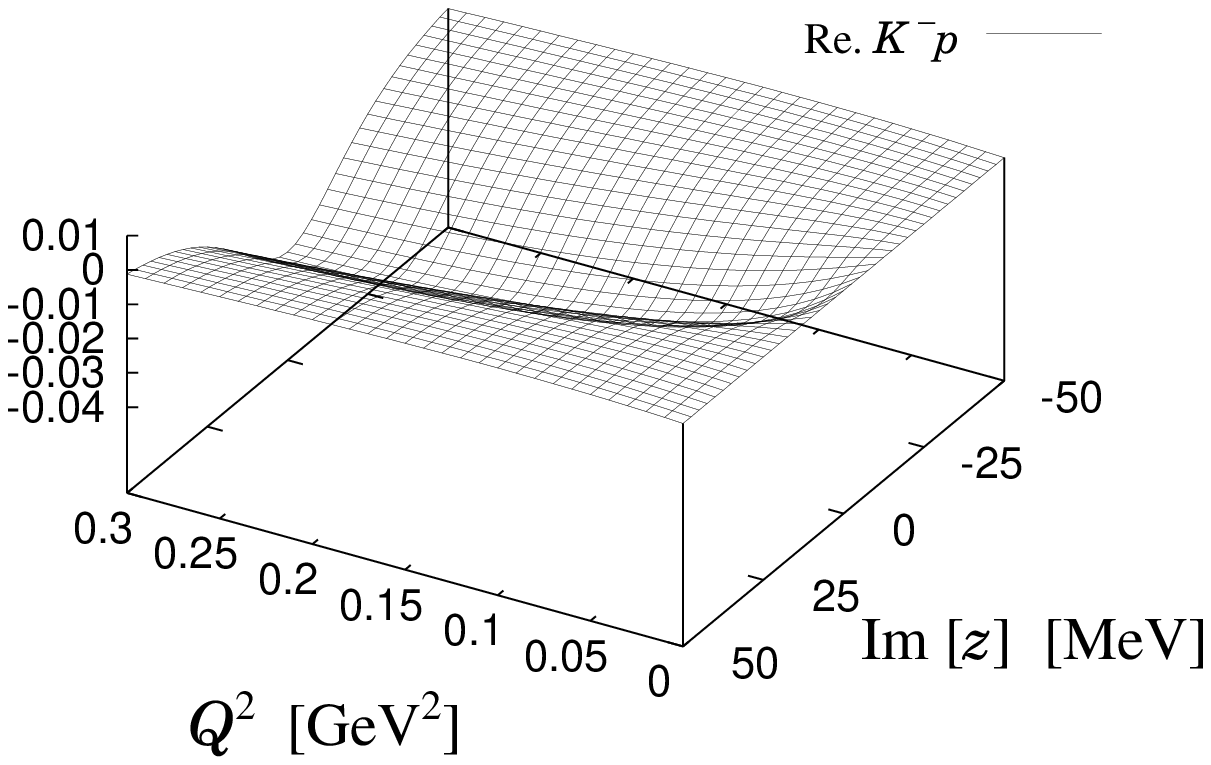} &
    \includegraphics[width=8.6cm]{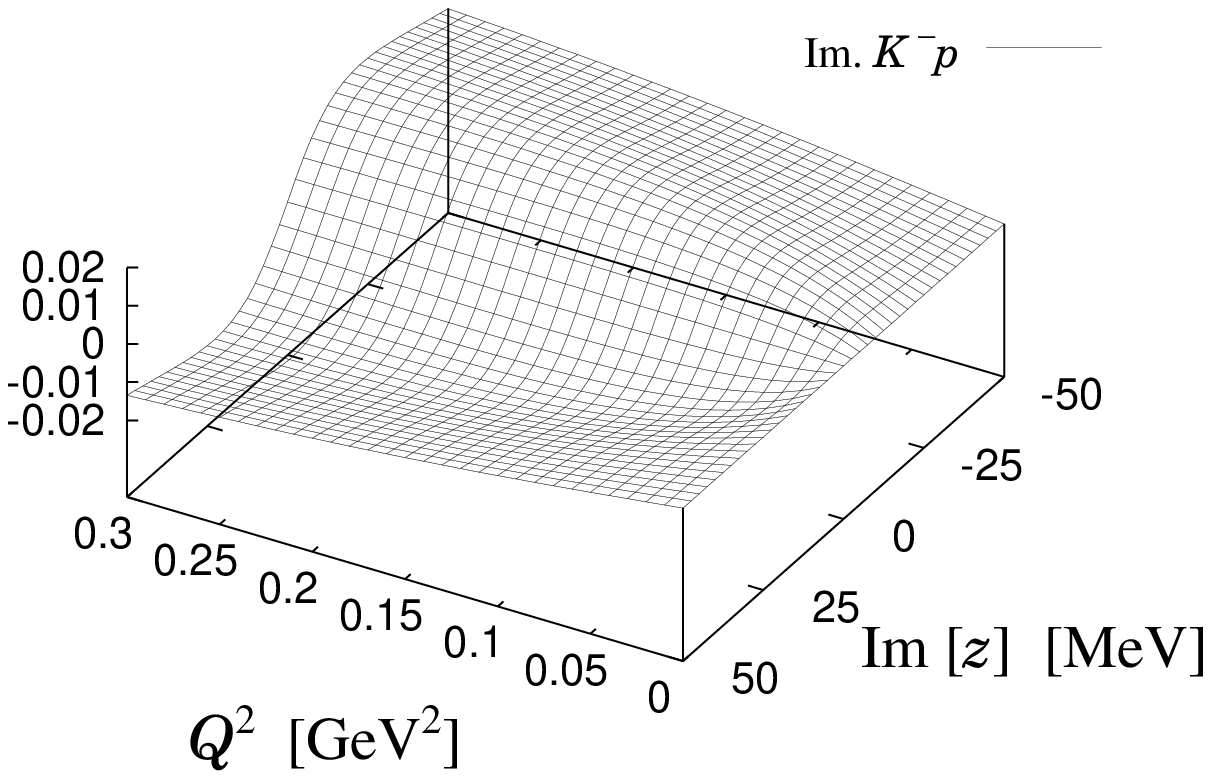} \\
    \includegraphics[width=8.6cm]{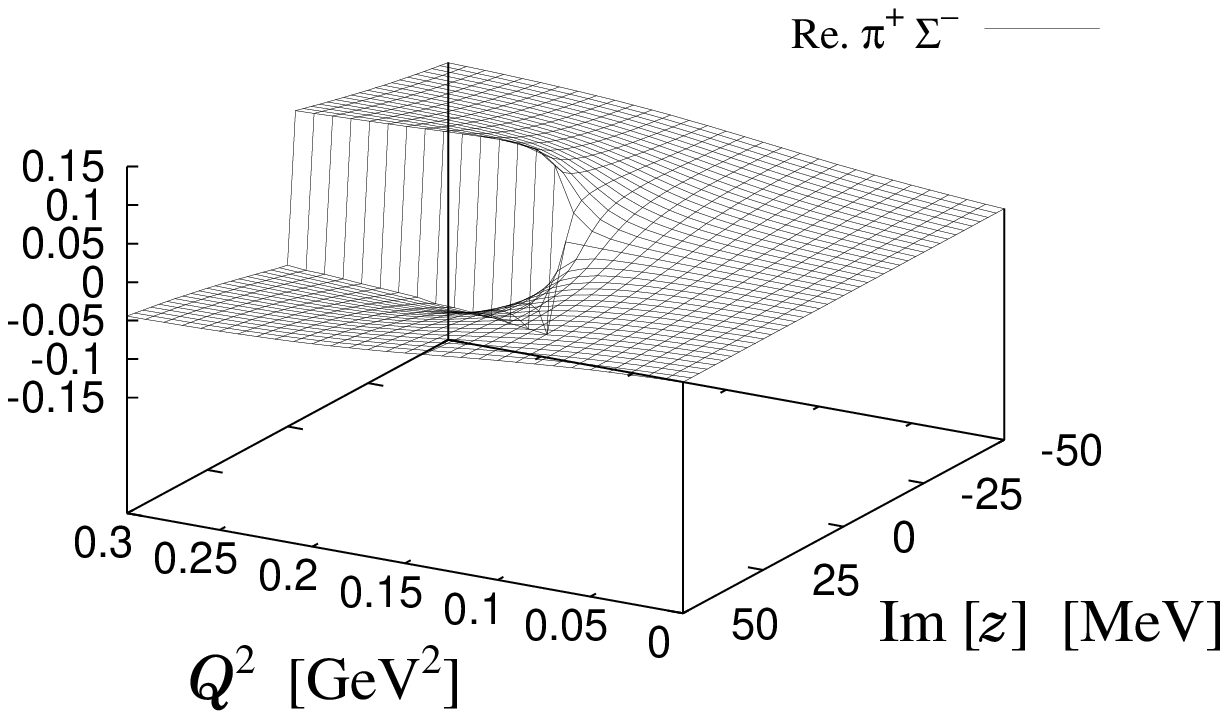} &
    \includegraphics[width=8.6cm]{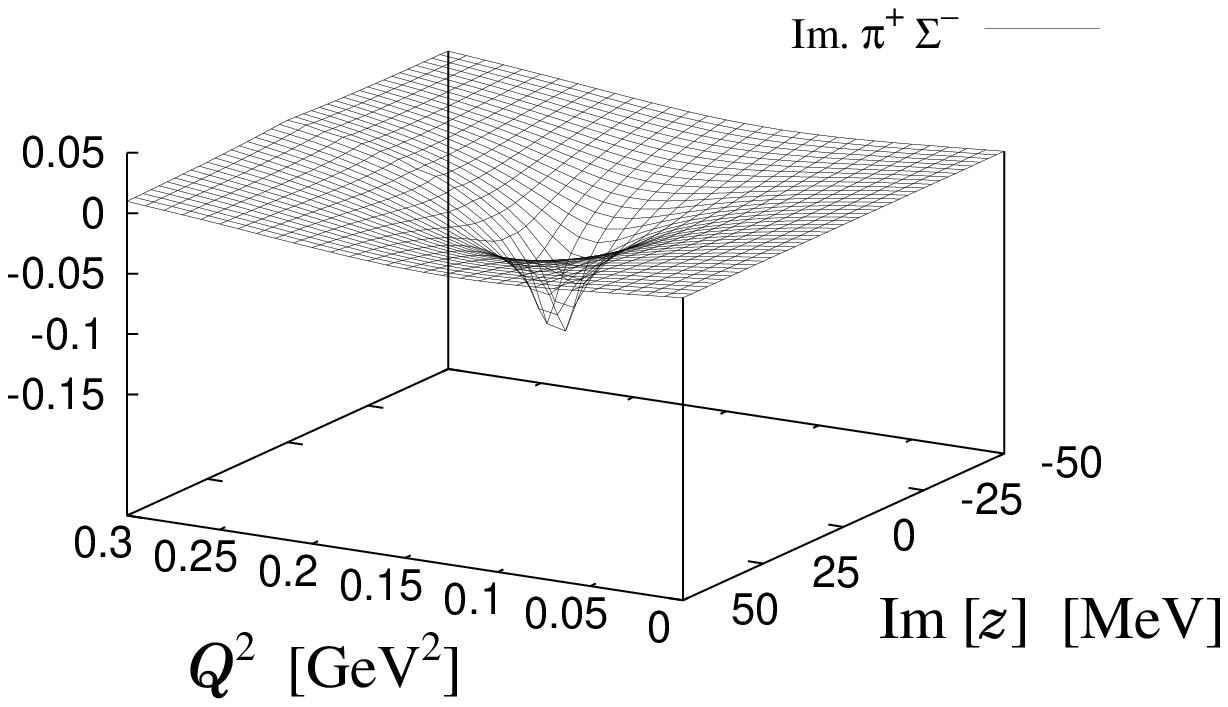} 
  \end{tabular*}
  \caption{The real part (left) and the imaginary (right) parts of the
    loop integrals $D_{\text{M}}^{0}+D_{\text{B}}^{0}$ of the $K^-p$
    channel (upper) and the $\pi^+\Sigma^-$ channel (lower) as
    functions of $\text{Im}[z]$ and $Q^{2}$. The real part of the
    energy is chosen to be $\text{Re}[z]=1420 \mev$. }
\label{fig:app-Loop}
\end{figure*}

Integrating over the four-momentum $q_{1}^{\mu}$ in
Eqs.~\eqref{eq:DloopM} and \eqref{eq:DloopB} in the Breit frame for 
the $\LamFOF$, these functions are
explicitly written with the Feynman parameters $x$ and $y$ as
\begin{widetext}
\begin{align}
D_{\text{M}_{k}}^{0} \! \left( \sqrt{s};\, Q^2 \right) 
&=- \frac{1}{(4 \pi )^2} \int_{0}^{1} d x \int_{0}^{1} d y 
\frac{4 x (1 - x) Q_{\text{M}_{k}} M_{k} \sqrt{s + Q^{2}/4}}
{x m_{k}^2 + (1 - x) M_{k}^2 - x (1 - x) s + 
x^{2} y (1 - y) Q^{2} - i \epsilon } \nonumber \\
&=- \frac{Q_{\text{M}_{k}}}{(4 \pi )^2} \int_{0}^{1} d y 
\frac{4 M_{k} \sqrt{s + Q^{2}/4}}
{s + y (1 - y) Q^{2}} 
\Bigg [ -1 + \frac{m_{k}^{2} - M_{k}^{2} + y (1 - y) Q^{2}}
{2 ( s + y (1 - y) Q^{2} )} 
\ln \left( \frac{m_{k}^{2} + y (1 - y) Q^{2}}{M_{k}^{2}} \right) \nonumber \\
& \phantom{= - (4 \pi )^{2} \int _{0}^{1} } 
+ \frac{4 \tilde{q}_{k}^{2} s - 4 y (1 - y) Q^{2} M_{k}^{2} 
- (s + y (1 - y) Q^{2}) (s - M_{k}^{2} - m_{k}^{2}) }
{2 (s + y (1 - y) Q^{2} ) 
\sqrt{ 4 \tilde{q}_{k}^{2} s - 4 y (1 - y) Q^{2} M_{k}^{2} + i \epsilon}}
L \left( \sqrt{s} ; \, Q^{2} ; \, M_{k}, \, m_{k} \right)
\Bigg ], 
\label{eq:app-loopDM0} 
\end{align}
\begin{align}
D_{\text{B}_{k}}^{0} \! \left( \sqrt{s};\, Q^2 \right) 
&= - \frac{1}{(4 \pi )^2} \int_{0}^{1} d x \int_{0}^{1} d y 
\frac{4 x (1 - x) Q_{\text{B}_{k}} M_{k} \sqrt{s + Q^{2}/4}}
{x M_{k}^2 + (1 - x) m_{k}^2 - x (1 - x) s + x^{2} 
y (1 - y) Q^{2} - i \epsilon }, \nonumber \\
&=- \frac{Q_{\text{B}_{k}}}{(4 \pi )^2} \int_{0}^{1} d y 
\frac{4 M_{k} \sqrt{s + Q^{2}/4}}
{s + y (1 - y) Q^{2}} 
\Bigg [ -1 + \frac{M_{k}^{2} - m_{k}^{2} + y (1 - y) Q^{2}}
{2 ( s + y (1 - y) Q^{2} )} 
\ln \left( \frac{M_{k}^{2} + y (1 - y) Q^{2}}{m_{k}^{2}} \right) \nonumber \\
& \phantom{= - (4 \pi )^{2} \int _{0}^{1} } 
+ \frac{4 \tilde{q}_{k}^{2} s - 4 y (1 - y) Q^{2} m_{k}^{2} 
- (s + y (1 - y) Q^{2}) (s - M_{k}^{2} - m_{k}^{2}) }
{2 (s + y (1 - y) Q^{2} ) 
\sqrt{ 4 \tilde{q}_{k}^{2} s - 4 y (1 - y) Q^{2} m_{k}^{2} + i \epsilon}}
L \left( \sqrt{s} ; \, Q^{2} ; \, m_{k}, \, M_{k} \right)
\Bigg ], 
\label{eq:app-loopDB0} 
\end{align}
\begin{align}
D_{\text{B}_{k}}^{a} \! \left( \sqrt{s};\, Q^2 \right) 
&= - \frac{1}{(4 \pi ) ^2} \int_{0}^{1} d x \int_{0}^{1} d y 
\frac{4 x \mu _{k} M_{k}^{2}}
{x M_{k}^2 + (1 - x) m_{k}^2 - x (1 - x) s + x^{2} 
y (1 - y) Q^{2} - i \epsilon }  \left( \frac{i \bm{\sigma} \times 
\bm{q} }{2 M_{\text{p}}} \right) ^{a} , \nonumber \\
&=- \frac{\mu _{k}}{(4 \pi )^2} \left( \frac{i \bm{\sigma} \times 
\bm{q} }{2 M_{\text{p}}} \right) ^{a} 
\int_{0}^{1} d y 
\frac{2 M_{k}^{2}}{s + y (1 - y) Q^{2}} 
\nonumber \\
& \phantom{= - (4 \pi )^{2} } 
\times \Bigg [ \ln \left( \frac{M_{k}^{2} + y (1 - y) Q^{2}}{m_{k}^{2}} \right) 
- \frac{(s + m_{k}^{2} - M_{k}^{2}) }
{\sqrt{ 4 \tilde{q}_{k}^{2} s - 4 y (1 - y) Q^{2} m_{k}^{2} + i \epsilon}}
L \left( \sqrt{s} ; \, Q^{2} ; \, m_{k}, \, M_{k} \right)
\Bigg ], 
\label{eq:app-loopDBa} 
\end{align}
\begin{align}
L \left( \sqrt{s} ; \, Q^{2} ; \, M_{k}, \, m_{k} \right)
\equiv 
& 
\ln \left(s + M_{k}^{2} - m_{k}^{2} 
+ \sqrt{4 \tilde{q}_{k}^{2} s - 4 y (1 - y) Q^{2} M_{k}^{2} + i \epsilon} \right) 
\nonumber \\ &
+ \ln \left( s - M_{k}^{2} + m_{k}^{2} + 2 y (1 - y) Q^{2} 
+ \sqrt{4 \tilde{q}_{k}^{2} s - 4 y (1 - y) Q^{2} M_{k}^{2} + i \epsilon} \right) 
\nonumber \\ & 
- \ln \left( - s - M_{k}^{2} + m_{k}^{2} 
+ \sqrt{4 \tilde{q}_{k}^{2} s - 4 y (1 - y) Q^{2} M_{k}^{2} + i \epsilon} \right) 
\nonumber \\ & 
- \ln \left( - s + M_{k}^{2} - m_{k}^{2} - 2 y (1 - y) Q^{2} 
+ \sqrt{4 \tilde{q}_{k}^{2} s - 4 y (1 - y) Q^{2} M_{k}^{2} + i \epsilon} \right) . 
\label{eq:app-loglog}
\end{align}
\end{widetext}
Here $Q_{\text{M}_{k}}$, $Q_{\text{B}_{k}}$, and $\mu _{\text{B}_{k}}$ are the charge 
of the meson and baryon, and magnetic moment of the baryon in $k$ channel, 
respectively, and $\tilde{q}_{k}$ is defined in Eq.~\eqref{eq:q_k} so that $4 \tilde{q}_{k}^{2} s$ is simply written as, 
\be
4 \tilde{q}_{k}^{2} s = \lambda (s, \, m_{k}^{2},\, M_{k}^{2}) , 
\ee
with the K\"{a}llen function 
$\lambda (x,\, y,\, z)=x^{2}+y^{2}+z^{2}-2xy-2yz-2zx$. 
We note that spatial components of the loop integrals in which 
photon couples to mesons vanish in the Breit frame, 
$D_{\text{M}}^{a}=0\, (a=1, \, 2, \, 3)$, 
since the pseudoscalar mesons do not have the magnetic moments. 
In the 
above equations the $x$ integrations are analytically performed; 
the $y$ integrations can in principle be done analytically, 
but we perform the numerical integration in the practical calculation 
with keeping $i \epsilon$ to be a small finite value, which ensures 
the correct boundary condition during the analytic continuation. 

As one can see from Eqs.~\eqref{eq:app-loopDM0}--\eqref{eq:app-loglog}, 
the integrands in $D_{\text{M},\text{B}}^{\mu}$ are obtained in 
the analytic form for the energy $\sqrt{s}$ and the squared photon momentum 
$Q^{2}$, hence we can make an analytic continuation from the real 
$\sqrt{s}$ to the complex value, $\sqrt{s} \to z$, for the integrands. 

In our study we perform the analytic continuation of $D_{\text{M},\text{B}}^{\mu}$ 
in a following way. At first, we fix $Q^{2}=0$ and make $\sqrt{s}$ a complex 
value, $\sqrt{s} \to \sqrt{s} + i b=z$ ($b$: real) so that 
$D_{\text{M},\text{B}}^{\mu}$ is continuous with respect to $b$. In this 
condition, we are in a first (second) Riemann sheet with respect to 
$\sqrt{s}$ with $b>0$ ($b<0$) if the energy $\sqrt{s}$ is above the 
meson-baryon threshold. Then, we make $Q^{2}$ vary from $0$ to finite real 
values so that $D_{\text{M},\text{B}}^{\mu}$ are continuous with respect 
to $Q^{2}$. 

In order to visualize this, we plot the sum of 
the loop integrals $D_{\text{M}}^{\mu}+D_{\text{B}}^{\mu}$, which appears in the 
amplitude $T_{\gamma}^{\mu}$ as in Eqs.~\eqref{eq:Tgamma}--\eqref{eq:Tgamma2}, 
in the $Q^2$-$\text{Im}[z]$ plane.  In Fig.~\ref{fig:app-Loop} 
we show time component $D_{\text{M}}^{0}+D_{\text{B}}^{0}$ in $K^{-}p$ 
and $\pi ^{+} \Sigma ^{-}$ channels with the
real part of the energy being fixed as $\text{Re}[z]=1420 \mev$,
around the pole position of the $\LamFOF$.  Here we note that in 
all 
meson-baryon channels for total charge-zero system, the functions 
$D_{\text{M}}^{0}+D_{\text{B}}^{0}$ vanish at $Q^{2}=0$ at any complex 
energies $z$, because at $Q^{2}=0$ 
the relations~\eqref{eq:DMGM} and \eqref{eq:DBGB} lead 
to $D_{\text{M}}^{0}+D_{\text{B}}^{0}\propto Q_{\text{EM}}=0$.

Now we discuss the behavior of the $K^{-}p$ loop integral.  One notice that
$\text{Im}[D_{\text{M}}^{0}+D_{\text{B}}^{0}]=0$ with real-valued
energy, that is, $\text{Im}[z]=0$. This is interpreted as the fact
that $K^{-}$ and $p$ cannot go on-shell simultaneously with the energy
below the $K^{-}p$ threshold, $\sqrt{s} < m_{K^{-}} +
M_{\text{p}}$. With the real-valued energy ($\text{Im}[z]=0$), the
real part $\text{Re}[D_{\text{M}}^{0}+D_{\text{B}}^{0}]$ decreases as
we increase $Q^{2}$.  This property of decreasing 
$\text{Re}[D_{\text{M}}^{0}+D_{\text{B}}^{0}]$ is
visible until around $\text{Im}[z] \simeq \pm 20 \mev$ in the analytic
continuation to imaginary energy, $\sqrt{s} \to z$.  
For the $K^{-}p$ channel, $D_{\text{M}}^{0}+D_{\text{B}}^{0}$ 
is smooth in the $Q^2$-$\text{Im}[z]$ plane. 

The $\pi ^{+}\Sigma ^{-}$ channel of the loop integral, on the other
hand, has some nontrivial behaviors. One is the divergence of the loop
integral at finite $Q^{2}$ with real $\sqrt{s}$, which is clearly seen
in the imaginary part $\text{Im}[D_{\text{M}}^{0}+D_{\text{B}}^{0}]$, and 
another is the discontinuity in the real part 
$\text{Re}[D_{\text{M}}^{0}+D_{\text{B}}^{0}]$ running above the same $Q^{2}$.  
The divergences of $D_{\text{M}}^{0}$ and $D_{\text{B}}^{0}$ take place, as 
will be discussed later, at
\be
Q^{2}=\frac{4 \tilde{q}_{\pi \Sigma}^{2}\, s}{M_{\Sigma}^{2}} , 
\quad 
\frac{4 \tilde{q}_{\pi \Sigma}^{2}\, s}{m_{\pi}^{2}} , 
\ee
respectively. 
Hence, the loop integral with $Q^{2}\ge 0$ diverges only when 
the real-valued energy $\sqrt{s}$ is above the two-body threshold, 
so that $4 \tilde{q}^{2} s>0$. 
Furthermore, this divergence point corresponds to the 
$t$-channel threshold; with fixed $\sqrt{s}$ above the threshold, 
all the three propagating particles can simultaneously go 
on-shell in $D_{\text{M}}$ ($D_{\text{B}}$) for
$Q^{2}<4 \tilde{q}^{2} s/M^{2}$ 
($Q^{2}<4 \tilde{q}^{2} s/m^{2}$), 
while for
$Q^{2}>4 \tilde{q}^{2} s/M^{2}$ 
($Q^{2}>4 \tilde{q}^{2} s/m^{2}$) two 
propagating mesons (baryons) cannot simultaneously go
on-shell in $D_{\text{M}}$ ($D_{\text{B}}$) if we constrain the baryon 
(meson) to go on-shell. 

The origin of the divergences of $D_{\text{M}_{k}}$ and 
$D_{\text{B}_{k}}$ is the factor 
$1/\sqrt{4\tilde{q}_{k}^{2}s-4y(1-y)Q^{2}\bar{m}_{k}^{2}+i\epsilon}$ in the 
integrands of Eqs.~\eqref{eq:app-loopDM0}--\eqref{eq:app-loopDBa}, 
where $\bar{m}_{k}$ represents the meson mass 
$m_{k}$ (the baryon mass $M_{k}$) when the photon couples to the baryon (meson) 
propagator. By dropping the irrelevant factors, which do not cancel 
singularity from 
$1/\sqrt{4\tilde{q}_{k}^{2}s-4y(1-y)Q^{2}\bar{m}_{k}^{2}+i\epsilon}$, 
and performing the $y$ integration, we obtain, 
\begin{align}
I_{k}^{\text{div}} & \equiv \int _{0}^{1} d y \frac{2 Q \bar{m}_{k}}
{\sqrt{4\tilde{q}_{k}^{2}s-4y(1-y)Q^{2}\bar{m}_{k}^{2} + i \epsilon}} 
\nonumber \\ & 
= \ln 
\left( \frac{2 \tilde{q}_{k} \sqrt{s} + Q \bar{m}_{k}}
{2 \tilde{q}_{k} \sqrt{s} - Q \bar{m}_{k}} \right) , 
\label{eq:app-Idiv}
\end{align}
which contains the logarithmic divergence at 
$Q\bar{m}_{k}=\pm 2\tilde{q}_{k}\sqrt{s}$. 
Therefore, the loop integral $D_{\text{M}_{k}}$ ($D_{\text{B}_{k}}$) 
logarithmically 
diverges at $Q^{2}=4\tilde{q}_{k}^{2} s/M_{k}$ 
($Q^{2}=4\tilde{q}_{k}^{2} s/m_{k}$) with $\sqrt{s}>m_{k}+M_{k}$. 

\begin{figure*}[!Ht]
  \centering
  \begin{tabular*}{\textwidth}{@{\extracolsep{\fill}}cc}
    \includegraphics[width=8.6cm]{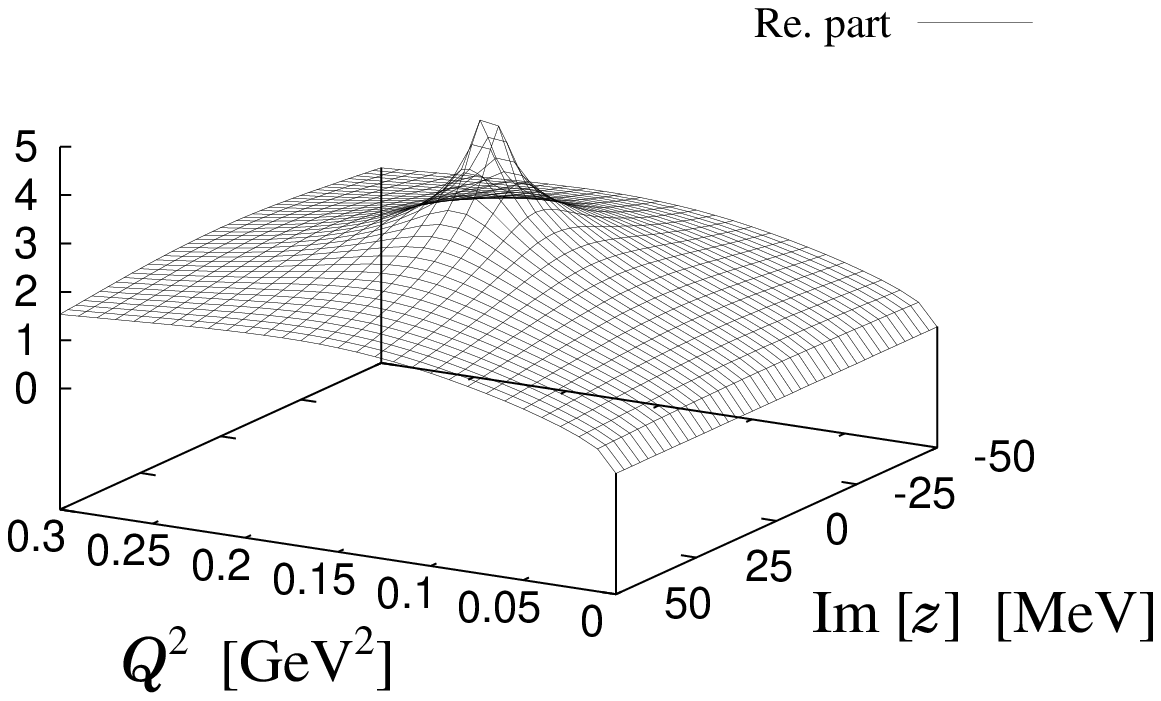} &
    \includegraphics[width=8.6cm]{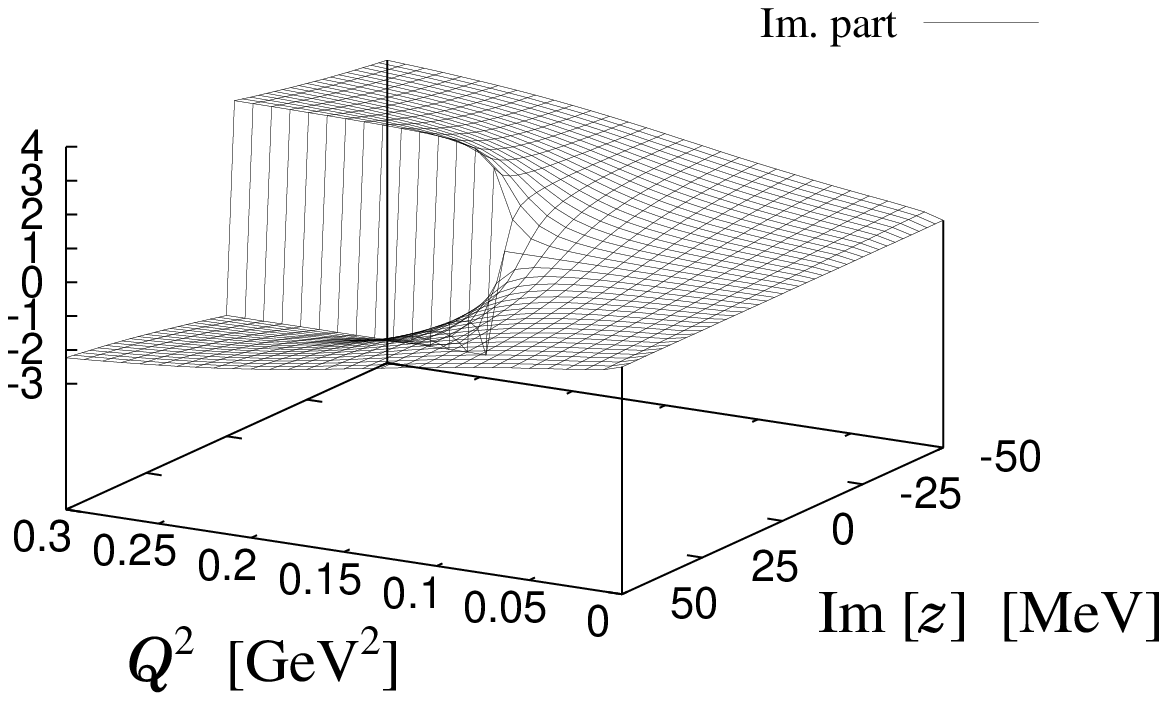} 
  \end{tabular*}
  \caption{The real part (left) and the imaginary (right) parts of the
    $I_{k=\pi \Sigma}^{\text{div}}$ \eqref{eq:app-Idiv} as
    functions of $\text{Im}[z]$ and $Q^{2}$. The real part of the
    energy and $\bar{m}_{k}$ are chosen to be $\text{Re}[z]=1420 \mev$ and 
    $\bar{m}_{k}=M_{\Sigma}$, respectively. }
\label{fig:app-Idiv}
\end{figure*}

Now it is instructive to investigate the behavior of $I_{k}^{\text{div}}$ 
with respect to $\text{Im}[z]$ and $Q^{2}$. In Fig.~\ref{fig:app-Idiv} 
we plot $I_{k=\pi \Sigma}^{\text{div}}$ as a function of $\text{Im}[z]$ and $Q^{2}$ 
with conditions $\bar{m}_{k}=M_{\Sigma}$ and $\text{Re}[z]=1420 \mev$. Here 
analytic continuation is performed in the same way as for $D_{\text{M}}$ 
and $D_{\text{B}}$. From Fig.~\ref{fig:app-Idiv} we can see the real 
part of $I_{k=\pi \Sigma}^{\text{div}}$ 
contains divergence point at $\text{Im}[z]=0$ 
and $Q^{2}= 4 \tilde{q}_{\pi \Sigma}^{2}\, s/M_{\Sigma}^{2}\simeq 0.16 \gev ^{2}$, 
whereas the imaginary part of $I_{k=\pi \Sigma}^{\text{div}}$ shows discontinuity 
along the real energy axis above the same 
$Q^{2}$ value ($\simeq 0.16 \gev ^{2}$). These properties 
are explained by using analytic properties of a logarithm function 
$\ln (x)$. Namely, taking 
limit $(2 \tilde{q}_{\pi \Sigma}\, \sqrt{s} - Q M_{\Sigma}) \to 0$ from 
any complex energy $z$ and any real $Q$, we have, 
\be
\ln (2 \tilde{q}_{\pi \Sigma}\, \sqrt{s} - Q M_{\Sigma}) 
= \ln |2 \tilde{q}_{\pi \Sigma}\, \sqrt{s} - Q M_{\Sigma}| 
+ i \theta , 
\ee
with argument of $(2 \tilde{q}_{\pi \Sigma}\, \sqrt{s} - Q M_{\Sigma})$, 
$\theta$, 
which is finite, hence the divergence appears only in the real part of 
$I_{k=\pi \Sigma}^{\text{div}}$. 
Further, 
in order to see the behavior of the imaginary part of 
$I_{k= \pi \Sigma }^{\text{div}}$, we first fix the energy with an infinitesimal 
imaginary part as $\sqrt{s} \pm i \epsilon$ ($\epsilon >0$). Then, 
from $\sqrt{s}$ the divergence point of $I_{k= \pi \Sigma }^{\text{div}}$ is 
fixed as $Q_{0} = \lambda _{\pi \Sigma}^{1/2} (\sqrt{s}) / M_{\Sigma}$ with 
the square root of the K\"{a}llen function 
$\lambda _{\pi \Sigma}^{1/2} (\sqrt{s}) = 2 \tilde{q}_{\pi \Sigma} \sqrt{s}$. 
Now let us consider $I_{k= \pi \Sigma }^{\text{div}}$ with the energy 
$\sqrt{s} \pm i \epsilon $ and momentum $Q=Q_{0}+\Delta Q$, where 
$\Delta Q$ is finite positive value.  Since the K\"{a}llen function 
gives a relation, 
\be
\lambda _{\pi \Sigma}^{1/2} (\sqrt{s} \pm i \epsilon ) 
= \lambda _{\pi \Sigma}^{1/2} (\sqrt{s}) \pm i \delta , 
\ee
where $\delta$ is an infinitesimal positive value determined by 
the $\sqrt{s}$ and $\epsilon$ and has a property $\lim _{\epsilon \to 0} 
\delta = 0$, the numerator and denominator inside the logarithm 
in Eq.~\eqref{eq:app-Idiv} become: 
\begin{align}
& \lambda _{\pi \Sigma}^{1/2} (\sqrt{s} \pm i \epsilon ) 
+ Q M_{\Sigma} = ( 2 Q_{0} + \Delta Q ) M_{\Sigma} \pm i \delta , 
\\
& \lambda _{\pi \Sigma}^{1/2} (\sqrt{s} \pm i \epsilon ) 
- Q M_{\Sigma} = - \Delta Q M_{\Sigma} \pm i \delta , 
\end{align}
respectively.  Therefore, $I_{k= \pi \Sigma }^{\text{div}}$ has the limit, 
\begin{align}
I_{k=\pi \Sigma}^{\text{div}} = & 
\ln \left ( \frac{(2 Q_{0} + \Delta Q) M_{\Sigma} \pm i \delta}
{- \Delta Q M_{\Sigma} \pm i \delta} \right ) 
\nonumber \\ 
\to & \ln \left | \frac{2 Q_{0} + \Delta Q}{\Delta Q} \right | 
\mp i \pi \quad (\epsilon , \, \delta \to 0), 
\end{align}
hence the discontinuity appears in the imaginary part of 
$I_{k=\pi \Sigma}^{\text{div}}$. 

As a consequence of the factor in the integrands, 
$1/\sqrt{4\tilde{q}_{k}^{2}s-4y(1-y)Q^{2}\bar{m}_{k}^{2}+i\epsilon}$, 
$D_{\text{M}_{k}}$ and $D_{\text{B}_{k}}$ in the channels with energy 
above the threshold contains divergence in the imaginary part at 
$Q^{2}=4\tilde{q}_{k}^{2}s/\bar{m}_{k}^{2}$ and the discontinuity in the real 
part along the real energy axis above the same $Q^{2}$ value. 
Here we note that the $L$ function~\eqref{eq:app-loglog} is 
pure imaginary so that the discontinuity (divergence) of the 
imaginary (real) part of $I_{k}^{\text{div}}$ is manifested in the 
real (imaginary) part of the loop integrals $D_{\text{M}_{k}, \text{B}_{k}}$. 

\begin{figure*}[t]
  \centering
  \begin{tabular*}{\textwidth}{@{\extracolsep{\fill}}cc}
    \includegraphics[width=8.6cm]{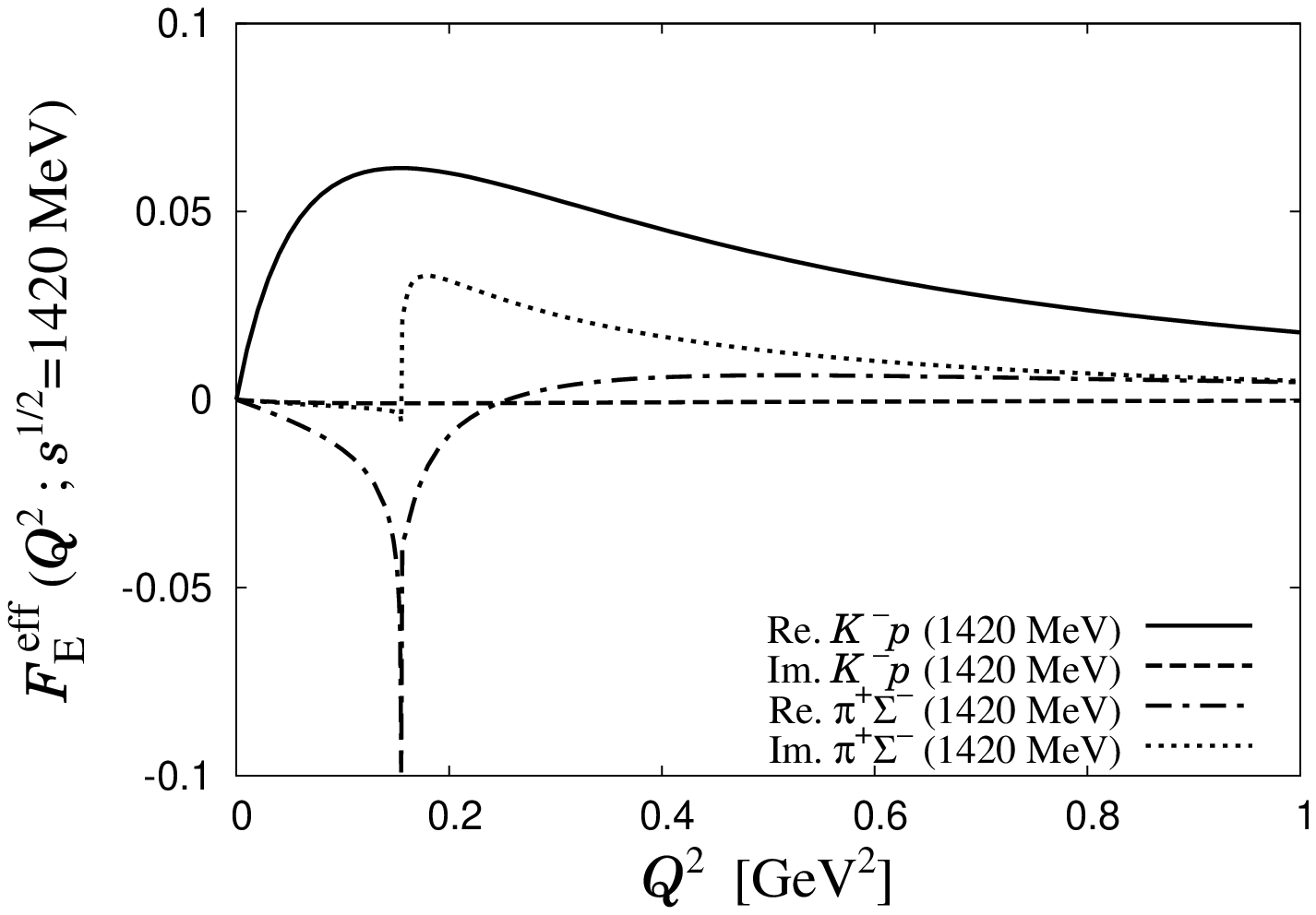} &
    \includegraphics[width=8.6cm]{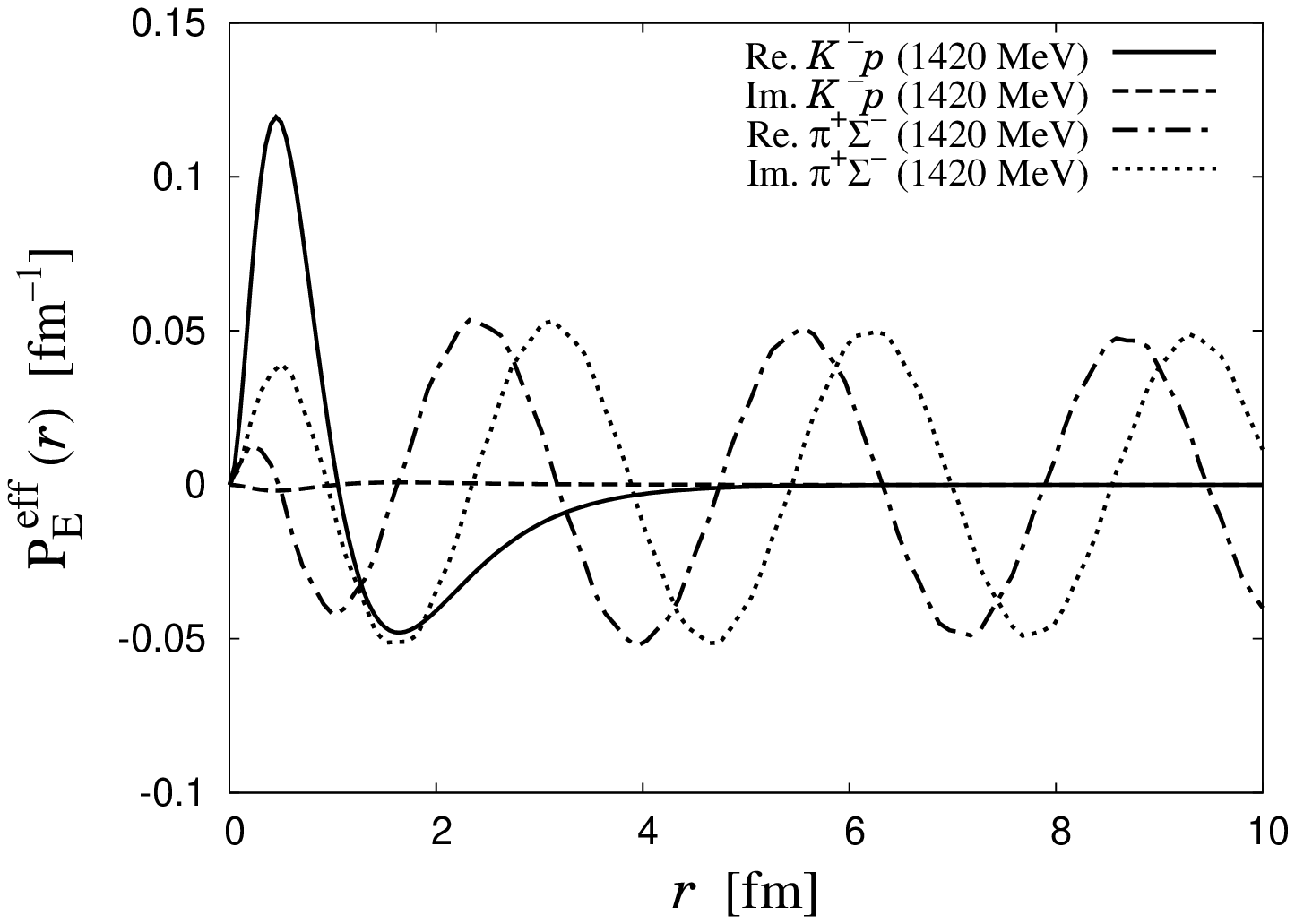} 
  \end{tabular*}
  \caption{Meson-baryon components of the 
    effective electric form factors ($\FEeff$, left) and 
    effective electric density distribution ($\PE ^{\text{eff}}$, right) 
    on the real energy axis. 
    Calculations are performed with the center-of-mass energy 
    $\sqrt{s}=1420 \mev$. 
  }
  \label{fig:app-eff}
\end{figure*}

These diverging structures in the loop functions
$D_{\text{M},\text{B}}$ eventually lead to the divergences of the effective 
form factors~\eqref{eq:Feffective} at $Q^{2}=4\tilde{q}^{2}s/\bar{m}^{2}$, 
and to the peak structures of the form factors~\eqref{eq:Res_scheme} 
around that $Q^{2}$. Then, the ``peaks'' of the (effective) 
form factors in the momentum space lead to the oscillation components
in the density distributions in the coordinate space with period given
by $2 \pi / (2 \tilde{q} \sqrt{s}/\bar{m})=\pi \bar{m} / (\tilde{q} \sqrt{s})$ 
through the Fourier transformation. 
Therefore, for the channel whose threshold is lower than the total
energy $\sqrt{s}$, we will have an oscillation component of the
``decaying part'', which can be interpreted as the decay of the system
into the open channels 
with kicked meson and baryon to the on-shell 
through the photon coupling.

Here let us show the behaviors of $K^{-}p$ and $\pi ^{+} \Sigma ^{-}$ 
components of effective form factor and density 
distribution in electric probe with energy fixed as $\sqrt{s}=1420 \mev$. 
The form factor and density distribution on the resonance pole position 
are shown in Sec.~\ref{subsec:on_pole}. As you can see, the effective 
electric form factor in $\pi ^{+} \Sigma ^{-}$ channel shows divergence 
at $Q^{2}=4 \tilde{q}_{\pi \Sigma}^{2}\, s/M_{\Sigma}^{2} 
\simeq 0.16 \gev ^{2}$, which generates an oscillation behavior 
in the effective electric density distribution with the period close to 
$\pi M_{\Sigma} / (\tilde{q} \sqrt{s}) \simeq 3.1 \fm$.  

Finally we note that the oscillation part will be clearly seen only in the case 
that the probe current couples to the mesons. In this case 
$\bar{m}$ corresponds to the baryon mass $M$, hence the value 
$4 \tilde{q}^{2} s/M^{2}$ appears around typical hadronic scale 
$Q^{2} \lesssim 1 \, \text{GeV}^{2}$, which will be important for the dynamics 
of the hadrons. This is not the case when the probe current 
couples to baryons, since in this case the value $4 \tilde{q}^{2} s/m^{2}$ 
becomes much larger than $1 \gev ^{2}$, due to the light meson 
masses $m$, especially for pions, $m=m_{\pi}$, and such a high $Q^{2}$ 
coupling should be strongly suppressed by form factors of constituent 
hadrons.


\end{document}